\documentclass[11pt,a4paper]{article}

\pdfoutput=1
\usepackage{cite}
\usepackage{graphicx}
\usepackage{jheppub}
\usepackage{amsmath}
\usepackage{amssymb}
\usepackage{enumerate}
\usepackage{multirow}
\usepackage{lscape}

\newcommand{\dilog}[2]{\ensuremath{{\rm Li}_2\left(-\frac{#1}{#2}\right)}}

\def\beq{\begin{equation}}
\def\eeq{\end{equation}}
\def\e{\epsilon}
\def\bsp#1\esp{\begin{split}#1\end{split}}
\def\beqa{\begin{eqnarray}}
\def\eeqa{\end{eqnarray}}
\def\eqn#1{eq.~(\ref{#1})}

\def\sect#1{Sec.~{\ref{#1}}}

\def\eqn#1{eq.~(\ref{#1})}

\def\Eqns#1#2{eqs.~(\ref{#1}) and~(\ref{#2})}

\title{Multiple Gluon Exchange Webs}

\author[a]{Giulio Falcioni,}
\author[b]{Einan Gardi,}
\author[b]{Mark Harley,}
\author[a]{Lorenzo Magnea,}
\author[c]{Chris D. White}

\affiliation[a]{Dipartimento di Fisica, Universit{\`a} di Torino, and \\ INFN, 
Sezione di Torino, Via P. Giuria 1, I-10125 Torino, Italy}
\affiliation[b]{Higgs Centre for Theoretical Physics, School of Physics 
and Astronomy,\\ The University of Edinburgh, Edinburgh EH9 3JZ, Scotland, UK}
\affiliation[c]{School of Physics and Astronomy, 
Scottish Universities Physics Alliance, University of Glasgow, \\
Glasgow, G12 8QQ, Scotland, UK}

\emailAdd{Giulio.Falcioni@unito.it}
\emailAdd{Einan.Gardi@ed.ac.uk}
\emailAdd{Mark.Harley@ed.ac.uk}
\emailAdd{Lorenzo.Magnea@unito.it}
\emailAdd{Christopher.White@glasgow.ac.uk}

\abstract{Webs are weighted sets of Feynman diagrams which build up
  the logarithms of correlators of Wilson lines, and provide the
  ingredients for the computation of the soft anomalous dimension.  
  We present a general analysis of multiple gluon exchange webs 
  (MGEWs) in correlators of semi-infinite non-lightlike Wilson lines, 
  as functions of the exponentials of the Minkowski cusp angles,
  $\alpha_{ij}$, formed between lines~$i$ and~$j$. We compute a range
  of webs in this class, connecting up to five Wilson lines through
  four loops, we give an all-loop result for a special class of diagrams,
  and we discover a new kind of relation between webs connecting 
  different numbers of Wilson lines, based on taking collinear 
  limits. Our results support recent conjectures, stating that
  the contribution of any MGEW to the soft anomalous dimension is a
  sum of products of polylogarithms, each depending on a single cusp
  angle, and such that their symbol alphabet is restricted to
  $\alpha_{i j}$ and $1 - \alpha_{i j}^2$.  Finally, we construct a
  simple basis of functions, defined through a one-dimensional
  integral representation in terms of powers of logarithms, which has
  all the expected analytic properties. This basis allows us to
  compactly express the results of all MGEWs computed so far, and 
  we conjecture that it is sufficient for expressing all MGEWs at any
  loop order.}

\keywords{Gauge theories, perturbative QCD, soft singularities, anomalous 
dimensions.}

\begin{document}

\begin{flushright}
Edinburgh 2014/12
\vspace*{-25pt}
\end{flushright}
\maketitle
\allowdisplaybreaks


\section{Introduction}
\label{Intro}

Infrared singularities of scattering amplitudes in non-Abelian gauge theories 
are essential for understanding the physics of the strong interactions. Determining 
infrared singularities is necessary in order to combine real and virtual corrections 
in cross section calculations. These singularities also dictate the structure of large 
logarithmic contributions in perturbation theory, which must in many cases be 
resummed in order to obtain reliable phenomenological predictions. Besides 
their significance to phenomenology, infrared singularities are very interesting from 
a theoretical perspective: first of all, they have a similar structure across all gauge 
theories; furthermore, their relative simplicity~\cite{Catani:1998bh,Sterman:2002qn,
Aybat:2006wq,Aybat:2006mz,Becher:2009cu,Becher:2009qa,Gardi:2009qi}, their 
exponentiation properties~\cite{Yennie:1961ad,Sterman:1981jc,Gatheral:1983cz,
Frenkel:1984pz,Magnea:1990zb,Magnea:2000ss,Gardi:2010rn,Mitov:2010rp,Gardi:2011wa,
Gardi:2011yz,Dukes:2013wa,Dukes:2013gea,Gardi:2013ita}, and their relation 
with the renormalization properties of Wilson line correlators~\cite{Polyakov:1980ca,
Arefeva:1980zd,Dotsenko:1979wb,Brandt:1981kf,Korchemsky:1985xj,Korchemsky:1985xu,
Korchemsky:1987wg} allow us to explore multi-loop and even all-order properties, 
which are beyond reach when studying complete scattering amplitudes. In ${\cal N} = 4$ 
supersymmetric Yang-Mills theory, where much contemporary focus is on the planar 
limit, infrared singularities give access to non-planar effects (see, for example, 
Refs.~\cite{Naculich:2009cv,Naculich:2011pd,Naculich:2013xa}), and they provide 
a link between the weak and strong coupling regimes~\cite{Alday:2007hr,Basso:2007wd,
Correa:2012nk,Henn:2012qz,Correa:2012at,Henn:2013wfa,Erdogan:2011yc,
Cherednikov:2012yd,Cherednikov:2012qq}.

It is by now well understood that infrared divergent contributions to scattering 
amplitudes are universal, and can be mapped~\cite{Korchemsky:1985xj} to 
ultraviolet divergences of correlators of Wilson line operators, which have been
extensively studied both in gauge and gravity theories~\cite{Korchemsky:1985xu,
Korchemsky:1987wg,Korchemsky:1985xj,Brandt:1981kf,Korchemsky:1992xv,
Korchemsky:1993uz,Korchemskaya:1994qp,Korchemskaya:1992je,Akhoury:2011kq,
Naculich:2011ry,Miller:2012an,White:2011yy,Melville:2013qca}. For fixed-angle 
scattering amplitudes, which are the focus of the present paper, these correlators 
involve semi-infinite Wilson lines, pointing in the directions defined by the classical 
trajectories of the hard partons participating in the scattering. The renormalisation 
properties of such operators are encoded in the \emph{soft anomalous 
dimension}, which is a matrix in the space of colour configurations available 
for the scattering process at hand. In processes involving only two Wilson 
lines, this reduces to the well-known cusp anomalous dimension, which is 
a function of the Minkowski angle between the two lines. In QCD, this function 
has been calculated up to two-loop order~\cite{Korchemsky:1987wg} (see
also~\cite{Kidonakis:2009ev}), and progress towards a three loop result was 
reported in Ref.~\cite{Grozin:2014axa}. In ${\cal N}=4$ Super-Yang-Mills theory
it is known to three loops~\cite{Correa:2012nk,Correa:2012at}, and partial results 
have been recently obtained at four loops~\cite{Henn:2012qz,Henn:2013wfa}.
For multi-parton amplitudes, the soft anomalous dimension has been calculated 
up to two-loop order for both massless~\cite{Aybat:2006wq, Aybat:2006mz} and
massive~\cite{Ferroglia:2009ep,Ferroglia:2009ii,Mitov:2010xw,Chien:2011wz} 
Wilson lines (see also~\cite{Mitov:2009sv,Becher:2009kw,Beneke:2009rj,
Czakon:2009zw,Chiu:2009mg,Ferroglia:2010mi,Gardi:2013saa}). 

In the case of massless partons, the two-loop result~\cite{Aybat:2006wq, 
Aybat:2006mz} turns out to be remarkably simple, replicating the colour-dipole 
structure of the one-loop correction. This motivated further investigations, which 
led to a compact ansatz (the \emph{dipole formula}) for the all-order structure of 
infrared singularities in massless gauge theories~\cite{Becher:2009cu,
Becher:2009qa,Gardi:2009qi}. The dipole formula can only receive corrections 
starting at three loops, with at least four hard partons.  The nature of these 
corrections was investigated in~\cite{Dixon:2008gr,Gardi:2009qi,Dixon:2009gx,
Becher:2009cu,Becher:2009qa,Dixon:2009ur,Gardi:2009zv,Gehrmann:2010ue,
Bret:2011xm,DelDuca:2011ae,Ahrens:2012qz,Naculich:2013xa,Caron-Huot:2013fea}, 
where they were shown to be highly constrained by non-Abelian exponentiation and 
Bose symmetry, as well as by the known behaviours of amplitudes in collinear limits 
and in the Regge limit~\cite{Bret:2011xm,DelDuca:2011ae,DelDuca:2013ara}; recently, 
the first evidence for a non-vanishing correction to the dipole formula at the four-loop 
level was inferred in Ref.~\cite{Caron-Huot:2013fea}, using the Regge limit.

In the massive case, much of this simplicity is lost, as `colour tripole' corrections, which 
are forbidden to all loops in the massless case, are present already at the two-loop
level~\cite{Ferroglia:2009ep,Ferroglia:2009ii,Mitov:2010xw}.  The result for the soft 
anomalous dimension retains however interesting features, in particular it has a 
factorized form as a function of the Minkowski angles between Wilson lines. The 
massless limit, where the tripole correction vanishes, involves a subtle cancellation 
between different diagrammatic contributions, which have different analytic behaviour 
in the massive case.

In order to understand the structure of infrared divergences of both massive and massless 
partons at three loops and beyond, it is ultimately necessary to tackle the direct calculation 
of correlators of multiple Wilson lines at the multi-loop level. Such correlators exponentiate, 
and it is then clearly beneficial to be able to calculate the exponent of the correlator 
\emph{directly}. This problem has been addressed over the past few years in a series of 
papers~\cite{Laenen:2008gt,Mitov:2010rp,Gardi:2010rn,Gardi:2011wa,Gardi:2011yz,
Gardi:2013ita,Gardi:2013saa} (see also~\cite{Vladimirov:2014wga}), which have developed 
a diagrammatic approach to the non-Abelian exponentiation of products of Wilson lines. It is 
then also necessary to develop techniques for carrying out the relevant Feynman integrals. 
Recently, there has been significant progress in this direction using configuration-space 
techniques both for strictly lightlike Wilson lines~\cite{Erdogan:2011yc} and for non-lightlike 
lines~\cite{Henn:2012qz,Henn:2013wfa,Gardi:2013saa,Grozin:2014axa}. The first three-loop 
results involving \emph{multiple} Wilson lines have been published in Ref.~\cite{Gardi:2013saa}, 
where a class of diagrams contributing to the four-line soft anomalous dimension were evaluated. 
This also constitutes a step towards the determination of the three-loop soft anomalous dimension. 
In parallel progress has been made also on other three-loop  diagrams~\cite{Gardi:2014kpa} 
and the complete result in the lightlike limit will soon be available.

The diagrammatic approach to exponentiation can be summarized as follows. Correlators 
of multiple Wilson lines exponentiate, and their logarithms can be computed directly in terms 
of \emph{webs}. The concept of web was introduced in formulating the non-Abelian 
exponentiation theorem in the simpler case of a Wilson loop (or two Wilson lines meeting 
at a cusp) in Refs.~\cite{Gatheral:1983cz,Frenkel:1984pz,Sterman:1981jc}. In this case
the diagrams that contribute to the exponent are \emph{irreducible}: their colour factors 
cannot be decomposed into products of colour factors of their subdiagrams. Importantly, 
irreducible diagrams are also free of ultraviolet subdivergences associated with the 
renormalization of the cusp. For multiple Wilson lines this classification does not hold: 
reducible diagrams do contribute to the exponent. Consequently, the concept of web has 
to be substantially  generalised. This was done in Refs.~\cite{Gardi:2010rn,Mitov:2010rp,
Gardi:2011wa,Gardi:2011yz,Gardi:2013ita,Gardi:2013saa}, which generalised the non-Abelian 
exponentiation theorem to the multi-line case. Each web $W$ then comprises a set of 
Feynman diagrams $\{D\}$, which is closed under permutations of the gluon attachments 
to the Wilson lines; writing each diagram $D$ as a product ${\cal F}(D) C(D)$, where $C(D)$ 
is the colour factor and ${\cal F}(D)$ is the kinematic integral, one finds that the contribution 
of each web to the logarithm of the Wilson line correlator is given by 
\beq 
  W \, = \, \sum_{D,D' \in W} {\cal F}(D) \, R_{DD'} \, C(D') \, ,
\label{Wdef}
\eeq 
where the sum is over all pairs of diagrams $D$ and $D'$ belonging to the web 
$W$. Kinematic and colour factors of individual diagrams mix with each other 
according to a \emph{web-mixing matrix} $R_{DD'}$, whose entries are rational 
numbers of purely combinatorial origin\footnote{The combinatorics of web-mixing 
  matrices have recently been related to the properties of order-preserving maps on
  partially-ordered sets~\cite{Dukes:2013wa,Dukes:2013gea}.}. Furthermore, as 
shown in Ref.~\cite{Gardi:2013ita}, the linear combinations of colour factors generated 
through the action of the mixing matrix in \eqn{Wdef} always correspond to connected 
graphs. This is the essence of the non-Abelian exponentiation theorem for multiple 
Wilson lines.

The calculation of the logarithm of a Wilson-line correlator at a given order in 
perturbation theory proceeds then by classifying the possible webs, identifying 
which connected colour factors they contribute to, and finally computing the 
corresponding kinematic integrals. The goal of the calculation is identifying the 
ultraviolet singularities which contribute to the anomalous dimension. This, in 
turn, requires introducing an infrared regulator, which can be achieved using
an exponential suppression of long-distance interactions with the Wilson lines, 
as proposed in~\cite{Gardi:2011wa,Gardi:2013saa}. Renormalization, and
the non-commuting nature of the colour factors, imply that each web enters 
the anomalous dimension together with counterterms which remove its 
subdivergences~\cite{Mitov:2010rp,Gardi:2010rn,Gardi:2011wa,Gardi:2011yz}. 
The relevant combinations that determine the anomalous dimension were called 
\emph{subtracted webs} in~\cite{Gardi:2013saa}; these integrals have just a single 
ultraviolet pole, and they play the same role that individual webs  play in the two-line
case. As a consequence, subtracted webs are independent of the infrared regulator, and 
they share the same symmetries as the anomalous dimension~\cite{Gardi:2013saa}.  
A notable example of this is crossing symmetry, which is broken by the infrared 
regulator for non-subtracted webs, but is restored for subtracted webs. As a 
consequence, subtracted webs evaluate to much simpler analytic functions as 
compared to the individual diagrams (or the non-subtracted webs) which build 
them. There is therefore a marked advantage in organising the calculation in such 
a way that the last integrations be performed at the level of subtracted webs, as 
advocated in Ref.~\cite{Gardi:2013saa}.

In this paper we will discuss in detail the simplest class of webs contributing to 
multiple Wilson line correlators. These were dubbed ``Multiple Gluon Exchange 
Webs'' (MGEWs) in Ref.~\cite{Gardi:2013saa}, where some of their properties 
were studied, and two non-trivial three-loop examples were computed. MGEWs 
can be characterized in general as those webs that arise when the Wilson line 
correlator is computed using only the quadratic part of the quantum Yang-Mills 
Lagrangian in the path integral. In diagrammatic terms, they consist of graphs 
where all gluons attach directly to the Wilson lines, with no interaction vertices 
located off the Wilson lines. The graphs generated in this way are Abelian-like, in 
the sense that they would also appear in QED, however there is an essential difference 
between the two cases: in QED the order of emission from the Wilson lines is 
immaterial, and one can easily show that MGEWs collectively reconstruct powers 
of the one-loop result; indeed, according to the exponentiation theorem, MGE 
diagrams are not part of the exponent in QED. In contrast, in a non-Abelian
theory the ordering of gluon emission is crucial, and, as a consequence, 
MGEWs contribute to the exponent, where they collectively generate fully 
connected colour factors. 

Some key features of MGEWs were uncovered in Ref.~\cite{Gardi:2013saa},
based on a general analysis of the structure of their kinematic integrals, along with 
some physically motivated considerations concerning their analytic properties.  
Most notably, it was found that subtracted MGEWs can be expressed as sums 
of products of functions depending on individual cusp angles. The analytic structure 
of these functions is best elucidated by introducing, for each pair of Wilson lines 
$\{i,j\}$, variables $\alpha_{ij}$, corresponding to the exponential of the cusp 
angle $\xi_{ij}$ between the two lines, so that
\beq
  \xi_{ij}\, = \, \ln \alpha_{ij} \, = \, \cosh^{-1} \left( - \frac{\gamma_{ij}}{2} \right)
  \, , \qquad \qquad  \gamma_{ij} \, = \, - \, \alpha_{ij} - \frac{1}{\alpha_{ij}} \, ,
\label{alphadef}
\eeq
where $\gamma_{ij}$ is defined as the normalized scalar product between the 
4-velocities of the two Wilson lines, $\beta_i$ and $\beta_j$, 
\beq
  \gamma_{ij} \, \equiv \, \frac{2 \beta_i \cdot \beta_j + {\rm i} \varepsilon}{
  \sqrt{\beta_i^2 - {\rm i} \varepsilon} \sqrt{\beta_j^2 - {\rm i} \varepsilon}}
  \, = \, \frac{2 p_i \cdot p_j + {\rm i} \varepsilon}{\sqrt{p_i^2 - {\rm i} \varepsilon}
  \sqrt{p_j^2 - {\rm i} \varepsilon}} \, .
\label{gamma_ij}
\eeq
Here in the second equality we restored the dimensionful momenta of the partons 
represented by the Wilson lines, and we specified how kinematic invariants should 
be analytically continued.

The results for the subtracted webs considered in Ref.~\cite{Gardi:2013saa} were 
expressed in terms of a highly constrained set of functions, consisting of products of 
polylogarithms, each depending on a single $\alpha_{ij}$.  The analytic structure of 
these functions has been elucidated using the symbol map~\cite{Goncharov.A.B.:2009tja,
Goncharov:2010jf,Duhr:2011zq,Duhr:2012fh}: it was conjectured that the symbols of 
the functions entering subtracted MGEWs are built out of the restricted alphabet
$\{\alpha_{ij} , \eta_{ij} \equiv \alpha_{ij}/(1 - \alpha_{ij}^2)\}$. This alphabet, in 
particular, realises crossing symmetry, which in this case is expressed by the 
reflection $\alpha_{ij} \to - \alpha_{ij}$. It is clear that, whilst all relevant integrals 
can be evaluated without going through the symbol map, its use is nevertheless 
invaluable in simplifying results and understanding their analytic structure. 

The primary aim of this paper is to study the all-order structure of MGEWs in further 
detail, and to test the conjectures of Ref.~\cite{Gardi:2013saa} in a broader range of 
examples. Specifically, the subset of webs computed in Ref.~\cite{Gardi:2013saa}, 
connecting the maximal number of Wilson lines accessible at two and three loops, 
yields integrals which are less entangled than certain MGEWs connecting a smaller 
number of lines at the same order; the latter, more entangled ones, are computed 
and analysed here in order to confirm the conjectures. With the more complete 
understanding of MGEWs we gain here we are able to construct an ansatz for an
all-order basis of functions. These are defined through \emph{one-dimensional} 
integrals of powers of logarithms only. This yields a very restricted set of harmonic 
polylogarithms~\cite{Remiddi:1999ew} satisfying the alphabet conjecture and other 
constraints. We show that this basis is sufficient to express all the subtracted webs 
we compute in a compact manner, and we argue that it should be sufficient for 
MGEWs at higher orders as well. 

The structure of the paper is as follows. \sect{Webs} begins by
reviewing the method to compute the soft-anomalous dimension using
subtracted webs; we keep this discussion brief and we refer the reader
to Refs.~\cite{Gardi:2011yz, Gardi:2013ita,Gardi:2013saa} for a more
detailed account. In the final part of \sect{Webs} we discuss the
colour structure of webs using the effective vertex formulation
developed in Ref.~\cite{Gardi:2013ita}, and identify a new type of
relations between webs comprising different numbers of Wilson lines,
through a process of \emph{collinear reduction}. Examples of this
procedure will be given later on in the paper, where it is used as a
check of the results of specific webs.  Then, in \sect{MGEWs}, we
provide a general characterization of MGEWs: we give an integral
representation valid for any web in this class, before subtraction, in
terms of variables with a transparent physical interpretation, and we
review the conjectures proposed in~\cite{Gardi:2013saa}.  In
\sect{Basis}, we explain how the basis of functions for MGEWs is
constructed, and provide the necessary definitions, which will be used
in what follows to express the results of the various MGEWs we
compute. In the subsequent sections, we provide explicit calculations
of MGEWs to substantiate our arguments; the results are also important
as ingredients for the computation of the soft anomalous dimension at
three loops and beyond. In Sections~\ref{ToCusp} and~\ref{Threeloop}
we consider three-loop webs connecting two and three Wilson lines: the
two-line case, already studied in~\cite{Henn:2013wfa}, is interesting
in this context since it provides an example of maximal entanglement
of gluon insertions at this order. The results of Sect.~\ref{Threeloop} 
constitute another significant step forward in constructing the complete 
three-loop soft anomalous dimension, as well as providing an interesting 
comparison with the four line case of Ref.~\cite{Gardi:2013saa}.  In 
\sect{Fourloop}, we provide for the first time the complete calculation 
of a four-loop subtracted web, connecting five Wilson lines. The result 
is in complete agreement with the conjectured all-order properties of 
MGEWs.  Finally, in \sect{Escher}, we show that a specific class of highly 
symmetric diagrams contributing to $n$-line webs can be explicitly 
computed for any $n$, obtaining a remarkably simple result that further
substantiates our conjectures. This all-order calculation of kinematic
factors further allows us to prove that a specific colour structure
arising from these webs has a vanishing coefficient for any $n$. We
discuss our results and conclude in \sect{Conclu}, while some
technical details concerning the calculation of the subtracted webs
that we have presented are collected in appendices.


\section{Webs and the soft anomalous dimension}
\label{Webs}

We begin by summarising the formalism which will be used in what follows to 
determine the soft anomalous dimension from webs. For a complete discussion of 
this material, the reader is referred to Refs.~\cite{Gardi:2011yz,Gardi:2013saa}. 
In \sect{sec:Webs_colour} we discuss the colour structure of webs using the 
effective vertex formulation developed in Ref.~\cite{Gardi:2013ita}. This section 
already contains some new results: we identify there a new type of relation between 
webs comprising different numbers of Wilson lines, through a process of collinear 
reduction. These relations will provide non-trivial checks of explicit calculations 
in the following sections. 


\subsection{From webs to the soft anomalous dimension}
\label{Webs1}

We consider the vacuum expectation value of a product of $L$ semi-infinite
Wilson lines operators, of the form
\beq
  {\cal S} \left( \gamma_{ij}, \alpha_s(\mu), \e, \frac{m}{\mu} \right) \, \equiv \, 
  \left\langle 0 \left| \Phi_{\beta_1}^{(m)} \otimes \Phi_{\beta_2}^{(m)} \otimes
  \ldots \otimes \Phi_{\beta_L}^{(m)} \right| 0 \right\rangle \, ,
\label{Sdef}
\eeq
where we have introduced an infrared regulator $m$, suppressing gluon emission
at large distances along each Wilson line, by defining~\cite{Gardi:2011yz,Gardi:2013saa}
\beq
  \Phi_{\beta_i}^{(m)} \, = \, {\cal P} \exp \left[ {\rm i} g \mu^\e 
  \int_0^\infty d \lambda \, \beta_i \cdot A \left( \lambda \beta_i \right) \, 
  {\rm e}^{- {\rm i} m \lambda \sqrt{\beta_i^2 - {\rm i} \varepsilon}}
  \right] \, ,
\label{Phidef}
\eeq
so that one recovers the unregulated Wilson line as $m \to 0$. With this definition, the 
correlator ${\cal S}$ is finite in $d = 4 - 2 \e$, for $\e > 0$: potential collinear divergences 
are regulated by keeping $\beta_i^2 \neq 0$ (spacelike or timelike, as discussed in 
Ref.~\cite{Gardi:2013saa}); infrared divergences are regulated by the exponential 
cutoff $m$, and ultraviolet divergences show up as poles in $\e$ as $\e \to 0^+$. 
The Wilson lines in ${\cal S}$ correspond to the classical trajectories of $L$ partons, 
emanating from a hard interaction vertex at the origin. As is well-known, the ultraviolet 
singularities of ${\cal S}$ correspond to the infrared divergences of the corresponding 
multiparton scattering amplitude~\cite{Korchemsky:1987wg}, which is the reason 
why the conventional ultraviolet anomalous dimension of ${\cal S}$ is referred to 
as the {\it soft anomalous dimension}. In order to compute it, we start by recalling that 
${\cal S}$ is multiplicatively renormalisable~\cite{Polyakov:1980ca,Arefeva:1980zd,
Dotsenko:1979wb,Brandt:1981kf}, which means that we can define the renormalised 
correlator
\beq
  {\cal S}_{\rm ren.} \left( \gamma_{ij}, \alpha_s(\mu^2), \e, \frac{m}{\mu} 
  \right) \, = \, {\cal S} \left( \gamma_{ij}, \alpha_s(\mu^2), \e, \frac{m}{\mu} 
  \right) \, Z \left( \gamma_{ij}, \alpha_s(\mu^2), \e \right) \, ,
\label{Srendef}
\eeq
which is finite as $\e\rightarrow 0$. The ultraviolet divergences of the regulated correlator 
${\cal S}(m)$ are the same as those that would arise in ${\cal S}(0)$, which is independent 
of $\mu$ (indeed, the unregulated bare correlator ${\cal S}(0)$ is simply the unit matrix in 
colour space, since it is scale-less, so that all contributing Feynman diagrams vanish in 
dimensional regularization). This leads to the renormalisation group equation
\beq
  \mu \frac{d }{d \mu} Z \left( \gamma_{ij}, \alpha_s(\mu^2), \e \right) \, = \, - \, 
  Z \left( \gamma_{ij}, \alpha_s(\mu^2), \e \right) 
  \Gamma \left(\gamma_{ij}, \alpha_s(\mu^2) \right) \, ,
\label{Zeq}
\eeq
which defines the soft anomalous dimension $\Gamma$; the latter, being finite, compactly 
encodes the ultraviolet singularities of $Z$, and thus of ${\cal S}$. Note that the ordering on 
the right-hand side of \eqn{Zeq} is important: both $Z$ and $\Gamma$ are matrix valued, 
and therefore do not commute in general. The solution of \eqn{Zeq} can be written in 
exponential form~\cite{Gardi:2011yz}, as
\beqa
\label{Zexp}
  Z \left(\alpha_s, \e \right) & = & \exp \Bigg\{ \alpha_s \,
  \frac{1}{2 \e} \, \Gamma^{(1)} + \, \alpha_s^2 
  \left( \frac{1}{4 \e} \Gamma^{(2)} - \frac{b_0}{4 \e^2} \Gamma^{(1)} \right) \\ 
  & & + \, \alpha_s^3 \left( \frac{1}{6 \e} \Gamma^{(3)} + 
  \frac{1}{48 \e^2} \left[ \Gamma^{(1)}, \Gamma^{(2)} \right] - 
  \frac{1}{6 \e^2} \left(b_0 \Gamma^{(2)} + b_1 \Gamma^{(1)} \right) + 
  \frac{b_0^2}{6 \e^3} \Gamma^{(1)} \right) \nonumber \\
  & & + \, \alpha_s^4 \left( \frac{1}{8 \e} \Gamma^{(4)} + 
  \frac{1}{48 \e^2} \left[ \Gamma^{(1)}, \Gamma^{(3)} \right] - 
  \frac{b_0}{8 \e^2} \Gamma^{(3)} + \frac{1}{8 \e^2} \left( 
  \frac{b_0^2}{\e} - b_1 \right) \Gamma^{(2)} \right. \nonumber \\ 
  & & \hspace{5mm} \left. - \frac{1}{8 \e^2}  \left( \frac{b_0^3}{\e^2} - \frac{2 b_0 b_1}{\e}  
  + b_2 \right) \Gamma^{(1)} - \frac{b_0}{48 \e^3}
  \left[ \Gamma^{(1)}, \Gamma^{(2)} \right] \right) + {\cal O} \left( \alpha_s^5 \right)
  \Bigg\} \, , \nonumber 
\eeqa
where we did not display the dependence on $\gamma_{ij}$ for simplicity, we 
expanded the soft anomalous dimension $\Gamma (\alpha_s)$ in powers of 
$\alpha_s$, and $b_n$ is the $n^{\rm th}$-order coefficient of the $\beta$-function. 
As discussed already extensively in Refs.~\cite{Gardi:2011yz,Gardi:2013saa}, the 
matrix nature of $\Gamma^{(i)}$ entails the presence of higher-order poles in the 
exponent of \eqn{Zexp}, involving commutators of lower-order contributions, even 
in a conformal theory where $\beta (\alpha_s) = 0$. At ${\cal O}(\alpha_s^n)$, the 
genuinely new information enters in the coefficient of the single $1/\epsilon$ pole, 
$\Gamma^{(n)}$. This can be directly computed from the unrenormalized webs as 
follows. First, one may write the unrenormalized soft function as
\beq
  {\cal S} \left(\alpha_s, \e \right) \, = \, \exp \Big[ w \left( \alpha_s, \e \right) \Big]
  \, = \, \exp \left[ \, \sum_{n = 1}^\infty \, \sum_{k = - n}^\infty  \alpha_s^n \, \e^k \, 
  w^{(n, k)} \right] \, ,
\label{Sunren}
\eeq
where again we omitted for simplicity the dependence on  $\gamma_{ij}$ and on 
the infrared cutoff $m$: the dependence on $m$ will in any case cancel at the level 
of the anomalous dimension. Note that, while the physically relevant matrix $Z$ is 
a pure counterterm, {\it i.e.} it contains only poles in $\e$, the infrared-regularized, 
unrenormalized correlator ${\cal S}$ has also non-singular dependence on $\e$, 
which plays a non-trivial role. Indeed, in the notation of \eqn{Sunren}, the first 
few perturbative coefficients of the soft anomalous dimensions can be written 
as~\cite{Gardi:2011yz}
\beqa
\label{Gamres}
  \Gamma^{(1)} & = & - 2 w^{(1,-1)} \, ,\nonumber \\
  \Gamma^{(2)} & = & - 4 w^{(2,-1)} - 2 \left[ w^{(1,-1)}, w^{(1,0)} \right] \, ,\nonumber \\
  \Gamma^{(3)} & = & - 6 w^{(3,-1)} + \frac{3}{2} b_0 \left[ w^{(1,-1)}, w^{(1,1)} \right]
  + 3 \left[ w^{(1,0)}, w^{(2,-1)} \right] + 3 \left[ w^{(2,0)}, w^{(1,-1)} \right] \nonumber \\
  & & \hspace{-1cm} 
  + \left[ w^{(1,0)}, \left[w^{(1,-1)}, w^{(1,0)} \right] \right] - \left[ w^{(1,-1)}, 
  \left[w^{(1,-1)}, w^{(1,1)} \right] \right] \, , \\
  \Gamma^{(4)} & = & - 8 w^{(4,-1)} + \frac{4}{3} \, b_0^2 \, \left[ w^{(1,2)}, w^{(1,-1)} \right]
  + b_0 \left( - 2 \left[ w^{(2,1)}, w^{(1,-1)} \right] - \frac{8}{3} \left[ w^{(1,1)}, w^{(2,-1)} 
  \right] \right. \nonumber \\ & & 
  \hspace{-1cm} \left. + \left[ w^{(1,1)}, \left[ w^{(1,0)}, w^{(1,-1)} \right] \right]
  - \frac{2}{3} \left[ w^{(1,0)}, \left[w^{(1,-1)}, w^{(1,1)} \right] \right]
  + \frac{4}{3} \left[ w^{(1,-1)}, \left[w^{(1,-1)}, w^{(1,2)} \right] \right] \right) \nonumber \\
  & & \hspace{-1cm}
  + \frac{4}{3} \, b_1 \left[ w^{(1,-1)}, w^{(1,1)} \right] + 4 \left[ w^{(1,0)}, w^{(3,-1)} \right]
  + 4 \left[ w^{(3,0)}, w^{(1,-1)} \right] + 4 \left[w^{(2,0)}, w^{(2,-1)} \right] \nonumber \\
  & & \hspace{-1cm}
  + 2 \left[ w^{(1,1)}, \left[w^{(2,-1)}, w^{(1,-1)} \right] \right]
  + \frac{8}{3} \left[ w^{(1,-1)}, \left[ w^{(1,1)}, w^{(2,-1)} \right] \right]
  - \frac{4}{3} \left[ w^{(2,0)}, \left[ w^{(1,0)}, w^{(1,-1)} \right] \right] \nonumber \\ 
  & & \hspace{-1cm}
  - \frac{4}{3} \left[ w^{(1,0)}, \left[w^{(2,0)}, w^{(1,-1)} \right] \right]
  + \frac{4}{3} \left[ w^{(1,-1)}, \left[ w^{(2,1)}, w^{(1,-1)} \right] \right]
  - \frac{4}{3} \left[ w^{(1,0)}, \left[ w^{(1,0)}, w^{(2,-1)} \right] \right] \nonumber \\
  & & \hspace{-1cm}
  - \frac{1}{3} \left[ w^{(1,-1)}, \left[ w^{(1,-1)}, \left[ w^{(1,0)}, w^{(1,1)} \right] \right] \right]
  - \frac{1}{3} \left[ w^{(1,-1)}, \left[ w^{(1,0)}, \left[ w^{(1,-1)}, w^{(1,1)} \right] \right] \right]
  \nonumber  \\
  & & \hspace{-1cm}
  + \left[ w^{(1,0)}, \left[ w^{(1,-1)}, \left[ w^{(1,-1)}, w^{(1,1)} \right] \right] \right]
  + \frac{1}{3} \left[ w^{(1,0)}, \left[ w^{(1,0)}, \left[ w^{(1,0)}, w^{(1,-1)} \right] \right] \right]
  \nonumber \\
  & & \hspace{-1cm}
  - \frac{1}{3} \left[ w^{(1,-1)}, \left[ w^{(1,-1)}, \left[ w^{(1,-1)}, w^{(1,2)} \right] \right] 
  \right] \, , \nonumber   
\eeqa
which is sufficient to calculate the soft anomalous dimension up to four-loops. Notice
that the exponent $w (\alpha_s, \e)$ in \eqn{Sunren} is given by a sum of regularized
webs $w_i$. Similarly, all commutator subtraction in \eqn{Gamres} can be organized on 
a web-by-web basis: one must subtract from each web all appropriate commutators 
constructed from subdiagrams of the diagrams comprising the original web. The 
contributions to the soft anomalous dimension are then given by the simple pole 
of the chosen web, plus all simple-pole contributions from the commutator 
counterterms\footnote{Note that the commutators involve also coefficients of positive
powers of $\e$ in the lower order webs. The overall power of $\e$ associated with each
commutator is however $\e^{-1}$.}. This combination of simple poles was called a 
{\it subtracted web} in~\cite{Gardi:2013saa}. For example, at the three-loop level, and 
taking into account the absence of $\beta$-function contributions for MGEWs, subtracted 
webs have the structure
\beqa
\label{websub}
  \overline{w}^{(3,-1)} & = & w^{(3,-1)} - \frac{1}{2} \left[ w^{(1,0)}, w^{(2,-1)} \right]
  - \frac{1}{2} \left[ w^{(2,0)}, w^{(1,-1)} \right] \nonumber \\
  & & - \frac{1}{6} \left[ w^{(1,0)}, \left[ w^{(1,-1)}, w^{(1,0)} \right] \right]
  - \frac{1}{6} \left[ w^{(1,-1)}, \left[ w^{(1,1)}, w^{(1,-1)} \right] \right] \, .
\eeqa
While the separate contributions of non-subtracted webs and the corresponding commutator
counterterms have higher-order ultraviolet poles, making them sensitive to the infrared 
regulator used to calculate the integrals, subtracted webs, which directly contribute to the 
soft anomalous dimension $\Gamma^{(n)}$, are free of these artifacts~\cite{Gardi:2013saa}. 
Subtracted webs are the direct analogue of the webs appearing in colour-singlet two-line 
correlators, as originally defined in~\cite{Gatheral:1983cz,Frenkel:1984pz,Sterman:1981jc},
which individually have just a single ultraviolet pole.


\subsection{The colour structure of webs and collinear reduction\label{sec:Webs_colour}}
\label{Webs2}
  
In order to compute the anomalous dimension coefficients at a given order using
\eqn{Gamres}, one must classify the independent colour factors that arise, and 
then determine the contributions of every web to each colour factor. The first 
observation is that contributions to $\Gamma^{(n)}$ may involve up to $(n+1)$ 
Wilson lines, namely
\beq
  \Gamma^{(n)}\, = \, \sum_{k = 2}^{n + 1} \, \Gamma_k^{(n)} \, ,
\label{numlegs}
\eeq
where, for example, $\Gamma^{(n)}_2$ are the coefficients of the cusp anomalous 
dimension. Contributions involving different numbers of Wilson lines are in principle
independent, and each of them constitutes a separately gauge-invariant physical 
quantity. At three-loops, Ref.~\cite{Gardi:2013saa} computed MGEWs contributing 
to $\Gamma_4^{(3)}$, while in the present paper we will compute those contributing 
to $\Gamma_k^{(3)}$ for $k \leq 3$. While $\Gamma_k^{(n)}$ are distinct physical 
quantities, we will see below that certain contributions of webs that span a non-maximal 
number of Wilson lines, $k \leq n$, can be deduced from webs that span a larger number 
of lines through a process we name \emph{collinear reduction}. 

Let us start by discussing the colour structure of webs as seen through the properties 
of the mixing matrix $R$ in \eqn{Wdef}. Using the idempotence property~\cite{Gardi:2010rn,
Gardi:2011wa}, a web $W$, connecting $L$ Wilson lines, can be conveniently expressed 
as 
\beq
  W \, = \, \sum_{j = 1}^r \left( \sum_D {\cal F}(D) \, Y_{D, j}^{-1} \right) \, 
  \left(\sum_{D'} Y_{j,D'} \, C(D') \right) \, = \, 
  \sum_{j = 1}^r {\cal F}_{W, j} \,  {c_j^{(L)}} \,\, ,
\label{W_diag}
\eeq
where $r$ is the rank of $R$ (which is always smaller than its dimension $d$) and 
$Y$ is the diagonalising matrix, $Y R Y^{-1} = {\rm diag}(\lambda_1, \lambda_2, \ldots, 
\lambda_d)$, with $\lambda_i = 1$ for $i \leq r$ and $\lambda_i = 0$ otherwise. Thus 
the first $r$ eigenvectors of $R$, all corresponding to unit eigenvalue, determine $r$ 
linear combinations of colour factors, $c_j^{(L)} = \sum_{D'} Y_{j,D'} C(D')$, each of 
which is associated with a particular linear combination of  kinematic integrals 
${\cal F}_{W,j}=\sum_D {\cal F}(D) Y_{D, j}^{-1}$ formed out of the diagrams 
in the web.

As mentioned above, an important property of webs is that their colour factors $c_j^{(L)}$ 
correspond to connected graphs~\cite{Gardi:2013ita}.  A convenient basis for these 
colour factors follows naturally from the effective vertex formalism developed in 
Ref.~\cite{Gardi:2013ita}, and we will adopt this basis in the present paper. In this 
formalism, $V_K^{(l)}$ is an effective vertex representing $K$ gluon emissions from a 
given Wilson line $l$, and involving $K - 1$ nested commutators. In general, $V_K^{(l)}$ 
contains $(K - 1)!$ independent colour factors $C_{K, j}$, which are enumerated by 
the index~$j$. For example, $V_2^{(l)}$, describing a double emission from Wilson 
line $l$, has a unique colour factor\footnote{We use the colour-insertion operator 
notation~\cite{Bassetto:1984ik,Catani:1996vz} by which $T_i^a$ represents a colour 
generator on Wilson line $i$ (in the appropriate representation) with adjoint index $a$.}.
\beq
  C_{2,1} \, = \, \big[ T^a,T^b \big] \, = \, {\rm i} f^{abc} T_c \, ,
\label{V2col}
\eeq
while for $V_3^{(l)}$, describing triple emission from the Wilson line, there are two 
independent colour factors involving fully-nested commutators with different permutations,
\beqa
\label{C_3}
  C_{3,1} & = & \left[ \big[ T^a, T^b \big], T^c \right] \, = \, 
  f^{abd} f_d^{\phantom{d} ec} \, T_e \, , \\
  C_{3,2} & = & \left[ \big[ T^a, T^c \big], T^b \right] \, = \, 
  f^{acd} f_d^{\phantom{d} eb} \, T_e \, ;
\eeqa
the third permutation is related to the previous two by the Jacobi identity.  Note that the 
attachment of the effective $K$-gluon-emission graph to the Wilson line involves a single 
generator. 

Recall that the diagrams we are considering (contributing to MGEWs) correspond to 
the emission of individual gluons directly from the Wilson line. Connected colour factors 
emerge from linear combinations of these diagrams: each effective vertex $V_K^{(l)}$ 
picks antisymmetric combinations of the corresponding $K$ colour generators on line 
$l$, through $K - 1$ nested commutators. The effective vertex $V_K^{(l)}$ also associates 
with each colour factor $C_{K, j}$ a specific $K$-fold parameter integral along the Wilson 
line, involving Heaviside functions that determine the order of attachments of the $K$ 
gluons to the Wilson line. Explicit expressions for these effective vertex integrals may 
be found in Ref.~\cite{Gardi:2013ita}. In the following, we will not make direct use of 
these integrals. Rather, we will use the fact that they end up generating linear combinations 
of Feynman integrals ${\cal F} (D)$ corresponding to the various diagrams $D$ in the web: 
these are precisely the linear combinations appearing in \eqn{W_diag}, which are 
determined by the corresponding web mixing matrix. For specific webs, we shall use 
the results for the mixing matrices, and the corresponding eigenvectors entering
\eqn{W_diag}, which are summarised in Appendix A of Ref.~\cite{Gardi:2013ita}.

For our present purposes the vertex formalism will be useful in fixing the basis of colour 
factors $c_j^{(L)}$. We will further see that in this language one may readily identify 
relations between webs involving different numbers of Wilson lines $L$. As explained 
in Ref.~\cite{Gardi:2013ita}, connected graphs in the vertex formalism may involve one 
or more effective vertices on each Wilson line. When a given line features several effective 
vertices, their order is taken to be fully symmetrised, defining
\beq
  \left\{C_1 \, C_2 \, \ldots \, C_n \right\}_+ \, \equiv \, \frac{1}{n!} \sum_{\pi \in S_n} 
  C_{\pi_1} \, C_{\pi_2} \, \ldots \, C_{\pi_n} \, .
\label{sumperm}
\eeq
A web is characterised by a fixed number of emissions, $n_l$, from line $l$. These 
$n_l$ emissions may be distributed between different effective emission vertices, 
and different possibilities result in different web colour factors $c_j^{(L)}$. 
Some examples are provided by figures 
\ref{fig:33_effective_vertices},
\ref{fig:222tot} and
\ref{fig:123eff}
below. It should be noted that further multiplicity of the web colour factors originates 
in the fact that each vertex $V_K^{(l)}$ has $(K - 1)!$ different colour factors $C_{K, j}$ 
(as exemplified by \eqn{C_3} for $K=3$). In general, web colour factors can be written 
in terms of the effective-vertex colour factors as
\beq
  c_j^{(L)} \, = \, \prod_{l = 1}^L \left\{C_{K_1, j_1}^{(l)} \, C_{K_2, j_2}^{(l)} \, \ldots \, 
  C_{K_{n_l},j_{n_l}}^{(l)}\right\}_+ \,\, ,
\label{c_j_vertices}
\eeq
where the product is an outer product between colour factors on different lines, and 
curly brackets indicate symmetrization, according to \eqn{sumperm}.

Here comes an important new observation: contributions to the web in \eqn{W_diag} in 
which a given line $l$ contains $v_l > 1$ effective vertices can be related to webs with 
a larger number of Wilson lines, where line $l$ is replaced by $v_l$ collinear Wilson 
lines, each of which carries one of the $v_l$ effective vertices. This conclusion follows 
from the fact that there is \emph{no ordering} between the effective vertices, so the
relevant Feynman integrals over the positions of these vertices all extend along the 
ray from the hard interaction to infinity.  This is exactly what happens in the situation 
where these vertices appear on different Wilson lines. We note that in colour space 
the two situations are distinct, in the sense that the colour generators of vertices that 
occur on different lines carry different indices, while if they occur on the same line
they must be in the same representation, and they multiply each other; according to 
the Feynman rules of Ref.~\cite{Gardi:2013ita}, one then takes the symmetrized 
product as in \eqn{sumperm}. 

This observation implies that one can make a precise identification between contributions 
corresponding to particular colour structures in webs involving different numbers of Wilson 
lines. Let us consider a simple case: consider a web with $L$ Wilson lines, where 
two lines $l_1$ and $l_2$ feature, respectively, a single vertex each, $V_{K_1}^{(l_1)}$ 
and $V_{K_2}^{(l_2)}$; consider then the collinear limit, where the velocity vectors of the 
two lines coincide; this yields a contribution to the corresponding web with $L - 1$ Wilson 
lines, where the two vertices are placed on the same line, with the colour factor replacement
\beq
  C_{K_1, j_1} \otimes C_{K_2, j_2} \otimes \ldots \, \longrightarrow \,
  \left\{ C_{K_1, j_1}, C_{K_2, j_2} \right\}_{+} \otimes \ldots \, ,
\label{coll_red}
\eeq
where the dots stand for the contribution to the colour factor from the rest of the web, 
involving $L-2$ Wilson lines. If the symmetry factor of the vertex diagram corresponding 
to the original graph differs from that of the final graph, this needs to be taken into account 
(an example will be given in \sect{Threeloop}). This process, which we call \emph{collinear 
reduction}, may be generalised to the identification of multiple lines. As we will see in the 
following sections, it provides non-trivial checks of the final results for webs which span 
less than the maximal number of lines at a given order. 

A corollary to this result is that starting with webs that span the largest number of Wilson 
lines at a given order (at three loops these are the ones connecting four legs, which were 
computed in Ref.~\cite{Gardi:2013saa}), and moving towards more entangled webs, where 
the same number of gluons connect fewer Wilson lines, the kinematic integrals corresponding 
to many of the colour factors $c_j^{(L)}$ in \eqn{W_diag} would already be known in advance. 
In fact, the only contribution of a given MGEW with $n_l$ attachments to leg $l$ which cannot 
be deduced from other MGEWs in which the same number of gluons connects a larger 
number of Wilson lines, is the one corresponding to having \emph{a single} effective vertex, 
$V_{n_l}^{(l)}$, on each line. 

In the remainder of this paper we focus on the calculation of the kinematic functions for 
MGEWs, whose study was started in~\cite{Gardi:2013saa}. We begin in the next section 
by  discussing the general structure of these integrals.


\section{General structure of MGEW integrals}
\label{MGEWs}

Multiple gluon exchange webs are the simplest class of webs contributing to 
the multi-particle soft anomalous dimension. As mentioned above, despite the
Abelian-like appearance of their Feynman graphs, in a non-Abelian gauge theory 
they contribute to the same colour structures as do fully connected webs containing 
the maximal number of gluon self-interactions. Understanding MGEWs is therefore 
a necessary step to compute the soft anomalous at high orders;  on the other hand,
the relative simplicity of MGEWs makes it possible to tackle multi-loop corrections, 
shedding light on the general structure of infrared singularities.

A simple way to characterize MGEWs is the following; they are the webs 
obtained when the Wilson line correlator in \eqn{Sdef} is evaluated with
a path integral in which the full gauge theory Lagrangian is replaced
with its free counterpart, given by the set of terms that are quadratic in the gauge 
fields. One may write
\beq
  \left. {\cal S} \left( \gamma_{ij}, \alpha_s(\mu), \e, \frac{m}{\mu} \right) 
  \right|_{\rm MGEW} \, \equiv \, \int \left[ D A \right] \Phi_{\beta_1}^{(m)} 
  \otimes \Phi_{\beta_2}^{(m)} \otimes \ldots \otimes \Phi_{\beta_L}^{(m)}
  \, \exp \Big\{ {\rm i} S_0 [A] \Big\} \, ,
\label{formdef}
\eeq
where $S_0 [A]$ comprises the classical gauge kinetic term, and the quadratic
contribution to the chosen gauge fixing (we will work in Feynman gauge). Terms
quadratic in matter fields and ghost fields are not included. As a consequence, 
$\beta$ function contributions are absent in MGEWs, and we are effectively working 
in a conformally invariant sector of the theory.


\subsection{Feynman integral for a MGE diagram}
\label{MGEDs}

It turns out to be possible to formally carry out a number of steps in
the calculation of Feynman diagrams contributing to MGEWs in complete
generality, as suggested in~\cite{Gardi:2013saa}. In order to do so we
need to introduce a precise characterization of the gluon configuration for a 
generic MGEW diagram. First, we introduce an ordering in the set of $L$ 
Wilson lines, $l = 1, \ldots, L$. A given MGEW will be denoted\footnote{In 
general, this notation does not uniquely identify a web, as it does not fully 
specify how the lines are connected. It will nevertheless be useful for the 
examples we consider.} by $(n_1, n_2, \ldots , n_L)$ where $n_i$ is the
number of gluon emissions from Wilson line $i$. As an example, the
diagram of Fig.~\ref{multi} is part of a (1,2,3,3,1) web. To further
characterise a specific diagram in the web, we need to identify the
order of gluon attachments to each Wilson line. Referring again to the
example of Fig.~\ref{multi}, we introduce an ordering in the set of
$n$ gluons contributing to the chosen $n$-loop Feynman diagram, in the
following way:
\begin{figure}[htb]
\begin{center}
\scalebox{0.33}{\includegraphics{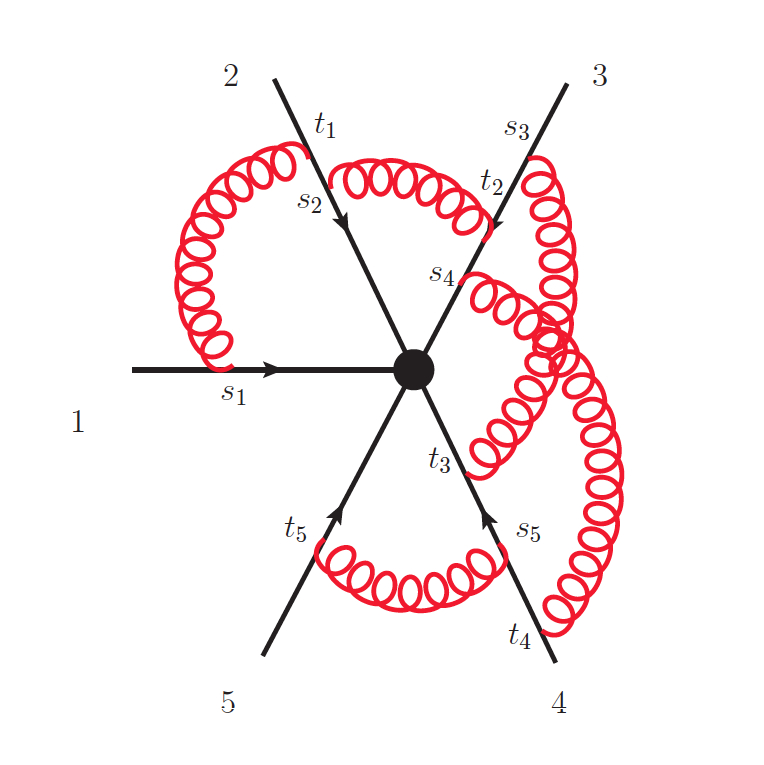}}
\caption{An example of a multiple gluon exchange diagram connecting five Wilson lines 
at five loops; it is part of the (1,2,3,3,1) web. The lines meet at a local effective vertex 
representing the hard interaction. For this diagram  $\Theta_D \big[ \left\{ s_k, t_k 
\right\} \big] \, = \, 
\theta(t_1 > s_2) \, \theta(s_3 > t_2 > s_4) \, \theta(t_4 > s_5 > t_3)$.}
\label{multi}
\end{center}
\end{figure}
we consider each of the Wilson lines in turn according to the chosen order, 
moving along each line starting at the far end and reaching the origin; gluons 
are assigned the ordering with which they are encountered in this procedure. 
A given gluon $k$ is counted only once, as it is first encountered. With this assignment, 
we say that the $k$-th gluon is emitted from the Wilson line in the direction $\beta_{i(k)}$ 
at point $s_k$, and is absorbed by the Wilson line in the direction $\beta_{f(k)}$ at point 
$t_k$. Using the coordinate space Feynman gauge gluon propagator
\beq
  D_{\mu \nu}^{a b} (x) \, = \, - \, \frac{\Gamma(1 - \e)}{4 \, \pi^{2 - \e}} \, 
  g_{\mu \nu} \, \delta^{a b} \, \left( - x^2 + {\rm i} \varepsilon \right)^{-1 + \e} \, ,
\label{glupprop}
\eeq
and expanding the Wilson line operators in \eqn{Phidef} in powers of the coupling,
one easily finds that the most general $n$-gluon MGE Feynmam diagram $D$, 
contributing to \eqn{formdef} at $n$ loops, can be written as
\beqa
\label{sigmatau}
  {\cal F}^{(n)} \left( D \right) & = & \left( g^2 \mu^{2\e} \, 
  \frac{\Gamma(1 - \e)}{4 \pi^{2 - \e}} \right)^n \, \prod_{k = 1}^n 
  \beta_{i(k)} \cdot \beta_{f(k)} \, \int_0^\infty \prod_{k = 1}^n
  d s_k \, d t_k  \, \Theta_D \big[ \left\{ s_k, t_k \right\} \big]
  \label{gendiag1} \\ & & \!\!\!\! \!\!\!\!\!\!\!
  \prod_{k = 1}^n \left[ - \left( \beta_{i(k)} s_k - \beta_{f(k)} t_k
  \right)^2 \right]^{-1 + \e} \exp \left[ - \, {\rm i} m \sum_{k =
  1}^n \left( s_k \sqrt{\beta_{i(k)}^2-{\rm i} \varepsilon} + t_k
  \sqrt{\beta_{f(k)}^2-{\rm i} \varepsilon} \right) \right] \, . 
\nonumber 
\eeqa 
Clearly the ordering of the attachments of the gluons on each Wilson line is essential: 
it is given by the function $\Theta_D \left[ \left\{ s_k, t_k \right\} \right]$, which is a
product of Heaviside functions assigned to each Wilson line, with $p - 1$ independent 
$\theta$ functions on a line with $p$ gluon attachments\footnote{Notice that 
  \eqn{gendiag1} applies also to the case of gluons being emitted and absorbed by 
  the same Wilson line, corresponding to $i(k) = f(k)$, but we will not compute such
  webs here.}. 

Following Ref.~\cite{Gardi:2013saa}, we proceed by rescaling the 
Wilson line coordinates by defining\footnote{For simplicity here we take the Wilson 
  lines to be timelike, $\beta_i^2 > 0$ for all $i$. Note that in this case the $- {\rm i}
  \varepsilon$ prescription is important for the infrared regulator, and it can be 
  implemented by taking $m \to m - {\rm i} \varepsilon$, as we do below. A similar 
  rescaling can be done for spacelike Wilson lines where $\sqrt{\beta_i^2} = - {\rm i}
  \sqrt{\left|\beta_i^2\right|}$, and we then define $\sigma_k \, = \,s_k \sqrt{\vert 
  \beta_{i(k)}^2\vert}$, ending up with the same final result, \eqn{gendiag3} below.}  
\beq 
  \sigma_k \, = \,s_k
  \sqrt{\beta_{i(k)}^2} \, , \qquad \tau_k \, = \, t_k
  \sqrt{\beta_{f(k)}^2} \, ,
\eeq
and furthermore we change variables by writing
\beq
  \sigma_k \, = \, x_k \lambda_k \, , \qquad 
  \tau_k \, = \, (1 - x_k) \lambda_k \, .
\label{lamx}
\eeq
In this way, $\lambda_k$ is a measure of the overall distance of the $k$-th gluon 
from the origin, whereas $x_k$ is an `angular' variable, measuring the degree 
of collinearity of the $k$-th gluon to either the emitting (as $x_k \to 1$) or the 
absorbing (as $x_k \to 0$) Wilson lines. In terms of these variables one finds
\beqa
\label{gendiag2} 
  {\cal F}^{(n)} \left( D \right) & = &\left(\frac{1}{2} \, g^2 \mu^{2\e} \, 
  \frac{\Gamma(1 - \e)}{4 \pi^{2 - \e}} \right)^n \,  \times
  \, \int_0^\infty \prod_{k = 1}^n d \lambda_k \, \lambda_k^{- 1 + 2 \e} \, 
  {\rm e}^{ - {\rm i} (m - {\rm i} \varepsilon)  \sum_{k = 1}^n \lambda_k }
  \\ & &  \hspace{-5mm} \nonumber \times \, \,
  \int_0^1 \prod_{k = 1}^n d x_k \, \gamma_k\, 
  \Big[ - x_k^2 - (1 - x_k)^2 +  \gamma_k \, x_k (1 - x_k) \Big]^{- 1 + \e}
  \, \Theta_D \big[ \left\{ x_k, \lambda_k \right\} \big]  \, , 
\eeqa
where $\gamma_k \equiv \gamma_{i(k), f(k)}$ was defined in \eqn{gamma_ij}. Note 
that the distance variables $\lambda_k$ have been scaled out of the propagators. We 
keep using the symbol $\Theta_D$ for the product of Heaviside functions, although 
now they are expressed in terms of the new variables.

To proceed we now extract from the diagram the overall ultraviolet singularity 
arising from the region where all gluons are emitted and absorbed very close to the 
origin. In order to do so, we change variable again expressing the $\lambda_k$'s as
\beq
  \lambda_k \, = \,  \left( 1 - y_{k - 1} \right) \, \prod_{p = k}^n y_p \, ,
\label{defy}
\eeq
for $k = 1, \ldots, n$, where we define $y_0 = 0$. 
Note that with this definition $\sum_{i = 1}^k{\lambda_i} = \prod_{p = k}^n y_p$, and
in particular the regulator in \eqn{gendiag2}, which involves the sum of all the 
$\lambda_i$, depends only on $y_n$. The Jacobian of this change of variables is 
given by $J = \prod_k y_k^{k - 1}$, so, after performing the integral over $y_n$, we 
find
\beqa
\label{gendiag3} 
  {\cal F}^{(n)} \left( D \right) & = & \left( - \frac{1}{2} g^2 \left(\frac{\mu^2}{m^2}\right)^{\e} 
  \, \frac{\Gamma(1 - \e)}{4 \pi^{2 - \e}} \right)^n \,  \Gamma(2n\e)
  \\ & &  \hspace{+5mm} \times \, \, \nonumber \int_0^1 \prod_{k = 1}^n d x_k\, 
  \gamma_k \, \Big[  x_k^2  +(1 - x_k)^2 -  \gamma_k \, x_k (1 - x_k) \Big]^{- 1 + \e}
  \\ & &  \hspace{+5mm} \times \, \, \nonumber \int_0^1 \prod_{k = 1}^{n - 1} d y_k  
  \left( 1 - y_k \right)^{- 1 + 2 \e} \, y_k ^{- 1+ 2 k \e} \, \,\, 
  \Theta_D \big[ \left\{ x_k, y_k \right\}\big]
  \nonumber  \\  & = & \kappa^n \, \, \Gamma( 2 n \e ) 
  \, \int_0^1 \prod_{k = 1}^n \Big[ d x_k  \,\gamma_k \, P_\epsilon 
  \left( x_k, \gamma_k \right) \Big]  \, \phi_D^{(n)} \left( x_i; \e \right)  
  \nonumber \, , 
\eeqa
where we defined the coefficient
\beq
 \kappa \, \equiv - \frac{1}{2} \, g^2 \left(\frac{\mu^2}{m^2}\right)^{\e} \, 
  \frac{\Gamma(1 - \e)}{4 \pi^{2 - \e}} \, ,
\label{prefac}
\eeq
as well as the function
\beq
  P_\epsilon \left( x, \gamma \right) \, \equiv \, 
 \left[ x^2  +(1 - x)^2 -  x (1 - x)
 \gamma) \right]^{-1 + \e} \,,
\label{propafu}
\eeq
related to the gluon propagator, and the kernel
\beq
  \phi_D^{(n)}  \left( x_i; \e \right) \, = \, \int_0^1 \prod_{k = 1}^{n - 1} d y_k  
  \left( 1 - y_k \right)^{- 1 + 2 \e} \, y_k ^{- 1+ 2 k \e} \, \, 
  \Theta_D \big[ \left\{ x_k, y_k \right\} \big] \, .
\label{phi_D}
\eeq
The analysis of Ref.~\cite{Gardi:2013saa} shows\footnote{Similar conclusions were 
reached in Ref.~\cite{Henn:2013wfa}, working on two-line MGEWs and using different 
tools.} that $\phi_D^{(n)}  \left( x_i; \e \right)$ has a Laurent expansion in $\epsilon$, 
$\phi_D^{(n)} = \sum_k \phi_D^{(n,k)} \epsilon^k$, where each term $\phi_D^{(n,k)}$ 
is a pure transcendental function of uniform weight $n - 1 + k$, containing logarithms 
and polylogarithms, as well as Heaviside functions depending on \emph{ratios} of the 
variables $x_i$ or $1-x_i$. 


\subsection{Feynman integral for a MGE web}
\label{MGEWint}

The next observation~\cite{Gardi:2013saa} is that all diagrams $D$ in a given web 
$W$ have a common integral structure\footnote{It is important to note that in order 
to combine Feynman integrals corresponding to individual diagrams, as in \eqn{gendiag3}, 
into the web Feynman integral, \eqn{genweb} below, one must use a common set of 
parameters, so that $x_k$ is associated with a given cusp angle $\gamma_{i(k),f(k)}$ 
for all diagrams in the web. In practice one therefore selects one diagram $D$, 
based on which one defines the ordering of the gluons, as explained using the 
example of Fig.~\ref{multi}; for any other diagram in the web, one then uses the 
assigned order, where gluon $k$ is always exchanged between the same pair of 
Wilson lines.}  of the form of \eqn{gendiag3}: assuming that in all diagrams $D \in W$ 
gluon $k$ is exchanged between the same pair of Wilson lines, such diagrams 
are only distinguished by the Heaviside functions $\Theta_D \big[ \left\{ x_k, y_k \right\} 
\big]$ representing the order of gluon attachments to the Wilson lines, hence they only 
differ by their kernels $\phi_D^{(n)} \left( x_i; \e \right)$. Because a web $W$ is defined 
as a linear combination of the contributing diagrams $D \in W$, one deduces that the 
web as a whole takes a form similar to \eqn{gendiag3}. To see this in more detail, 
recall that, according to \eqn{W_diag}, every web can contribute to different colour 
structures $c_j^{(L)}$ building up the anomalous dimension, $W^{(n)} = \sum_{j = 1}^r 
{\cal F}_{W,j}^{(n)} \, {c_j^{(L)}}$, where the kinematic functions are specific linear 
combinations of the integrals corresponding to individual diagrams in the web, 
\beq
  {\cal F}_{W, \, j}^{(n)} \left( \gamma_{i j}, \e \right) \, = \, \sum_{D \in W} Y_{D,j}^{-1} \, 
  {\cal F}^{(n)} (D) \, ,
\label{webcolco}
\eeq
and where the numerical coefficients $Y_{D,j}^{-1}$ are fixed by the web mixing matrix. 
One concludes that the contribution of web $W$ to the $j$-th colour structure is given by 
an integral similar to \eqn{gendiag3}, 
\beqa
\label{genweb} 
  {\cal F}_{W, \, j}^{(n)} \left( \gamma_{i j}, \e \right) \, = \,  \kappa^n \, \Gamma( 2 n \e ) 
  \, \int_0^1 \prod_{k = 1}^n \Big[ d x_k  \,\gamma_k \, P_\epsilon \left( x_k, \gamma_k 
  \right) \Big]  \, \phi_{W, \, j}^{(n)} \left( x_i; \e \right)  \, , 
\eeqa
with a \emph{web kernel} given by
\beq  
  \phi^{(n)}_{W, \, j} \left( x_i; \e \right) \, = \, \sum_{D \in W} Y_{D,j}^{-1} \, \phi^{(n)}_D 
  \left( x_i; \e \right) \, .
\label{webker}
\eeq
Before proceeding, it is useful to contrast non-Abelian MGEWs with their much simpler 
Abelian counterparts. According to the Abelian exponentiation theorem only connected 
graphs enter the exponent, so in particular Abelian MGEWs are not part of the exponent. 
Instead, they are reproduced by expanding the exponential involving a single exchange 
between each pair of Wilson lines. Because in the Abelian theory ordering is immaterial, 
one simply sums all diagrams with \emph{equal} weights. This sum must yield a product 
of the relevant one-loop integrals. The result can readily be verified from \eqn{gendiag3}: 
indeed
\beqa
\label{Abelian_sum}
  \sum_{D \in W} {\cal F}^{(n)} \left( D \right) & = & \kappa^n \, \Gamma( 2 n \e ) 
  \, \int_0^1 \prod_{k = 1}^n \Big[ d x_k  \, \gamma_k \, P_\epsilon \left( x_k, \gamma_k 
  \right) \Big]  \, \sum_{D \in W} \phi_D^{(n)} \left( x_i; \e \right) \nonumber
  \\  & = & \Big( \kappa \, \Gamma(2 \epsilon) \Big)^n \, \, \int_0^1 \prod_{k = 1}^n 
  \Big[ d x_k  \,\gamma_k \, P_\epsilon \left( x_k, \gamma_k \right) \Big]  \, ,
\eeqa
where in the second line we used the fact that the sum of Heaviside functions for a MGEW
gives unity, so that one can use
\beq
  \sum_{D \in W} \phi_D^{(n)}  \left( x_i; \e \right) \, = \, \int_0^1 \prod_{k = 1}^{n - 1} d y_k  
  \left( 1 - y_k \right)^{- 1 + 2 \e} \, y_k ^{- 1+ 2 k \e} \, = \, 
  \frac{ \left( \Gamma(2 \epsilon) \right)^2}{ \Gamma(2 n \epsilon)} \, .
\label{Abelian_constraint}
\eeq
As expected, \eqn{Abelian_sum} is a product of one-loop integrals associated to the
relevant cusp angles. The unweighted sum in \eqn{Abelian_constraint} is a constraint 
on the web kernels of any non-Abelian MGEW, providing a valuable check of explicit 
calculations in what follows.


\subsection{Feynman integral for a MGE subtracted web}
\label{MGESW}

An important conclusion of the analysis in Ref.~\cite{Gardi:2013saa} is that the 
integration over the angular variables $x_k$ in \eqn{genweb} is vastly simplified for 
subtracted webs. To this end, each web is expanded in powers of $\epsilon$, as
\beq
  W^{(n)} \left( \gamma_{ij}, \epsilon \right) \, = \, \alpha_s^n 
  \sum_{k = - n}^\infty w^{(n,k)}  \left( \gamma_{ij} \right) \, \epsilon^k  \, ,
\label{W_expanded_alpha_epsilon}
\eeq
and then it is combined with the commutators of the webs composed by its subdiagrams,
according to the pattern forming the anomalous dimension in \eqn{Gamres}. Then, in 
the notation of \eqn{websub}, $\Gamma^{(n)} \, = \, - 2 n \, \sum_i \overline{w}^{(n, - 1)}_i$.
This step relies on the fact that the commutators build up the same colour factors and 
similar kinematic integrals as those of the web itself, \eqn{genweb}. The ${\cal O}(\alpha^n, 
\epsilon^{-1})$ subtracted web can then be written as
\beq
  \overline{w}^{(n, -1)} \left( \alpha_k \right) \, = \, 
  \left(\frac{1}{4\pi}\right)^n \, \sum_{j=1}^{r}  \, c_j^{(L)} \, F^{(n)}_{W, \, j}
  \big( \alpha_k \big) \, ,
\label{subtracted_web_form}
\eeq
where we chose to express the kinematic dependence in terms of $\alpha_k \equiv 
\alpha_{i(k),f(k)}$, as defined in \eqn{alphadef}, so that for each gluon $\gamma_{k} 
\, = - \, \alpha_{k} - 1/\alpha_{k}$. Following Ref.~\cite{Gardi:2013saa}, we then define
the functions of the $\alpha$ variables which arise from the gluon propagators as
\beq
  p_{\epsilon} \left(x, \alpha \right) \, \equiv \, - \, \left( \alpha + \frac{1}{\alpha} \right) \, 
  \Big[ q ( x, \alpha) \Big]^{- 1 + \e} \, ; \quad
  q \left(x, \alpha \right) \, \equiv \, x^2 + (1 - x)^2 +\left(\alpha+\frac{1}{\alpha}\right) \, 
  x (1 - x) \, .
\label{propafu2}
\eeq
Upon expanding $p_{\epsilon}(x, \alpha)$ in powers of $\epsilon$ we get 
\beq
  p_\e \left( x, \alpha \right) \, = \, p_0 \left( x, \alpha \right) \,
  \sum_{n = 0}^\infty \frac{\e^n}{n!} \Big[ \log  q \left( x, \alpha \right) \Big]^n \, ,
\label{expP}
\eeq
where 
\beq
  p_0 \left( x, \alpha \right) \, = \, - \, \left( \alpha + \frac{1}{\alpha} \right) \, 
  \frac{1}{q(x,\alpha)} \, = \, r (\alpha) \left[ \frac{1}{x - \frac{1}{1 - \alpha}} - 
  \frac{1}{x + \frac{\alpha}{1 - \alpha}} \right] \, ,
\label{p0_part_fracs}
\eeq
and where the rational function $r(\alpha)$ is given by 
\beq
  r (\alpha)\, = \, \frac{1 + \alpha^2}{1 - \alpha^2} \, .
\label{r_def}
\eeq
Using these notations, the subtracted web kinematic function $F^{(n)}_{W, \, j}$ can
be written as
\beqa
\label{subtracted_web_mge_kin}
  F^{(n)}_{W, \, j} \big( \alpha_i \big) & = &
  \int_0^1 \left[ \, \prod_{k = 1}^n d x_k \, p_0 (x_k, \alpha_k) \right]  \, 
  {\cal G}^{(n)}_{W, \, j} \Big(x_i, q(x_i, \alpha_i) \Big) \, \nonumber
  \\ & = &  \, \left(\prod_{k = 1}^{n} r (\alpha_k) \right)
  \int_0^1  \left[ \prod_{k = 1}^n d x_k \left( \frac{1}{x_k - \frac{1}{1 - \alpha_k}}
  - \frac{1}{x_k + \frac{\alpha_k}{1 - \alpha_k}} \right) \right] \, 
  {\cal G}^{(n)}_{W, \, j} \Big(x_i, q(x_i, \alpha_i) \Big)\, \nonumber
  \\ & \equiv &  \, \left(\prod_{k = 1}^{n} r (\alpha_k) \right)
  G^{(n)}_{W, \, j} \big(\alpha_i \big) \, ,
\eeqa
where in the first line we introduce the subtracted web kernel ${\cal G}^{(n)}_{W, \, j}$. 
This function contains the finite term\footnote{Recall that $\phi_{W, \, j}^{(n)} 
\left( x_i; \e \right)$ enters at  ${\cal O}(\epsilon^{- 1})$ due to the overall factor 
$\Gamma(2 n \epsilon)$ in \eqn{genweb}.}  of the web kernel $\phi_{W, \, j}^{(n)} 
\left( x_i; \e \right)$, along with related terms from the commutators of the relevant 
subdiagrams; additional contributions occur due to (multiple) poles of $\phi_{W, \, j}^{(n)} 
\left( x_i; \e \right)$, related to subdivergences, which are multiplied by appropriate 
powers of $\log  q \left( x, \alpha \right)$ from the expansion in \eqn{expP}. In the 
second expression in \eqn{subtracted_web_mge_kin} we used the partial fraction 
form of $p_0$ introduced in \eqn{p0_part_fracs}, which is the convenient form for 
performing the $x_k$ integrals. This also fixes the rational function associated with 
the web, which is simply a factor of $r (\alpha_k)$ for each gluon exchange. In the 
final expression in \eqn{subtracted_web_mge_kin} all the $x_k$ integrals are done, 
defining the function $G^{(n)}_{W, \, j} \big( \alpha_i \big)$.

As discussed in Ref.~\cite{Gardi:2013saa}, remarkable simplifications occur at the level 
of \eqn{subtracted_web_mge_kin}. These can most easily be described through the 
properties of the subtracted web kernel ${\cal G}^{(n)}_{W, \, j}$: this function depends 
on its arguments only through powers of the logarithms, $\ln(x_k)$, $\ln(1-x_k)$ and 
$\ln q(x_k,\alpha_k)$, and through Heaviside functions involving ratios of the variables 
$x_k$. The integrals in the second line of \eqn{subtracted_web_mge_kin} are of a $d\log$
form. Thus, the resulting function $G^{(n)}_{W, \, j} \big( \alpha_i \big)$ is a pure function 
of transcendental weight $2 n - 1$. Subtracted multi-particle webs thus share the properties 
of two-parton webs described in Ref.~\cite{Henn:2013wfa}. It should be stressed, however, 
that the route by one which calculates multiparton webs is substantially different, due to 
the combinatorics associated with the non-trivial colour structure, and to the presence 
of subdivergences.

We emphasise that the absence of polylogarithms in ${\cal G}^{(n)}_{W, \, j}$ 
is rather surprising: recall that polylogarithms of weight $n - 1$ do occur in the 
${\cal O}(\epsilon^0)$ term of the non-subtracted web kernel $\phi_{W, \, j}^{(n)} 
\left( x_i; \e \right)$. Ref.~\cite{Gardi:2013saa} argued that these polylog cancellations 
(and the related analytic properties of $G^{(n)}_{W, \, j} \big( \alpha_i \big)$, which 
we describe below) are linked with the restoration of crossing symmetry. The latter 
is lost at the level of non-subtracted webs, due the action of the infrared regulator 
in the presence of ultraviolet subdivergences, but it is recovered for subtracted webs. 

The most important consequence of the purely logarithmic nature of the subtracted 
web kernel is that the resulting integrated function, $G^{(n)}_{W, \, j} \big( \alpha_i \big)$, 
is a sum of products of polylogarithms, each depending on a single $\alpha_k$ variable. 
Furthermore, these polylogarithms are very specific; their symbol alphabet is restricted 
to $\alpha_k$ and $1 - \alpha_k^2$. The goal of the next section is to fully characterize 
these functions and obtain an explicit basis for them. 

It should be stressed that the properties just described have been conjectured to 
be general, but they have not been proven. Specifically,  all explicit calculations in 
Ref.~\cite{Gardi:2013saa} were of webs whose subtracted kernel is free of Heaviside 
functions, in which case the relation between the purely logarithmic nature of the kernel 
and the factorization property is obvious. Such a relation is less obvious when Heaviside 
functions occur in ${\cal G}^{(n)}_{W, \, j}$. The number of Heaviside functions appearing 
in a given subtracted web kernel depends on the level of entanglement of the web: webs 
spanning the maximal number, $n+1$, of  Wilson lines at $n$ loops are not entangled 
(so there is no Heaviside function after performing the $y_k$ integrals in \eqn{phi_D}) while 
those connecting fewer Wilson lines are entangled by up to $n - 1$ Heaviside functions. 
A central goal of the present paper is to verify that the factorization property does 
indeed hold even for entangled webs.


\section{A basis of functions for MGEWs}
\label{Basis}


\subsection{Constructing a basis}
\label{construba}

One of the conclusions of Ref.~\cite{Gardi:2013saa} is that the subtracted webs 
(1,2,2,1) and (1,1,1,3), connecting the maximum possible number of Wilson lines at 
three loops, can be written in terms of the simple set of integrals
\beqa
\label{RUSigma}
  R_0 (\alpha) & = & \frac{1}{r (\alpha)} \int_0^1 dx \, p_0(x, \alpha) \, , \nonumber \\
  \Sigma_2 (\alpha) & = & \frac{1}{2 \, r (\alpha)} \int_0^1 dx \, p_0(x, \alpha) 
  \ln^2 \bigg( \frac{x}{1 - x} \bigg) \, , \nonumber \\
  U_1(\alpha) & = & \frac{1}{r (\alpha)} \int_0^1 dx \, p_0(x, \alpha) 
  \ln \bigg( \frac{q (x,\alpha)}{x^2} \bigg) \, ,
  \label{eq:MaxLinesOldFns}  \\
  U_2 (\alpha) & = & \frac{1}{4 \, r (\alpha)} \int_0^1 dx \, p_0(x, \alpha) 
  \ln^2 \bigg( \frac{q (x,\alpha)}{x^2} \bigg) \, . \nonumber
\eeqa
These integrals have an integrand consisting only of logarithms, depend upon only a 
single cusp angle, and individually satisfy the alphabet constraints outlined above. 
It is natural to ask whether one may construct a basis for \emph{all} MGEWs, given 
the requirements of the alphabet and factorization conjectures, and the limited range 
of elements entering the subtracted web kernel in \eqn{subtracted_web_mge_kin}, 
namely
\beq
  \ln \frac{q (x, \alpha)}{x^2} \, = \, \ln \left( \frac{1}{x} + \alpha - 1 \right) 
  + \ln \left( \frac{1}{x} + \frac{1}{\alpha} - 1 \right) \, .
\label{allog}
\eeq
and $\ln\Big( \frac{x}{1 - x} \Big)$. A first attempt could be to consider the set of 
functions
\beq
  M_{k,m} (\alpha) \, = \, \frac{1}{r (\alpha)} \int_0^1 d x \, \, p_0 (x, \alpha) \, \ln^k 
  \bigg( \frac{q(x, \alpha)}{x^2} \bigg) \ln^{2 m} \bigg( \frac{x}{1 - x} \bigg) \, ,
\label{firsta}
\eeq
with $k$ and $m$ non-negative integers (note that $q(x,\alpha)$ is symmetric under 
$x \to 1-x$, and odd powers of $\ln\Big( \frac{x}{1 - x} \Big)$ can be eliminated in 
terms of ones with even powers). The functions in \eqn{firsta} have uniform weight 
$2 m  + k + 1$, and they satisfy the alphabet conjecture. In terms of these functions 
one finds
\begin{align}
 \begin{split}
  R_0 (\alpha) & = M_{0,0} (\alpha) \, , \quad \quad 
  U_1 (\alpha) = M_{1,0} (\alpha) \, , 
  \\
  \Sigma_2 (\alpha) & = \frac12 M_{0,1} (\alpha) \, , \quad \, \,
  U_2 (\alpha) = \frac14 M_{2,0} (\alpha) \, .
 \end{split}
\label{oldfun}
\end{align}
At least through three loops, the basis of \eqn{firsta} is sufficient to describe 
MGEWs that connect the maximum number of Wilson lines at a given loop order. 
We now wish to check whether more entangled webs, which connect fewer lines, 
and thus may have leftover Heaviside functions in their subtracted web kernel 
${\cal G}_W$, will belong to the span of this basis. To do this, we shall first consider 
the (2,2) web at two loops, and then the (3,3), (1,2,3), and (2,2,2) subtracted webs 
at three loops. The relevant combination of effective vertices are, respectively: 
$V^{(1)}_2 V^{(2)}_2$, $V^{(1)}_3 V^{(2)}_3$, $V^{(1)}_1 V^{(2)}_2 V^{(3)}_3$ 
and $V^{(1)}_2 V^{(2)}_2 V^{(3)}_2$.

\begin{figure}[htb]
\begin{center}
\scalebox{0.55}{\includegraphics{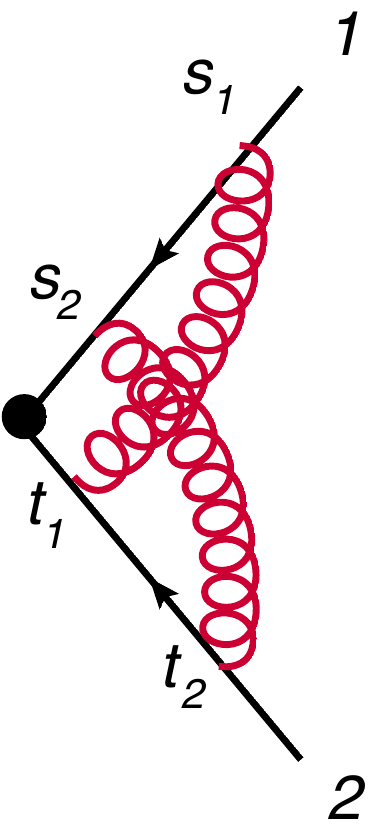}}\hspace*{30pt}
\scalebox{0.55}{\includegraphics{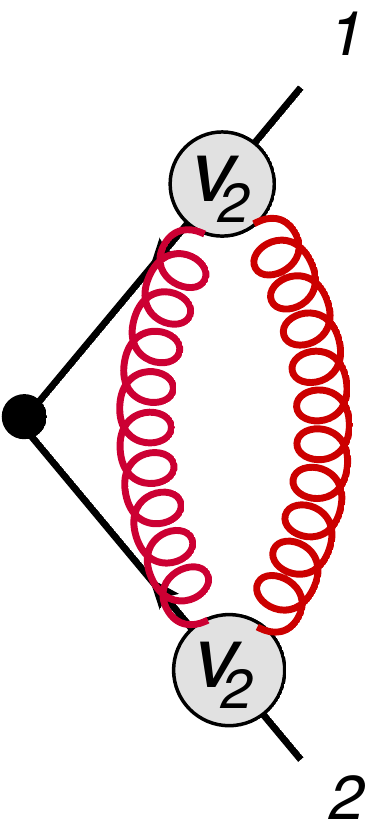}}
\caption{The two-loop crossed graph connecting two Wilson lines, and the 
corresponding effective vertex graph containing a double emission vertex 
$V_2$ on each of the two lines.}
\label{cross}
\end{center}
\end{figure}

Let us begin by considering the simplest example, the two-loop, two-line 2-2 web, 
which is of course well-known~\cite{Korchemsky:1987wg,Kidonakis:2009ev}. This 
web contains two diagrams: the ladder ($II$) and the crossed (${\rm X}$) one. It is 
immediately evident, however, according to the definition of webs for the colour-singlet 
case~\cite{Gatheral:1983cz,Frenkel:1984pz}, that only the kinematic integral of the latter, 
shown in Fig.~\ref{cross}, contributes. The web therefore evaluates to 
\beq
  W_{(2,2)}^{(2)} \, = \, \Big[ C(X) - C(II) \Big] {\cal F}(X) \, = \, 
  \frac12 [T_1^a,T_1^b] \, [T_2^b,T_2^a] \, {\cal F}(X) \, = \, \frac{N_c}{2} 
  T_1\cdot T_2\,{\cal F}(X) \, .
\label{22}
\eeq 
A derivation of this result using the effective vertex formalism was given in Section 3 of 
Ref.~\cite{Gardi:2013ita}, where it was shown that the only contribution arises from a 
double-emission vertex $V_2$ on each of the two Wilson lines, each vertex having a 
connected colour structure given by \eqn{V2col}. 
We now proceed to consider the integral ${\cal F}(X)$. Specifically we will be
interested in the simple pole of this function, 
\beq
  {\cal F}(X) \, = \, \left( \frac{\alpha_s}{4\pi} \right)^2 \left[ \frac{1}{\epsilon} \, 
  F^{(2)}_{(2,2)} (\alpha) + {\cal O}(\epsilon^0) \,\right] \, ,
\label{spol22}
\eeq
which depends on a single kinematic variable, $\alpha_{12} \equiv \alpha$. In the 
notation of \eqn{sigmatau}, one has four semi-infinite parameter integrals, two over 
$s_1$ and $s_2$ along line 1, and two over $t_1$ and $t_2$ along line 2, with the 
restrictions $\Theta_X = \theta(s_1 - s_2) \, \theta(t_2 - t_1)$. Following the steps 
leading to~\eqn{subtracted_web_mge_kin}, we find
\beq
  F^{(2)}_{(2,2)} (\alpha) \, = \,  \int_0^1 dx_1 \int_0^1 dx_2 \, p_0 (x_1, \alpha)  
  \, p_0 (x_2, \alpha) \, {\cal G}^{(2)}_{(2,2)} (x_1, x_2)  \, ,
\label{webkin22_initial}
\eeq
with the kernel 
\beq
  {\cal G}^{(2)}_{(2,2)} (x_1, x_2) \, = \, \theta(x_1 - x_2) \ln \bigg(\frac{x_1}{1 - x_1}
  \frac{1 - x_2}{x_2} \bigg) \, .
\label{webker22}
\eeq
Using $p_0 (1 - x, \alpha) = p_0 (x, \alpha)$, we can write the kinematic factor in
\eqn{webkin22_initial} as
\beq
  F^{(2)}_{(2,2)} (\alpha) \, = \, 2 \int_0^1 dx_1 \, p_0 (x_1, \alpha) 
  \ln \bigg( \frac{x_1}{1 - x_1} \bigg)  \int_0^1 dx_2 \, p_0 (x_2, \alpha) 
  \, \theta(x_1 - x_2) \, .
\label{webkin22}
\eeq
The second integral in \eqn{webkin22} does not yield an expression of the form of
\eqn{firsta}, so we shall be forced to extend our basis. Let us 
proceed as follows. First we define
\beq
  \ln \tilde{q} (x, \alpha) \, \equiv \, \frac{1}{r (\alpha)} \int_0^1 dy \, p_0 (y, \alpha) 
  \theta(x - y) \, = \ln \left(\frac{1}{x} + \alpha - 1 \right) - \ln \left(\frac{1}{x} + 
  \frac{1}{\alpha} - 1 \right) \, .
\label{lnqtil}
\eeq
We may then extend our basis by defining the set of functions
\beq
  M_{k,l,n} (\alpha) \, = \, \frac{1}{r (\alpha)} \int_0^1 dx \, p_0 (x, \alpha) \ln^k 
  \bigg( \frac{q (x, \alpha)}{x^2} \bigg) \ln^{l} \bigg( \frac{x}{1 - x} \bigg) \ln^n 
  \tilde{q}(x, \alpha) \, .
\label{eq:Mbasis}
\eeq
These functions have uniform weight ${\rm w} = k + l + n + 1$, and we will see below 
that $M_{k,l,n} (\alpha)$ still gives rise to the required alphabet composed of $\alpha$ 
and $\eta = \alpha/(1 - \alpha^2)$. Clearly, these functions are consistent with our 
constraints. Note also that $\ln \tilde{q} (x, \alpha)$ is a natural addition to the previous 
basis: as shown in \eqn{lnqtil}, it is precisely the difference of the same two logarithms 
whose sum is given by the factor in \eqn{allog}. Notice also that $\ln \tilde{q}(x, \alpha)$ 
is odd under $\alpha \to 1/\alpha$, while the full result for every web must be even, 
because of the relation between $\alpha$ and $\gamma$. Therefore if this logarithm 
appears raised to an odd power, the symmetry in the corresponding contribution will 
have to be restored by some other factor in the result, for example a factor of $r(\alpha)$, 
or multiplication by another function of the basis also odd under the same transformation.

We can now revisit \eqn{webkin22}, and express the result for the (2,2) web in terms of 
the basis in \eqn{eq:Mbasis}), as
\beq
  \overline{w}^{(2)}_{(2,2)} \, = \,\left(\frac{1}{4\pi}\right)^2 \frac{N_c}{2} \, \,
  T_1 \cdot T_2 \, \, F^{(2)}_{(2,2)} (\alpha) \, , \qquad \quad  
  F^{(2)}_{(2,2)} (\alpha) \, = \, 2 \, r^2 (\alpha) \, M_{0,1,1} (\alpha) \, .
\label{F22}
\eeq
By using the explicit expression for the function $M_{0,1,1}(\alpha)$ given in 
Appendix~\ref{appBasis}, we see that this result agrees with the calculation reported 
in~\cite{Korchemsky:1987wg,Kidonakis:2009ev}. 

Having fixed the basis, we can readily express all previously computed subtracted webs 
in terms of the first few basis functions. To begin with, the functions in \eqn{RUSigma} can 
be expressed in terms of \eqn{eq:Mbasis} as
\begin{align}
 \begin{split}
  R_0 (\alpha) & =  M_{0,0,0} (\alpha) \, , \quad \, \, \, \, \,
  U_1 (\alpha)  =  M_{1,0,0} (\alpha)\, ,  \\
  \Sigma_2 (\alpha)  &=  \frac12 M_{0,2,0}(\alpha) \, , \quad 
  U_2 (\alpha) \, = \, \frac14 M_{2,0,0}(\alpha) \, .
 \end{split}
\label{aloldfun}
\end{align}

\begin{figure}[htb]
\centering
\scalebox{0.5}{\includegraphics{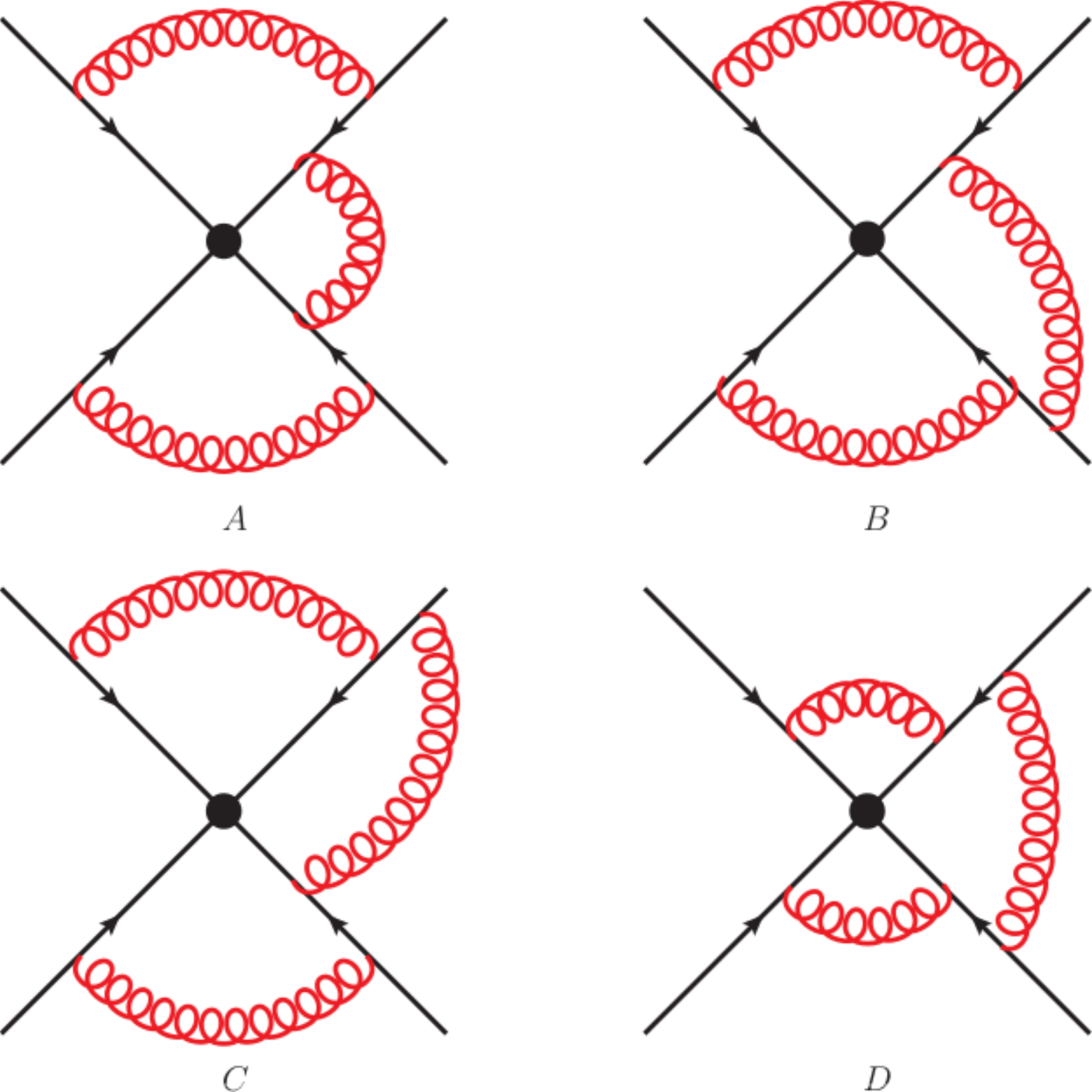}}
\caption{The (1,2,2,1) web, connecting four Wilson lines at three loops.}
\label{fig:1221}
\end{figure}
\begin{figure}[htb]
\begin{center}
\scalebox{0.8}{\includegraphics{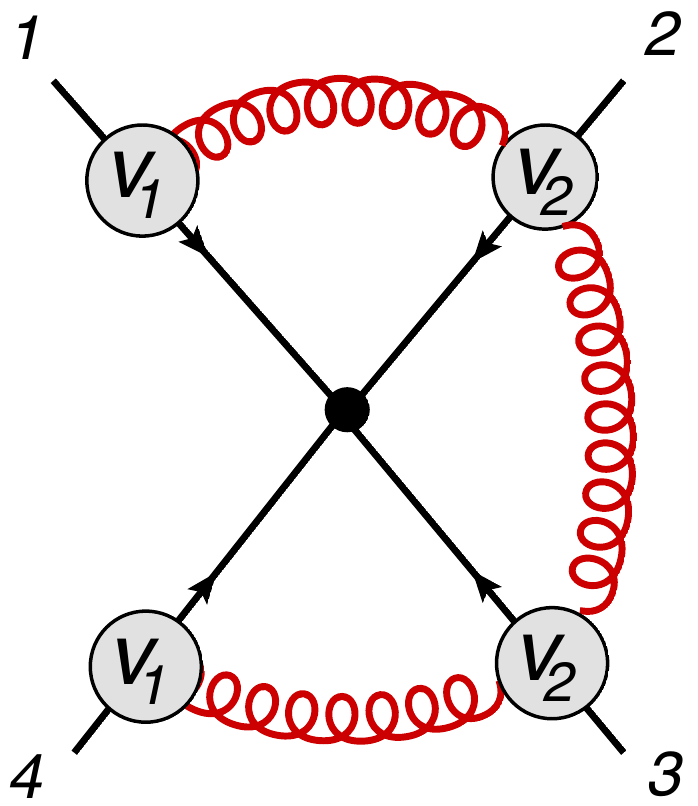}}
\caption{The (1,2,2,1) web in the effective vertex formalism.}
\label{effver1221}
\end{center}
\end{figure}
Further, the (1,2,1) subtracted web computed in~\cite{Ferroglia:2009ep,
Ferroglia:2009ii,Mitov:2010xw,Chien:2011wz,Gardi:2013saa} can be written as
\beqa
\label{two_loop_121_sum}
  \overline{w}_{121}^{(2,-1)} & = & - \, {\rm i} \, f^{abc} T_i^aT_j^bT_k^c \, 
  \left( \frac{1}{4\pi} \right)^2 \, \frac12 \, r(\alpha_{ij}) \, r(\alpha_{jk}) \,
  \\ & & \hspace{1cm} \times \, \Big(M_{0,0,0} (\alpha_{ij}) M_{1,0,0}(\alpha_{jk}) - 
  M_{0,0,0} (\alpha_{jk}) M_{1,0,0} (\alpha_{ij})\Big) \, ,
  \nonumber
\eeqa

Next, the results of the (1,2,2,1) and (1,1,1,3) subtracted webs, computed in 
Ref.~\cite{Gardi:2013saa}, can be expressed in the new basis as follows.
For the (1,2,2,1) web of Fig.~\ref{fig:1221}, whose colour structure in the 
effective vertex formalism is shown in Fig.~\ref{effver1221}, one finds
\beqa
\label{1221gen}
  \overline{w}^{(3)}_{(1,2,2,1)} \left( \alpha_{12}, \alpha_{23}, \alpha_{34} \right) & = &
  - \frac16  \, f_{abe} f^e_{\phantom{e}cd} T_1^a T_2^b T_3^c T_4^d \,\,
  \left( \frac{1}{4 \pi} \right)^3 r(\alpha_{12}) \, r(\alpha_{23}) \, r(\alpha_{34})
  \nonumber \\
  && \hspace{2cm} \times \,G_{(1,2,2,1)} \left( \alpha_{12}, \alpha_{23}, \alpha_{34} 
  \right) \, ,
\eeqa
where
\beqa
\label{1221genG}
  && \hspace{-1cm} G_{(1,2,2,1)} \left( \alpha_{12}, \alpha_{23}, \alpha_{34} \right) \, = \,  
  - \, \frac12 \, M_{2,0,0} (\alpha_{12})  M_{0,0,0} (\alpha_{23})  M_{0,0,0} (\alpha_{34}) 
  \label{1221}  \\  && 
  - \, \frac12 \, M_{2,0,0} (\alpha_{34}) M_{0,0,0} (\alpha_{12}) M_{0,0,0} (\alpha_{23})
  + M_{2,0,0} (\alpha_{23}) M_{0,0,0}(\alpha_{12}) M_{0,0,0}(\alpha_{34}) 
  \nonumber \\ &&
  - \, M_{0,0,0} (\alpha_{12}) M_{1,0,0} (\alpha_{23}) M_{1,0,0} (\alpha_{34})
  - M_{0,0,0} (\alpha_{34}) M_{1,0,0} (\alpha_{12}) M_{1,0,0} (\alpha_{23}) 
  \nonumber \\ &&
  + \, 2 \, M_{0,0,0} (\alpha_{23}) M_{1,0,0} (\alpha_{12}) M_{1,0,0} (\alpha_{34})
  - 4 \, M_{0,2,0} (\alpha_{23}) M_{0,0,0} (\alpha_{12}) M_{0,0,0} (\alpha_{34}) \, . 
  \nonumber
\eeqa

\begin{figure}[htb]
\begin{center}
\scalebox{0.6}{\includegraphics{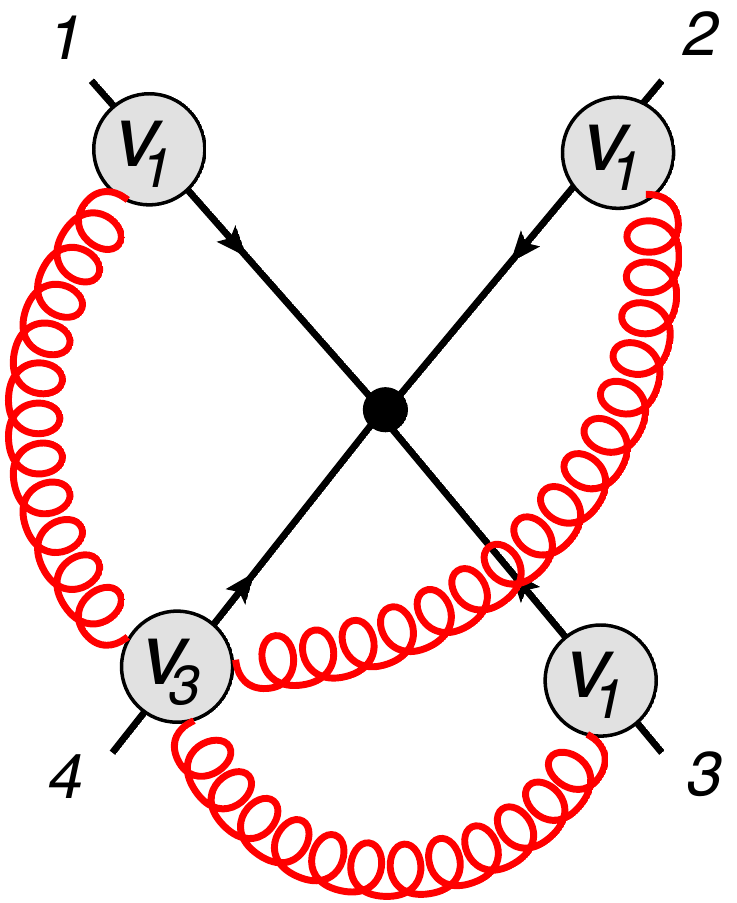}}
\caption{Effective vertex diagram for the (1,1,1,3) web. }
\label{fig:1113eff}
\end{center}
\end{figure}

For the (1,1,1,3) web, whose colour structure is depicted in figure~\ref{fig:1113eff}, one finds
\beqa
\label{1113gen}
  \overline{w}^{(3)}_{(1,1,1,3)} \left( \alpha_{14}, \alpha_{24}, \alpha_{34} \right)
  & = & - \frac16 \, T_1^a T_2^b T_3^c T_4^d  \, \left( \frac{1}{4 \pi} \right)^3
  r(\alpha_{14}) \, r(\alpha_{24}) \, r(\alpha_{34}) \nonumber \\
  && \hspace{-4cm} \times \, \Big[ f_{ade} f^e_{\phantom{e}bc} \, G_{(1,1,1,3)} 
  \left( \alpha_{14}, \alpha_{24}, \alpha_{34} \right) + f_{ace} f^e_{\phantom{e}bd} \,
  G_{(1,1,1,3)} \left( \alpha_{24}, \alpha_{14}, \alpha_{34} \right) \Big] \, ,
\eeqa
where
\beqa
 \label{1113}
  && \hspace{-1cm} G_{(1,1,1,3)} \left( \alpha_{14}, \alpha_{24}, \alpha_{34} \right) \, = \, 
  \frac12 \, M_{2,0,0} (\alpha_{14}) M_{0,0,0} (\alpha_{24}) M_{0,0,0} (\alpha_{34})
  \\ &&
  + \, \frac12 \, M_{2,0,0} (\alpha_{34}) M_{0,0,0} (\alpha_{14}) M_{0,0,0} (\alpha_{24})
  -  \, M_{2,0,0} (\alpha_{24}) M_{0,0,0} (\alpha_{14}) M_{0,0,0} (\alpha_{34})
  \nonumber \\ &&
  + \, M_{0,0,0} (\alpha_{14}) M_{1,0,0} (\alpha_{24}) M_{1,0,0} (\alpha_{34})
  + \, M_{0,0,0} (\alpha_{34}) M_{1,0,0} (\alpha_{14}) M_{1,0,0} (\alpha_{24})
  \nonumber \\ &&
  - \, 2 \, M_{0,0,0} (\alpha_{24}) M_{1,0,0} (\alpha_{14}) M_{1,0,0} (\alpha_{34}) \, .
  \nonumber
\eeqa
We note that, while the weight of $G$ is ${\rm w} = 2 n - 1$, these webs span $n+1$ 
Wilson lines at order $n$, involving $n$ individual gluons, each depending on a separate 
$\alpha_{i j}$. Thus the weight w is partitioned so that each term involves a function 
of ${\rm w} \geq 1$ for each of the $n$ gluons. Consequently, we only encounter functions 
up to weight ${\rm w} = n$. To explore the validity of the basis beyond weight 3 we need to 
either consider more entangled webs spanning fewer lines, where fewer, but higher weight 
functions enter, or explore higher loop corrections. In the following we will do both.  

As a first step, we need to generate the basis functions up to weight five (this will be 
sufficient for the calculations we present in this paper). Before doing so, however, we 
must note that not all functions $M_{k,l,n}$ are independent, and we must discuss the 
relevant degeneracies. As an example, in the $n = 0$ case one finds that
\beq
  M_{k, 2 \lambda + 1, 0} (\alpha)  \, = \, - \, \sum_{r = 1}^{k} \left(
  \begin{array}{c} k \\ r \end{array}
  \right) 2^{r - 1} \, M_{k - r, 2 \lambda + 1 + r, 0} (\alpha) \, , 
\label{con1}
\eeq
so, for $n = 0$, we can recursively express all the functions with odd values of $l$, 
in terms of those with even values of $l$. Similarly, we can find relations in the general 
case $n \neq 0$, by considering the symmetry under $x \leftrightarrow 1 - x$. One 
verifies that
\beqa
\label{Mid}
  M_{k,l,n} (\alpha) & = & \frac{1}{r (\alpha)} \int_0^1 d x \, p_0(x, \alpha) \,
  (- 1)^{l + n} \left[ \log \left( \frac{q(x, \alpha)}{x^2} \right) + 2 \log \left( \frac{x}{1 - x} \right) 
  \right]^k \nonumber \\ & & \hspace{2cm} \times \, \left[ \log \left( \frac{x}{1 - x} \right)
  \right]^l \Big[ \log( \tilde{q}(x, \alpha)) - 2 \log(\alpha) \Big]^n \, .
\eeqa
By expanding the integrand in \eqn{Mid} we obtain then the general relation
\beq
  M_{k,l,n} (\alpha) \, = \, (-1)^{l + n} \sum_{r = 0}^k \sum_{s = 0}^n 
  \left( \begin{array}{c} k \\ r \end{array} \right)
  \left( \begin{array}{c} n \\ s \end{array} \right)
  2^{s + r} (- 1)^s  \log^s (\alpha) \, M_{k - r,l + r,n - s} (\alpha) \, .
\label{con2}
\eeq
Once again, we can express $M_{k,l,n}$, with $l + n$ odd, in terms of the basis functions 
with $l + n$ even, and lower weights. Using these relations it is easy to derive the set 
of independent functions up to any desired weight. For example, at weight two, we have 
the relation
\beq
  M_{0,0,1} (\alpha) \, = \, \frac{1}{2} M_{0,0,0}^2(\alpha) \, ,
\label{relw2}
\eeq
and at weight three we find
\beqa
\label{relw3}
  M_{1,1,0} (\alpha) & = & - \, M_{0,2,0} (\alpha) \, , \nonumber \\
  M_{1,0,1} (\alpha) & = & - \, M_{0,1,1} (\alpha) + \frac{1}{2} \, M_{0,0,0} (\alpha) 
  M_{1,0,0}(\alpha) \, .
\eeqa
Using these relations to eliminate redundant entries, we give the basis functions up 
to weight ${\rm w} = 5$ in Table~\ref{tab:basis}. The table presents the symbol of 
each function, while explicit expressions in terms of classical and harmonic 
polylogarithms~\cite{Remiddi:1999ew} are given in Appendix~\ref{appBasis}. All 
functions have the required symbol alphabet; consequently, they can all be expressed 
in terms of harmonic polylogarithms with entries $0$ and $1$.
\begin{table}
\begin{center}
\begin{tabular}{|c|c|c|}
  \hline
  \multicolumn{3}{ | c | }{$M_{k, l,n} (\alpha)$} \\ 
  \hline 
  w & Name & Symbol \\ 
  \hline \hline
  1 & $M_{0,0,0}$ & $ 2 \, (\otimes \alpha )$ \\ 
  \hline \hline
  2 & $M_{1,0,0}$ & $ - 4 \,\alpha \otimes \eta$  \\ 
  \hline \hline
  \multirow{4}{*}{3} 
  & $M_{0,0,2}$ & $ 16 \, \alpha \otimes \alpha \otimes \alpha $ \\ 
  \cline{2-3}
  & $M_{0,1,1}$ & $ - 4 \, \alpha \otimes \eta \otimes \alpha$  \\ 
  \cline{2-3}
  & $M_{0,2,0}$ & $ 4 \, \alpha \otimes \alpha \otimes \alpha$  \\ 
  \cline{2-3}
  & $M_{2,0,0}$ & $ 16 \, \alpha \otimes \eta \otimes \eta$  \\ 
  \hline \hline
  \multirow{4}{*}{4} 
  & $M_{1,0,2}$ & $ - 32 \, \alpha \otimes \alpha \otimes \alpha \otimes \eta$  \\ 
  \cline{2-3}
  & $M_{1,1,1}$ & $ - 16 \, \alpha \otimes \alpha \otimes \alpha \otimes \alpha 
                                 + 8 \, \alpha \otimes \eta \otimes \alpha \otimes \eta 
                                 + 8 \, \alpha \otimes \eta \otimes \eta \otimes \alpha$  \\ 
  \cline{2-3}
  & $M_{1,2,0}$ & $ - 8 \, \alpha \otimes \alpha \otimes \alpha \otimes \eta 
                                - 8 \, \alpha \otimes \eta \otimes \alpha \otimes \alpha$  \\ 
  \cline{2-3}
  & $M_{3,0,0}$ & $ - 96 \, \alpha \otimes \eta \otimes \eta \otimes \eta$  \\ 
  \hline \hline
  \multirow{9}{*}{5} 
  & $M_{0,0,4}$ & $ 768 \, \alpha \otimes \alpha \otimes \alpha \otimes \alpha \otimes \alpha$ \\   
  \cline{2-3}
  & $M_{0,1,3}$ & $ - 96 \, \alpha \otimes \alpha \otimes \alpha \otimes \eta \otimes \alpha 
                                - 96 \, \alpha \otimes \alpha \otimes \eta \otimes \alpha \otimes \alpha 
                                - 96 \, \alpha \otimes \eta \otimes \alpha \otimes \alpha \otimes \alpha$  \\ 
  \cline{2-3}
  & $M_{0,2,2}$ & $   96 \, \alpha \otimes \alpha \otimes \alpha \otimes \alpha \otimes \alpha 
                               + 32 \, \alpha \otimes \eta \otimes \alpha \otimes \eta \otimes \alpha $ \\ 
  \cline{2-3}
  & $M_{0,3,1}$ & $ - 24 \, \alpha \otimes \alpha \otimes \alpha \otimes \eta \otimes \alpha 
                                - 24 \, \alpha \otimes \eta \otimes \alpha \otimes \alpha \otimes \alpha $ \\ 
  \cline{2-3}
  & $M_{0,4,0}$ & $   48 \, \alpha \otimes \alpha \otimes \alpha \otimes \alpha \otimes \alpha $ \\ 
  \cline{2-3}
  & $M_{2,0,2}$ & $ 128 \, \alpha \otimes \alpha \otimes \alpha \otimes \eta \otimes \eta$  \\ 
  \cline{2-3}
  & \multirow{2}{*}{$M_{2,1,1}$} & 
  $ 64 \, \alpha \otimes \alpha \otimes \alpha \otimes \alpha \otimes \eta 
  + 32 \, \alpha \otimes \alpha \otimes \eta \otimes \alpha \otimes \alpha 
  + 32 \, \alpha \otimes \eta \otimes \alpha \otimes \alpha \otimes \alpha$ \\ & &  
  $ - 32 \, \alpha \otimes \eta \otimes \alpha \otimes \eta \otimes \eta 
     - 32 \, \alpha \otimes \eta \otimes \eta \otimes \alpha \otimes \eta 
     - 32 \, \alpha \otimes \eta \otimes \eta \otimes \eta \otimes \alpha$  \\ 
  \cline{2-3}
  & \multirow{2}{*}{$M_{2,2,0}$} & 
  $ 32 \, \alpha \otimes \alpha \otimes \alpha \otimes \alpha \otimes \alpha 
  + 32 \, \alpha \otimes \alpha \otimes \alpha \otimes \eta \otimes \eta 
  + 32 \, \alpha \otimes \eta \otimes \alpha \otimes \alpha \otimes \eta$ \\ & & 
  $ + 32 \, \alpha \otimes \eta \otimes \eta \otimes \alpha \otimes \alpha$  \\ 
  \cline{2-3}
  & $M_{4,0,0}$ & $ 768 \, \alpha \otimes \eta \otimes \eta \otimes \eta \otimes \eta$ \\ 
  \hline
\end{tabular}
\caption{Symbols of all the linearly independent functions of the MGEW basis of 
\eqn{eq:Mbasis}, up to weight ${\rm w} = 5$. We use the shorthand notation 
$\eta = \alpha/(1 - \alpha^2)$.}
\label{tab:basis}
\end{center}
\end{table}
A crucial question at this point is whether further extensions of our proposed basis will
be required at higher orders, when more entangled webs are present. In the following, 
we present several examples of webs at three and four loops, providing evidence that 
the basis of functions in \eqn{eq:Mbasis} is indeed sufficient. We begin by looking at 
the most entangled three-loop web, the (3,3) web involving only two Wilson lines. 


\subsection{Testing the basis: a three-loop, two-line web}
\label{ToCusp}

\begin{figure}[htb]
\begin{center}
\scalebox{0.18}{\includegraphics{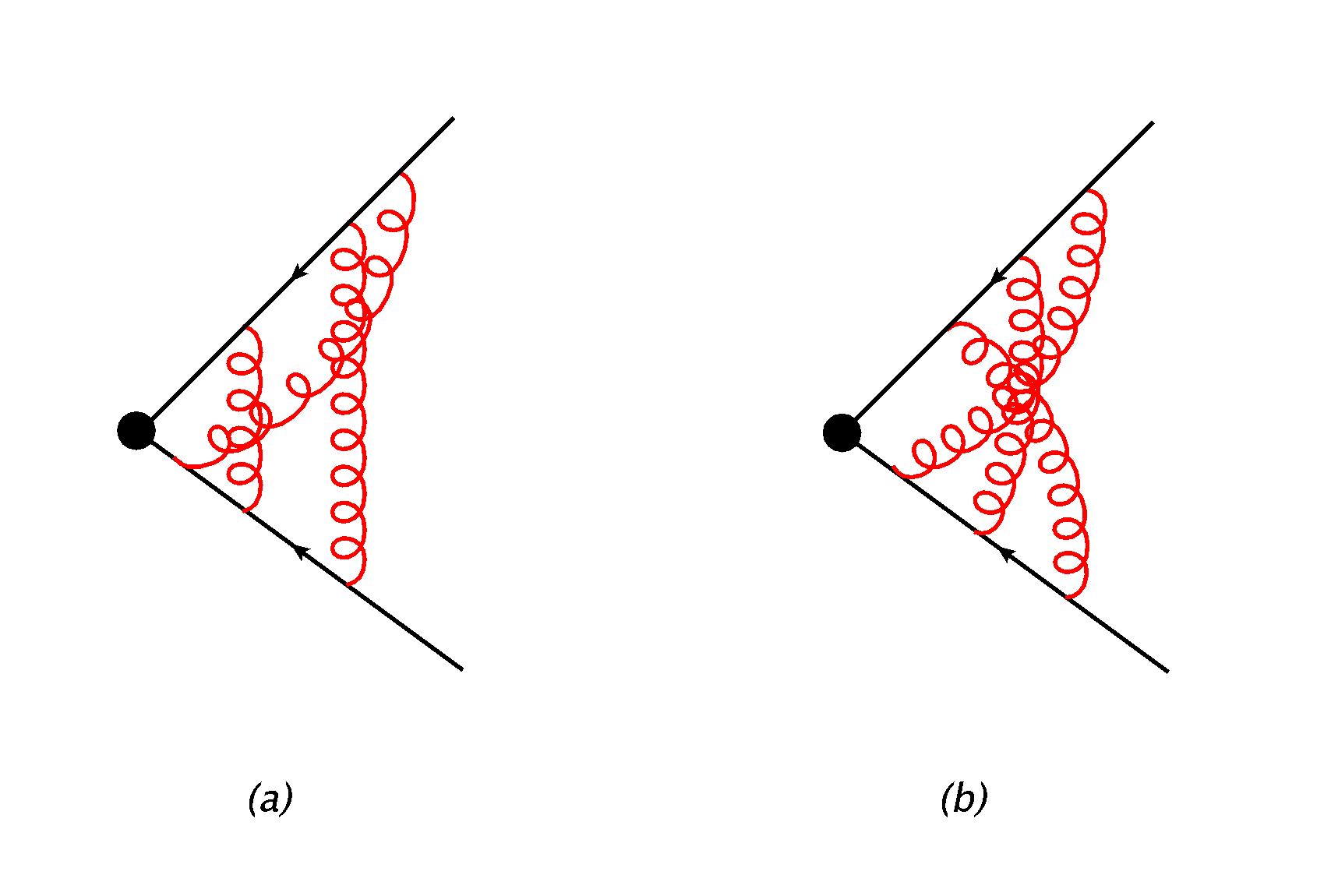}}
\caption{The two diagrams contributing to the (3,3) web at the three loop order. Diagram 
(a) has a twin under the symmetry swapping the two Wilson lines; its kinematic integral 
yields the same function as (a).}
\label{fig:33}
\end{center}
\end{figure}

In the colour singlet channel, the (3,3) web $W_{(3,3)}$ contributes to the three-loop cusp 
anomalous dimension, and it could easily be computed, for example, with the techniques 
of Ref.~\cite{Henn:2013wfa}. For open Wilson lines, the only minor complication is that
two independent colour structures arise, while of course the relevant kinematic integrals
are the same. The diagrams contributing to  $W_{(3,3)}$ are displayed in Fig.~\ref{fig:33}, 
and are denoted by $(a)$ and $(b)$ respectively.
Using the procedure outlined in \sect{MGEWs}, it is straightforward to evaluate the 
contributions of the two diagrams to the web. Since these diagrams are irreducible, 
they each have just a single ultraviolet pole, and we define
\beq
  {\cal F}(a) \, = \, \left( \frac{\alpha_s}{4\pi} \right)^3 \left[ \frac{1}{\epsilon} \, 
  F^{(3)}_{(3,3)} (a) + {\cal O}(\epsilon^0) \, \right] \, , \quad
  {\cal F}(b) \, = \, \left( \frac{\alpha_s}{4\pi} \right)^3 \left[ \frac{1}{\epsilon} \,
  F^{(3)}_{(3,3)} (b) + {\cal O}(\epsilon^0) \, \right] \, .
\label{33struct}
\eeq
In the absence of subdivergences (in the colour singlet case, each diagram separately 
is a `web' in the sense of Refs.~\cite{Gatheral:1983cz,Frenkel:1984pz}), no 
subtractions are needed. The entangled nature of the diagrams, which leads to the 
absence of  subdivergences, also implies, however, that their contributions to the web 
kernel involve two Heaviside functions, as we shall see explicitly below.
\begin{figure}[htb]
\begin{center}
\scalebox{0.7}{\includegraphics{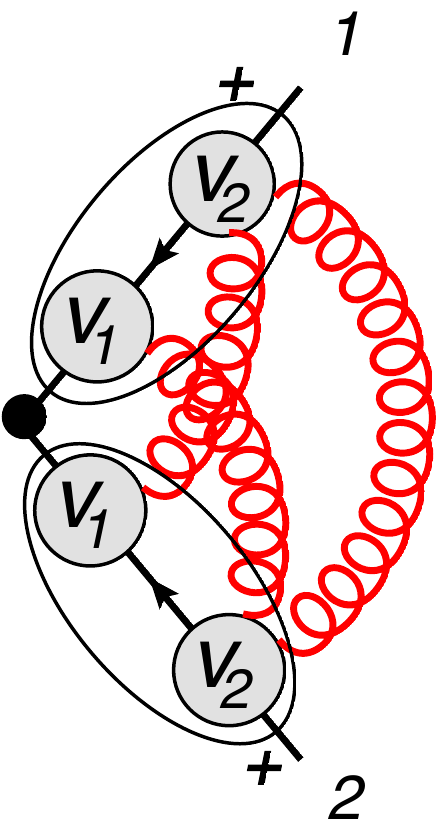}}\hspace*{50pt}
\scalebox{0.7}{\includegraphics{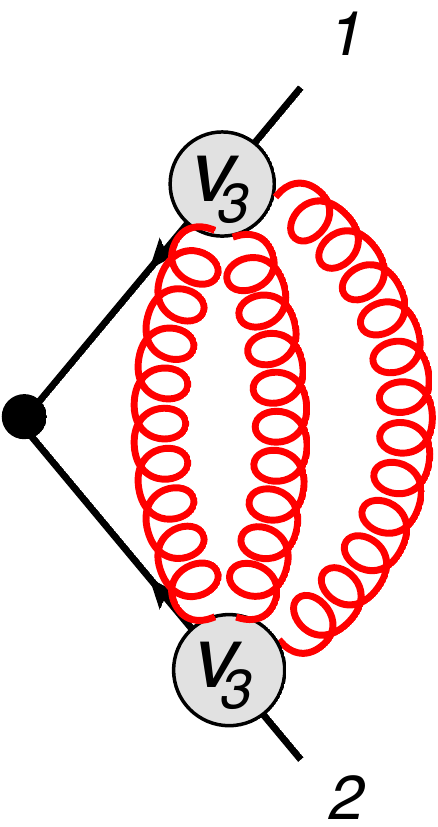}}
\caption{The elements of the (3,3) web in the effective vertex formalism. The diagram 
on the left hand side, where each Wilson line features a (symmetrized) pair of $V_2$ 
and $V_1$ vertices, can be obtained from the (1,2,2,1) web of figure~\ref{effver1221} 
upon taking collinear limits, as explained in the text.} 
\label{fig:33_effective_vertices}
\end{center}
\end{figure}

For open Wilson lines (that is to say, when the hard interaction vertex is not a colour 
singlet), one finds that the (3,3) web involves two independent colour structures. Working 
in the effective colour vertex basis, the result can be written as
\beqa
\label{decomp33}
  \overline{w}^{(3)}_{(3,3)} (\alpha)  & = & \left( \frac{1}{4 \pi} \right)^3 \, 
  \frac14 \,\bigg[ - f_{a b e} \, f^e_{\phantom{e} c d}  \Big\{T_i^a,  T_i^c\Big\} 
  \Big\{T_j^b,  T_j^d\Big\}  \, F^{(3)}_{(V_1 V_2)_+ (V_1 V_2)_+} (\alpha) 
  \nonumber \\ 
  & & \hspace{3cm} + \, {N_c^2} \, T_i \cdot T_j \, 
  F^{(3)}_{V_3 V_3} (\alpha) \bigg]  \, ,
\eeqa
where the first term involves the symmetric combination of a single emission vertex and a 
double emission vertex on each line, while the second term has one triple emission vertex 
per line. These two colour structures are depicted in Fig.~\ref{fig:33_effective_vertices}.
The linear combinations of kinematic integrals corresponding to each colour structure can 
be easily computed from the corresponding web mixing matrix~\cite{Gardi:2013ita}, and 
one finds
\begin{align}
\label{mix33}
\begin{split}
  F^{(3)}_{(V_1 V_2)_+ (V_1 V_2)_+} (\alpha) & =  2 \, F_{(3,3), a}^{(3)} (\alpha) + 
  F_{(3,3),b}^{(3)} (\alpha) \, ,  \\
  F^{(3)}_{V_3 V_3} (\alpha) & =  F_{(3,3), a}^{(3)} (\alpha)+\frac32 F_{(3,3), b}^{(3)} 
  (\alpha) \, , 
\end{split}
\end{align}
where $F^{(3)}_{(3,3), j} (\alpha)$ with $j = a, b$ denotes the contributions of the two
diagrams in Fig.~\ref{fig:33}. Following the steps described in \sect{MGEWs}, we get
\beq
  F^{(3)}_{(3,3), j} (\alpha) \, = \,  \int_0^1 dx \int_0^1 dy   \int_0^1 dz
\, p_0 (x, \alpha)  \, p_0 (y, \alpha)\, p_0 (z, \alpha)\, {\cal G}^{(3)}_{(3,3), j} (x, y, z)
\label{webkin33_initial}
\eeq
with the kernels 
\beqa
\label{kerab}
  \mathcal{G}_{(3,3), \, a}^{(3)} \left(x, y, z \right) & = & - \frac43 \ln^2 \bigg( \frac{x}{1 - x} 
  \frac{1 - z}{z} \bigg) \, \theta(z - x) \, \theta(y - z) \, , \nonumber \\
  \mathcal{G}_{(3,3), \, b}^{(3)} \left(x, y, z \right) & = & - \frac43 \ln \bigg( \frac{x}{1 - x} 
  \frac{1 - y}{y} \bigg) \ln \bigg(\frac{y}{1 - y} \frac{1 - z}{z} \bigg) \, 
  \theta(y - x) \, \theta(z - y) \, .
\eeqa
Using eqs.~(\ref{webkin33_initial}) and (\ref{kerab}) with the combination $F^{(3)}_{(V_1 
V_2)_+ (V_1 V_2)_+} (\alpha)$ from (\ref{mix33}),  yields
\beq
  F^{(3)}_{(V_1 V_2)_+ (V_1 V_2)_+} (\alpha) \, = \, - \frac23 \, r^3(\alpha) \, M_{0,2,0} 
  (\alpha) \, M_{0,0,0}^2(\alpha) \, ,
\label{finsym}
\eeq
which is the final answer for this component of the (3,3) web in \eqn{decomp33}.

According to the general reasoning outlined in \sect{sec:Webs_colour}, we expect this 
result to be reproduced by a two-fold collinear reduction process starting with the (1,2,2,1) 
web. Specifically, in Fig.~\ref{effver1221} we must take Wilson line 1 to be collinear 
to Wilson line 3, and line 4 to be collinear to line 2. It is clear that in this limit the diagram 
degenerates to reproduce the first configuration in Fig.~\ref{fig:33_effective_vertices}, 
provided we take the symmetrized product of the colour factors of the two vertices on 
each line according to \eqn{coll_red}. Considering \eqn{1221gen}, taking the limit requires
identifying $\alpha_{12}$ and $\alpha_{34}$ with $\alpha_{23}$, which we denote in the 
context of the (3,3) web as $\alpha$. This yields
\beqa
\label{double_coll_redu_of_1221gen}
  & & \overline{w}_{(1,2,2,1)} \left( \alpha_{12}, \alpha_{23}, \alpha_{34} \right)
  \begin{array}{c} 
  \\ \longrightarrow \\ _{1||3,\, 2||4}
  \end{array}
  - \, \frac16  \, f_{abe} f^e_{\phantom{e}cd} \, \, 
  \frac12 \big\{ T_1^a, T_1^c \big\} \,  
  \frac12 \big\{ T_2^b, T_2^d \big\}  \, \, \\
  & & \hspace{6cm} \times \left( \frac{1}{4 \pi} \right)^3 r^3(\alpha) \,G_{(1,2,2,1)} 
  \left( \alpha, \alpha, \alpha \right) \, , \nonumber 
\eeqa
where $G_{(1,2,2,1)}(\alpha,\alpha,\alpha)$ was defined in \eqn{1221genG}. It is 
easy to check that this collinear reduction result exactly reproduces the first term 
in \eqn{decomp33}, with the kinematic function obtained in \eqn{finsym} through 
a direct calculation. 

Notice that a direct calculation of the (subtracted) web yields in general a combination 
of polylogarithms that may not be immediately identified in terms of our basis functions. 
In order to express the results in terms of the basis, it is very useful to construct the symbol 
of the result, and then use the properties of the symbol map~\cite{Goncharov.A.B.:2009tja,
Goncharov:2010jf,Duhr:2011zq}, and more generally of the co-product structure described 
in Ref.~\cite{Duhr:2012fh}. We emphasise that the use of these algebraic methods to 
manipulate polylogarithmic functions is merely an intermediate step, as the final goal is 
always to find the result for the subtracted web as an analytic function, written as a 
sum of products of basis elements $M_{k,l,n}$ with numerical rational coefficients. In 
the case of \eqn{finsym}, the symbol is very simple
\beq
  \mathcal{S} \left[ \frac{1}{r^3(\alpha)}\, F^{(3)}_{(V_1 V_2)_+ (V_1 V_2)_+} (\alpha) 
  \right] \, = \,  -\frac{640}{3} \,
  \alpha \otimes \alpha \otimes \alpha \otimes \alpha \otimes \alpha \, \,.
\label{symsym}
\eeq
Note however that the identification of the result, at function level, in terms of the basis 
is not fully determined by the symbol: for example, in addition to the correct result in 
\eqn{finsym}, also $M_{0,0,0}^5(\alpha)$ has a symbol proportional to \eqn{symsym}. 
This illustrates the well known fact that the symbol is not sufficient to control lower-weight 
functions multiplied by transcendental constants such as $\zeta(n)$. Such terms however
can be easily recovered using the co-product technique, along with a numerical evaluation 
of the integrals. 

We can now turn to the more interesting case of the $V_3 V_3$ colour structure, which is 
novel, in the sense that it cannot be derived from collinear reduction of less entangled webs.
It is not obvious a priori that our proposed basis suffices for this kinematic function, 
since now two integrals over the `propagator' functions $p_0$ are cut off by the Heaviside 
functions appearing in \eqn{kerab}. Having two Heaviside functions, this web is clearly 
more entangled than the ones considered so far, thus providing a non-trivial test of the 
generality of the basis.

It is not difficult to perform the required integrals, yielding, as expected, a combination of 
polylogarithms of uniform weight ${\rm w} = 5$. In order to map the result to our basis, 
we compute the symbol, which is given by 
\beqa
\label{sym33}
  \mathcal{S} \bigg[ \frac{1}{r^3(\alpha)} \, F_{V_3 V_3}^{(3)} (\alpha) \bigg] 
  & = & - \frac{64}{3} \, \bigg[ \,
  4 \, \Big( \alpha \otimes \eta \otimes \eta \otimes \alpha \otimes \alpha + 
  \alpha \otimes \eta \otimes \alpha \otimes \eta \otimes \alpha \Big) \\ 
  & & \hspace{3cm} - \,  
  \alpha \otimes \alpha \otimes \alpha \otimes \alpha \otimes \alpha \, \bigg] \, .  
  \nonumber
\eeqa
Expressing the result in our basis is now an algebraic problem. We find that the 
$M_{k,l,n}$ basis is sufficient to express the $V_3V_3$ function, and the resulting 
expression is
\beqa
\label{eq:V3V3}
  F_{V_3 V_3}^{(3)} (\alpha) & = & -\frac43 \, r^3(\alpha) \, \bigg[ \frac{1}{4} M_{0,0,0}^2 (\alpha) 
  M_{2,0,0}(\alpha) - \frac{1}{4} M_{0,0,0} (\alpha) M_{1,0,0}^2 (\alpha) + M_{0,0,0} (\alpha) 
  M_{1,1,1}(\alpha) \nonumber \\ & & \hspace{-1cm} 
  - \, M_{0,1,1} (\alpha) M_{1,0,0} (\alpha) + \frac{3}{2} M_{0,2,2} (\alpha) 
- \frac14 M_{0,0,0}^2 (\alpha) M_{0,2,0} (\alpha) + \frac{1}{48} M_{0,0,0}^5 (\alpha) \bigg] \, .
\eeqa
For future reference, let us summarise the results we obtained for MGEWs in the two-line 
case, with arbitrary colour exchange at the cusp, at the level of the anomalous dimension.
We find
\begin{subequations}
 \begin{align}
  \left. \alpha_s \, \Gamma_2^{(1)} (\alpha) \right\vert_{(1,1)} 
  & = \frac{\alpha_s}{\pi} \, T_1 \cdot T_2 \, r(\alpha) \, \ln ( \alpha ) \\
  \left. \alpha_s^2 \, \Gamma_2^{(2)} (\alpha) \right\vert_{(2,2)}  
  &= - 4 \left( \frac{\alpha_s}{4 \pi} \right)^2 \, N_c \, T_1 \cdot T_2 \, 
  r^2(\alpha) \, M_{0,1,1} (\alpha) \\
 \begin{split}
  \left. \alpha_s^3 \, \Gamma_2^{(3)} (\alpha) \right\vert_{(3,3)}  
  & = - \frac32 \left(\frac{\alpha_s}{4\pi}\right)^3 \bigg[ -
   f_{a b e} \, f^e_{\phantom{e} c d}  \, \big\{T_1^a,  T_1^c \big\} \big\{T_2^b, T_2^d \big\} \, 
  F^{(3)}_{(V_1 V_2)_+ (V_1 V_2)_+} (\alpha) \\
  & \hspace{3cm} + {N_c^2} \, T_1 \cdot T_2 \, 
  F^{(3)}_{V_3 V_3} (\alpha) \bigg]  \, ,
 \end{split}
 \end{align}
\label{Gamma_2_line}
\end{subequations}
with the two functions in the three-loop result given in Eqs.~(\ref{finsym}) and 
(\ref{eq:V3V3}). The results agree with previous calculations. In particular, at the 
three-loop level, the result for the colour singlet projection of the (3,3) web can be 
read off from eq. (28) in Ref.~\cite{Correa:2012nk}: it is given by the coefficient of 
$\xi^3$ in that expression\footnote{The calculation in~\cite{Correa:2012nk} is done 
for ${\cal N} = 4$ Super Yang-Mills theory, with supersymmetric Wilson lines, but 
one may extract the Yang-Mills limit by choosing the directions of the scalar fields 
in the internal space on the two Wilson lines to be perpendicular to each other, in 
which case $\xi$ maps to our rational factor $r(\alpha)$. The highest power of 
$r(\alpha)$ is fully determined by MGEWs, and at three loops by the (3,3) web 
alone.}. In order to project our result (\ref{decomp33}) onto the colour singlet 
case, we simply need to substitute $T_2 = - T_1$, which guarantees colour 
conservation at the cusp. The three-loop result is
\beq
  \left. \alpha_s^3 \, \Gamma_2^{(3)} (\alpha) \right\vert_{(3,3)}  \, = \, 
  \frac32 \left( \frac{\alpha_s}{4 \pi} \right)^3 \, N_c^2 \, C_R \,
  \bigg[ \frac12 F^{(3)}_{(V_1 V_2)_+ (V_1 V_2)_+} (\alpha) +
  F^{(3)}_{V_3 V_3} (\alpha) \bigg] \, ,
\label{Gamma_2_line_colour_singlet}
\eeq
where $C_R$ is the quadratic Casimir eigenvalue of representation $R$, corresponding 
to $T_i \cdot T_i$. The result is in full agreement with Ref.~\cite{Correa:2012nk}.


\section{Results for three-loop, three-line webs}
\label{Threeloop}

\begin{figure}[htb]
\begin{center}
\scalebox{0.9}{\includegraphics{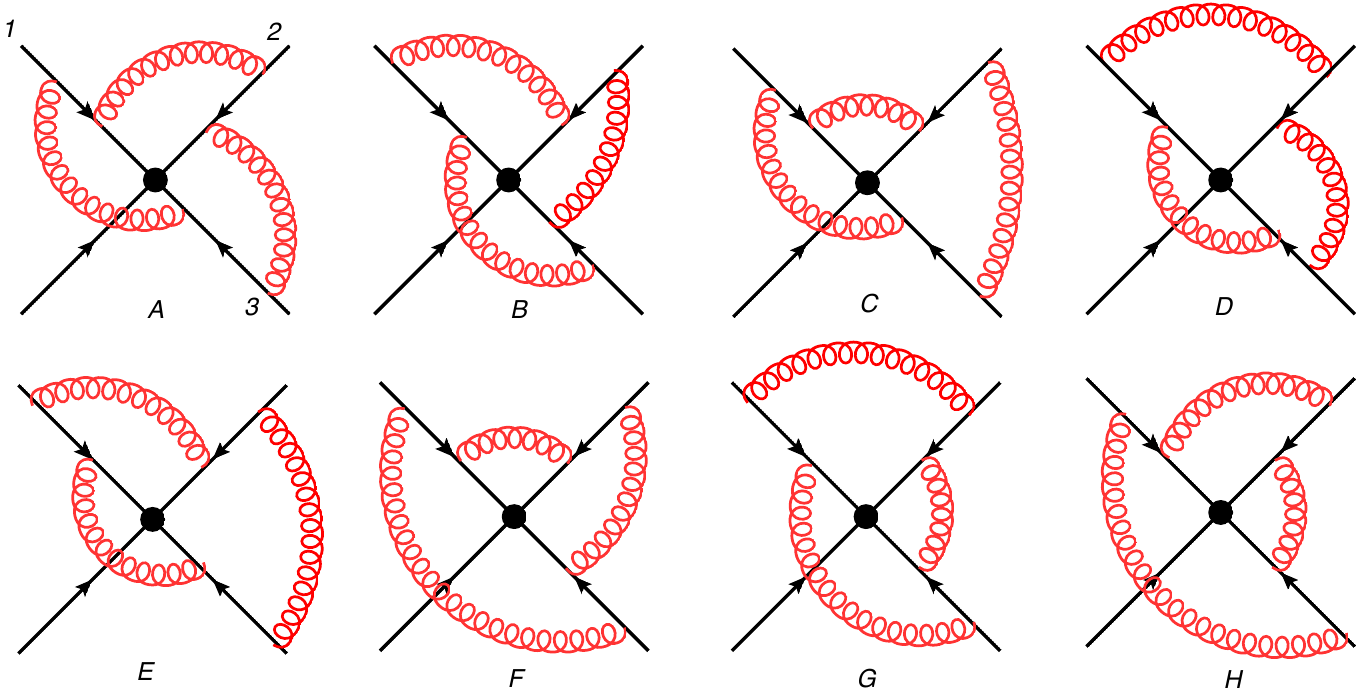}}
\caption{The (2,2,2) web connecting three Wilson lines at three-loop order.}
\label{fig:222}
\end{center}
\end{figure}

\begin{figure}[htb]
\begin{center}
\scalebox{0.7}{\includegraphics{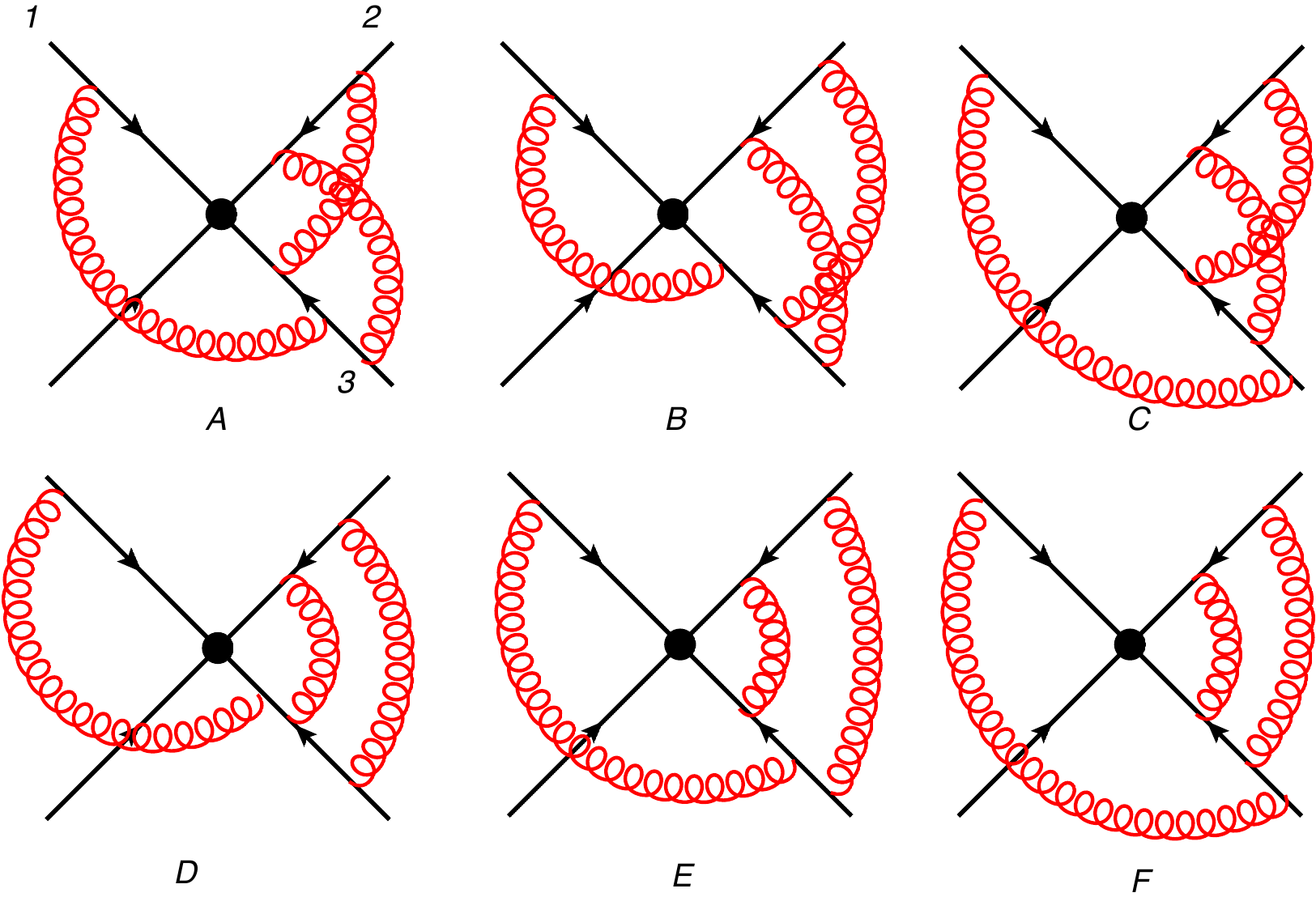}}
\caption{The (1,2,3) web connecting three Wilson lines at three-loop order.}
\label{fig:123}
\end{center}
\end{figure}

In this section, we present the calculation of the three-loop, three-line webs of 
Figs.~\ref{fig:222} and~\ref{fig:123}, and the corresponding contributions to the 
soft anomalous dimension. While the calculations are lengthy, they closely follow 
the steps described in \sect{Webs}: we can therefore concentrate on the results 
and on the role of the basis functions defined in \sect{Basis} above. The most 
important intermediate steps are summarised in two appendices, Appendix~\ref{App222} 
for the (2,2,2) web and Appendix~\ref{App123} for the (1,2,3) web. We choose 
our conventions so that both webs connect lines 1, 2 and 3, counting 
clockwise from top-left in Figs.~\ref{fig:222} and~\ref{fig:123}. A suitable basis for 
the colour factors of all three-loop three-line webs is~\cite{Gardi:2013ita}
\begin{align}
\label{colfacs3}
\begin{split}
  c^{(3)}_1 & =  \{T_1^a, T_1^b\} [T_2^b, T_2^c] [T_3^a, T_3^c] \, , \\
  c^{(3)}_2 & =  [T_1^a, T_1^b] \{T_2^b, T_2^c\} [T_3^a, T_3^c] \, , \\
  c^{(3)}_3 & =  [T_1^a, T_1^b] [T_2^b, T_2^c] \{T_3^a, T_3^c\} \, ,  \\
  c^{(3)}_4 & =  [T_1^a, T_1^b] [T_2^b, T_2^c] [T_3^a, T_3^c]  \, .
\end{split}
\end{align}
Note that terms with more than one anticommutator cannot occur because they 
would correspond to disconnected colour diagrams.


\subsection{The (2,2,2) web}
\label{222}

The (2,2,2) web potentially contributes to all four colour factors in the basis of 
\eqn{colfacs3}. As a consequence, one may write the unsubtracted web as
\beq
  W_{(2,2,2)} \left( \alpha_{12}, \alpha_{23}, \alpha_{13} \right) \, = \,
  \sum_{i = 1}^4 c_i^{(3)} \, {\cal F}_{(2,2,2); i} \left( \alpha_{12}, \alpha_{23},
  \alpha_{13},\epsilon \right) \, .
\label{W222res}
\eeq
The combinations of kinematic factors accompanying each colour factor are collected 
in Table~\ref{tab:222}. These form the web kernel, which  is then combined with 
appropriate commutators, to form the subtracted web. We present the details of this 
calculation in appendix~\ref{App222}, while here we discuss the results. 

Using the specified colour basis the subtracted web takes the form
\beq
  \overline{w}_{(2,2,2)} \left( \alpha_{12}, \alpha_{23}, \alpha_{13} \right) \, = \, 
  \left(\frac{1}{4\pi}\right)^3\,
  \sum_{i = 1}^4 \, c_i^{(3)} \, F_{(2,2,2); i} \left( \alpha_{12}, \alpha_{23},
  \alpha_{13} \right) \, ,
\label{W222res_sub}
\eeq
where
\beqa
\label{F222}
  F_{(2,2,2);i} \left( \alpha_{12}, \alpha_{23}, \alpha_{13} \right) & = &
  \int_0^1 dx_1 dx_2 dx_3 \, p_0(x_1,\alpha_{12}) \, p_0( x_2, \alpha_{23}) \,
  p_0(x_3,\alpha_{13})  \\ 
  & & \hspace{5mm} \times \, {\cal G}_{(2,2,2); i} \left(x_1, x_2, x_3, q(x_1,\alpha_{12}), 
  q(x_2,\alpha_{23}) , (x_3,\alpha_{13}) \right) \, . \nonumber
\eeqa
The subtracted web kernels ${\cal G}_{(2,2,2);i}$, defined in \eqn{subtracted_web_mge_kin}, 
will be given below. Note that, in contrast to the two-line cases analysed in the previous 
section, where ${\cal G}_W$ depends only on the $x_i$ variables, here the subtracted web 
kernel depends on $q (x_i, \alpha_{ij})$ as well: this dependence is related to the presence 
of subdivergences in these webs. We will see that the subtracted web kernels depend on 
their arguments via powers of logarithms only, as anticipated in Ref.~\cite{Gardi:2013saa}.  
This simple structure emerges  through the cancellation of all polylogarithms amongst the 
various diagrams and commutators, when the subtracted web kernel is assembled (see 
for example \eqn{FC222c} and eqs.~(\ref{comms1}) through (\ref{comms3}), respectively).
This simplification is responsible for the factorized structure of the final result.

To express the subtracted web kernels in compact form, we define the logarithmic functions
\beq
  L_{ij} \, \equiv \, \log \left( \frac{q (x_i, \alpha_{ij})}{x_i^2} \right)\,;
  \qquad 
  R_i \, \equiv \, \log \left( \frac{x_i}{1 - x_i} \right) \, .
\label{logdef}
\eeq
In terms of these functions, the subtracted web kernels associated to 
the first three colour factors are
\beqa
\label{calG222} 
  {\cal G}_{(2,2,2), \, 1}^{(3)}  & = &
  \frac{1}{3} \, \bigg[ R_2^2 - \frac{1}{4} L_{2 3}^2 + \frac{1}{8 }L_{1 2}^2 + 
  \frac{1}{8} L_{3 1}^2 + \frac{1}{4} L_{1 2} L_{2 3} - \frac{1}{2} L_{3 1} L_{1 2} + 
  \frac{1}{4} L_{2 3} L_{3 1} \bigg] \, , \nonumber \\
  {\cal G}_{(2,2,2), \, 2}^{(3)}  & = & 
  \frac{1}{3} \, \bigg[ R_3^2 - \frac{1}{4} L_{3 1}^2  + 
  \frac{1}{8} L_{2 3}^2 + \frac{1}{8 }L_{1 2}^2
+   \frac{1}{4} L_{2 3} L_{3 1} 
 - \frac{1}{2} L_{1 2} L_{2 3} 
+ \frac{1}{4} L_{3 1} L_{1 2}\bigg] \, , \\
  {\cal G}_{(2,2,2), \, 3}^{(3)}  & = & - \,
  \frac{1}{3} \, \bigg[ R_1^2 - \frac{1}{4} L_{1 2}^2 + \frac{1}{8 } L_{2 3}^2 + 
  \frac{1}{8} L_{3 1}^2 + 
  \frac{1}{4} L_{3 1} L_{1 2} 
 - \frac{1}{2} L_{2 3} L_{3 1} 
+ \frac{1}{4} L_{1 2} L_{2 3}\bigg] \, . \nonumber
\eeqa
As expected from Bose symmetry in \eqn{W222res_sub}, the three functions ${\cal G}_{(2,2,2), \, i}^{(3)}$ can be obtained from each other by permuting the relevant indices; the overall sign for $i=3$ compared to $i=1$ and $2$ reflects the symmetry properties of the corresponding colour factors in \eqn{colfacs3} under cyclic permutations.

In contrast to \eqn{calG222}, the contribution of the (2,2,2) web to the fully antisymmetric 
colour factor $c^{(3)}_4$ is found to vanish,
\beq
  {\cal G}_{(2,2,2), \, 4}^{(3)} \, = \, 0 \, .
\label{G2224}
\eeq
One sees explicitly that each subtracted web kernel consists of products of logarithms 
involving distinct kinematic invariants, consistent with the basis of functions defined in 
\sect{Basis}. It is now straightforward to integrate the results over the `angle' parameters 
$x_i$. In line with \eqn{subtracted_web_form}, we denote the integrated coefficient of 
each colour factor (with factors of $(4 \pi)^3$ removed) by $F_{(2,2,2), \, i}^{(3)}$; the 
result for the first kinematic factor is then
\begin{table}
\begin{center}
\begin{tabular}{|c|c|}
\hline
Colour Factor & Kinematic Feynman Integral ${\cal F}_{(2,2,2), \, i}$ \\
\hline
$c_1^{(3)}$ & $\frac{1}{12} (- 2 A - 2 B + C - 2 D + E - 2 F + G + H)$ \\
$c_2^{(3)}$ & $\frac{1}{12} (- 2 A - 2 B - 2 C + D + E + F - 2 G + H)$ \\
$c_3^{(3)}$ & $\frac{1}{12} (2 A + 2 B - C - D + 2 E - F - G + 2 H)$ \\
$c_4^{(3)}$ & $\frac{1}{2} (- A + B)$ \\
\hline
\end{tabular}
\caption{Kinematic integral associated with each colour factor in
  the (2,2,2) web of Fig.~\ref{fig:222}, where $A \equiv {\cal F}(A)$
  and similarly for $B, C$, etc.}
\label{tab:222}
\end{center}
\end{table}
\beqa
\label{gamints222}
  F^{(3)}_{(2,2,2), \, 1}  \left( \alpha_{12}, \alpha_{23}, \alpha_{13} \right) & = & 
  \frac13 \, r (\alpha_{12})  r (\alpha_{23}) r (\alpha_{13})\, \times
  \nonumber \\ & & \hspace{-3cm} 
 \bigg[- \, 
  M_{0,0,0} (\alpha_{12}) M_{0,0,0} (\alpha_{13}) \bigg( \frac14 \, M_{2,0,0} (\alpha_{23}) 
  - \, M_{0,2,0}(\alpha_{23}) \bigg)
\nonumber \\ & & \hspace{-3cm} 
 + \frac{1}{8} \, M_{0,0,0} (\alpha_{13}) 
  M_{0,0,0} (\alpha_{23}) M_{2,0,0} (\alpha_{12}) 
  \\ & & \hspace{-3cm} 
  + \, \frac{1}{8} \, M_{0,0,0}  (\alpha_{12}) M_{0,0,0} (\alpha_{23}) 
  M_{2,0,0} (\alpha_{13}) - \frac12 \, M_{0,0,0} (\alpha_{23}) M_{1,0,0} (\alpha_{12}) 
  M_{1,0,0} (\alpha_{13}) 
  \nonumber \\ & & \hspace{-3cm}
  + \, \frac{1}{4} \, M_{0,0,0} (\alpha_{13}) M_{1,0,0} (\alpha_{12}) 
  M_{1,0,0} (\alpha_{23}) + \frac{1}{4} \, M_{0,0,0} (\alpha_{12}) 
  M_{1,0,0} (\alpha_{13}) M_{1,0,0} (\alpha_{23}) \bigg] \, . \nonumber 
\eeqa
The second and third kinematic contributions may be obtained via
\beqa
\label{gamints222b}
  F^{(3)}_{(2,2,2), \, 2} \left( \alpha_{12}, \alpha_{23}, \alpha_{13} \right)
  & = & F^{(3)}_{(2,2,2), \, 1} \left( \alpha_{23}, \alpha_{13}, \alpha_{12} \right) \, ,
  \nonumber \\
  F^{(3)}_{(2,2,2), \, 3} \left( \alpha_{12} , \alpha_{23}, \alpha_{13} \right)
  & = & - \, F^{(3)}_{(2,2,2), \, 1} \left( \alpha_{13}, \alpha_{12}, \alpha_{23} \right) \, ,
\eeqa
as follows from the symmetry of the web, and the relabelling of the colour
factors in \eqn{colfacs3}. Finally, the fourth kinematic factor vanishes
\beq
  F^{(3)}_{(2,2,2), \, 4}  \, = \, 0 \, ,
\label{gamints222c}
\eeq
as is clear from the vanishing of the subtracted web kernel in \eqn{G2224}. One
may note that the subtracted kernels ${\cal G}_{(2,2,2), \, i}^{(3)}$ do not contain 
any Heaviside function, despite the fact that individual diagrams (given in 
Appendix~\ref{App222}) contain one for every Wilson line. As a consequence, 
only a subclass of the basis functions is relevant: those without any power of 
$\ln \widetilde{q}(x,\alpha)$.

\begin{figure}[htb]
\begin{center}
\vspace{5mm}
\scalebox{0.45}{\includegraphics{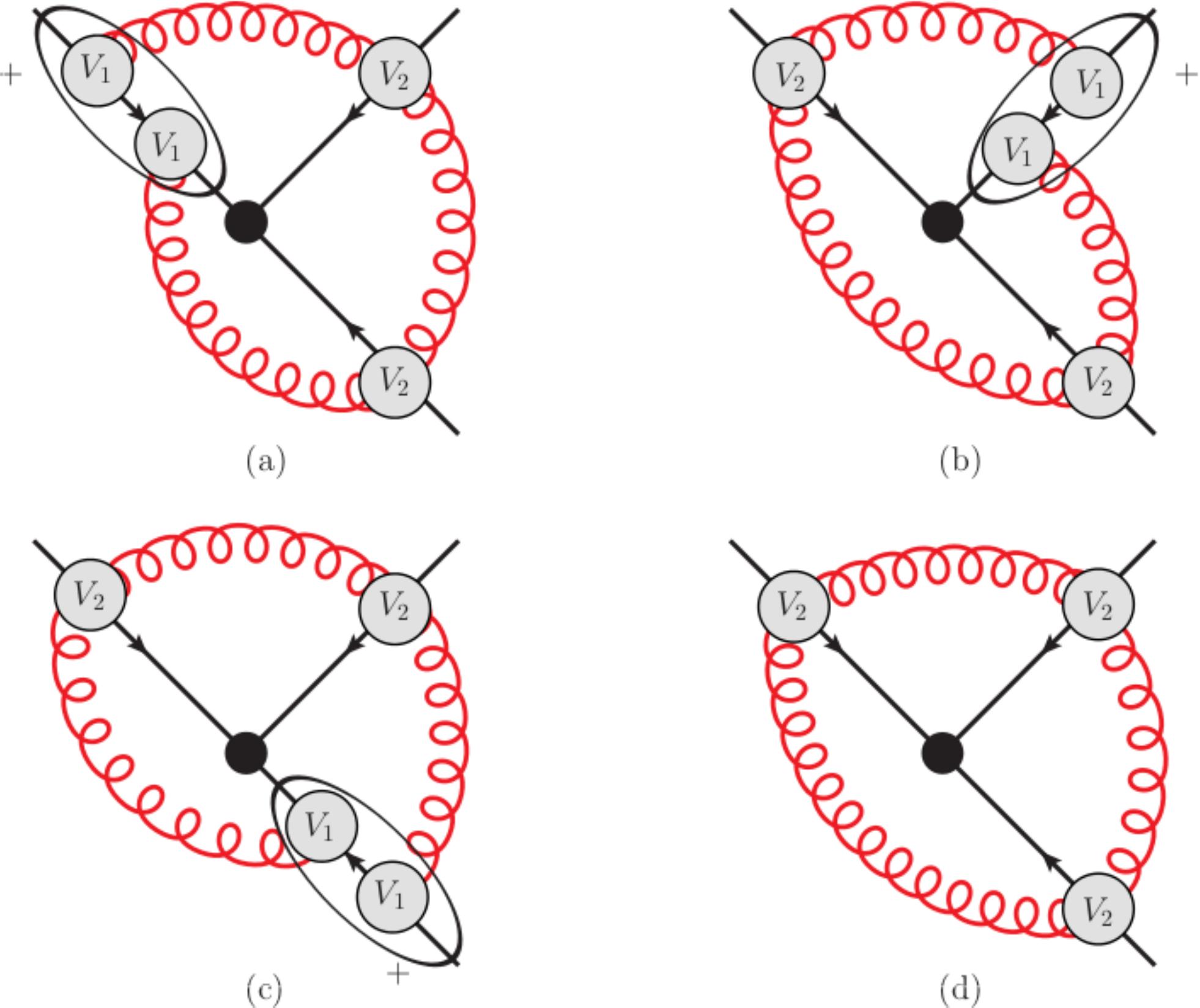}}
\caption{Effective vertex diagrams for the (2,2,2) web. The first three cases, $a,b$ and 
  $c$, can be obtained via collinear reduction from the (1,2,2,1) web of figure~\ref{effver1221} 
  and its permutations.}
\label{fig:222tot}
\end{center}
\end{figure}

Let us now discuss the collinear reduction process, following \sect{sec:Webs_colour}, 
in the context of the (2,2,2) web. We will see that the above results can be derived 
from the (1,2,2,1) web of Fig.~\ref{effver1221}. Indeed, upon taking external lines 
1 and 4 to be collinear, one ends up with the diagram of Fig.~\ref{fig:222tot}(a), 
involving a symmetric combination of one-gluon vertices on line 1. Applying the 
collinear reduction according to \eqn{coll_red}, the colour factor corresponds to 
$c_1^{(3)}$ of \eqn{colfacs3}. Eqns.~(\ref{1221gen}) and (\ref{1221}), with $\alpha_{34} 
\rightarrow \alpha_{13}$, then yield
\beqa
  \label{1221gen_into_222}
  & & \overline{w}^{(3)}_{(1,2,2,1)} \left( \alpha_{12}, \alpha_{23}, \alpha_{34} \right) 
  \begin{array}{c}
  \\ \longrightarrow \\ _{1||4}
  \end{array} 
  \, - \, \frac16  \, f_{abe} f^e_{\phantom{e}cd} \, \, \frac12 \big\{ T_1^a,T_1^d \big\} \, 
  T_2^b \, T_3^c  \,\, \left( \frac{1}{4 \pi} \right)^3 r(\alpha_{12}) \, r(\alpha_{23}) \, 
  r(\alpha_{31}) \nonumber \\
  & & \hspace{6cm} \times \, G_{(1,2,2,1)} \left( \alpha_{12}, \alpha_{23}, \alpha_{31} 
  \right) \, \\ & & \hspace{3cm} = \, - \, c_1^{(3)} \, \frac{1}{12} \left( \frac{1}{4 \pi} \right)^3 
  r(\alpha_{12}) \, r(\alpha_{23}) \, r(\alpha_{31})\, \,G_{(1,2,2,1)} 
  \left( \alpha_{12}, \alpha_{23}, \alpha_{31} \right) \, , \nonumber 
\eeqa
which indeed agrees with \eqn{gamints222}.
The contributions to the colour factors $c_2^{(3)}$ and $c_3^{(4)}$ arise from 
permuting external lines in Fig.~\ref{effver1221}, as was done in \eqn{gamints222b}, so 
clearly these can also be obtained via collinear reductions of the (1,2,2,1) web.
The only contribution that cannot be generated in this way is the effective vertex 
diagram of Fig.~\ref{fig:222tot}(d), which features a $V_2$ vertex on all three lines. 
As explained in \sect{Webs}, diagrams which feature a single effective vertex 
on each line constitute the genuinely new information in a given web, that cannot be 
obtained from collinear reductions of webs connecting more external lines. In the
present case, however, the colour factor of the diagram in Fig.~\ref{fig:222tot}(d) is 
the fully antisymmetric combination $c_4^{(3)}$, and we have seen above that the 
kinematic function associated with this colour structure vanishes. The reason for this 
is that the kinematic function associated with $c_4^{(3)}$ involves only diagrams 
$A$ and $B$ in Fig.~\ref{fig:222}, in the antisymmetric combination ${\cal F}(A) - 
{\cal F}(B)$. Diagrams $A$ and $B$, which were referred to as {\it Escher staircase} 
diagrams in Ref.~\cite{Gardi:2010rn}, are special for several reasons: they are highly 
symmetric, they are chiral enantiomers of each other, and they are fully irreducible: 
one cannot shrink any gluon to the origin without also pulling in the others. Therefore, 
they have no subdivergences\footnote{Because these diagrams do not have any 
subdivergences, their ultraviolet pole can be computed in isolation, yielding a 
regularization-independent result. The first computation of the staircase diagram 
A of the (2,2,2) web in Fig.~\ref{fig:222} was performed by Johannes Henn using a 
cutoff regularization~\cite{Henn:2013wfa}. We would like to thank Johannes Henn
for a very useful exchange in this regard.}, and they do not need any commutator 
counterterms. Finally, their kinematic parts are equal, so that the antisymmetric 
combination vanishes.

In \sect{Escher}, we will be able to construct the kinematic integrals of the Escher
staircases to all loop orders, and we will prove that a similar cancellation (though
with slightly different mechanisms for even and odd numbers of gluons) happens
for any number of gluons. More precisely, we will show that, out of $n + 1$ colour
structures sampled by the $(2^n)$ web, the only one which cannot be obtained
from collinear reduction of $(1, 2^{n-1}, 1)$ webs, which corresponds to a product of $n$
effective vertices of type $V_2$, receives contributions only from the two Escher staircases
which are present for any $n$, and these contributions cancel, so that the corresponding 
kinematic function vanishes. Note however that this discussion does not imply that 
staircase diagrams do not enter the exponent at all. Indeed, as can be seen in 
Table~\ref{tab:222}, they do contribute to the colour factors $c_i^{(3)}$, with 
$1 \leq i \leq 3$.
%


\subsection{The (1,2,3) web}
\label{123}

In this section we focus on the (1,2,3) web of Fig.~\ref{fig:123}. Analogously to the 
(2,2,2) web of the previous section, one may write the unsubtracted web as
\beq
  W_{(1,2,3)} \left( \alpha_{13}, \alpha_{23} \right) \, = \, 
  \sum_{i = 1}^4 c_i^{(3)} \, {\cal F}_{(1,2,3), \, i} \left( \alpha_{13},
  \alpha_{23},\epsilon\right) \, .
\label{W123res}
\eeq
The combinations of kinematic functions of individual diagrams required for 
each colour factor are collected in Table~\ref{tab:123}, and the details of the
calculation of the subtracted web may be found in Appendix~\ref{App123}. 
Here we quote the results.

Note first that according to Table~\ref{tab:123} this web, in the present basis, has 
no projection on the $c_1^{(3)}$ colour factor. We can therefore consider only the 
three components $c_i^{(3)}$ for $i = 2, 3, 4$. The subtracted web is given by
\beq
  \overline{w}_{(1,2,3)} \left( \alpha_{13}, \alpha_{23} \right) \, = \, \left(\frac{1}{4\pi}\right)^3\,
  \sum_{i = 2}^4 c_i^{(3)} \, F_{(1,2,3); \, i} \left( \alpha_{13},
  \alpha_{23} \right) \, ,
\label{W123res_sub}
\eeq
where
\beqa
  F_{(1,2,3); \, i} \left( \alpha_{13}, \alpha_{23} \right) & = &
  \int_0^1 dx_1 dx_2 dx_3 \,\, p_0 (x_1,\alpha_{13}) \, p_0 (x_2,\alpha_{23}) \,
  p_0(x_3,\alpha_{23}) \label{F123} \\
  & & \, \quad \times \,  {\cal G}_{(1,2,3); \, i} \left(x_1, x_2, x_3, 
  q(x_1, \alpha_{13}), q(x_2, \alpha_{23}), q(x_3, \alpha_{23}) \right) \, . \nonumber 
\eeqa
The subtracted kernels are
\begin{table}
\begin{center}
\begin{tabular}{|c|c|}
\hline
Colour Factor & Kinematic Feynman Integral ${\cal F}_{(1,2,3), \, i}$\\
\hline
  $c^{(3)}_1$ & $0$ \\
  $c^{(3)}_2$ & $\frac{1}{12} (2 A - B - C - D + 2 E - F)$ \\
  $c^{(3)}_3$ & $- \frac{1}{12} (4 A + B + C + D - 2 E + F)$ \\
  $c^{(3)}_4$ & $-\frac{1}{2} (B - C)$ \\
\hline
\end{tabular}
\caption{Kinematic Feynman integrals accompanying each connected colour factor for
the (1,2,3) web of Fig.~\ref{fig:123}, where $A \equiv{\cal F}(A)$, etc.}
\label{tab:123}
\end{center}
\end{table}
\beqa
\label{calG123}
  {\cal G}_{(123), \, 2}^{(3)} & = & \frac{1}{3}
  \bigg[ \frac{1}{8} L_{1 3}^2 - \frac{1}{8} L_{2 3}^2 + \frac{1}{4} L_{2 3} L_{3 2} 
  - \frac{1}{4} L_{1 3} L_{2 3} \bigg] \, , \nonumber \\
  {\cal G}_{(123), \, 3}^{(3)} & = & - \frac{1}{3}
  \bigg[ \frac{1}{4} L_{2 3} L_{1 3} - \frac{1}{4} L_{2 3} L_{3 2} + 
  \frac{1}{8} L_{2 3}^2 - \frac{1}{8} L_{1 3}^2 - R_2^2 \bigg] \, , 
   \\
  {\cal G}_{(1,2,3),4}^{(3)} & = & \frac{2}{3} \, \theta(x_2 - x_3)
  \bigg[ \, 2 \,L_{13} R_2 + L_{23} \left(R_3 - R_2 \right) \nonumber \\
  & & \hspace{1cm} - \, \log^2 \left( \frac{x_2}{x_3} \right) + \log \left( \frac{x_2}{x_3} 
  \right) \log \left( \frac{1 - x_2}{1 - x_3} \right) \bigg] \, , \nonumber
\eeqa
where we used the definitions in \eqn{logdef}. As a consequence of the presence of 
two entangled gluons spanning the cusp between lines 2 and 3, all subtracted 
kernels have a leftover Heaviside function; for the first two colour structures, it has been 
eliminated using symmetries of the integrand in the variables $(x_1, x_2, x_3)$, while 
this cannot be done for the coefficient of $c_4^{(3)}$. This not withstanding, the final 
integration can be performed, and the result can be expressed in terms of our basis 
functions. Indeed, given the above subtracted kernels, we find
\begin{subequations}
 \label{results123b}
 \beqa
  F^{(3)}_{(1,2,3), \, 2} (\alpha_{13},\alpha_{23}) & = & \frac{1}{12} \, r(\alpha_{13}) 
  r^2(\alpha_{23}) \bigg[ \frac12 \, M_{2,0,0} (\alpha_{13}) M_{0,0,0}^2 (\alpha_{23}) 
  \nonumber \\ & &
  - \, \frac12 \, M_{2,0,0} (\alpha_{23}) M_{0,0,0} (\alpha_{13}) M_{0,0,0} (\alpha_{23}) 
  + M_{0,0,0} (\alpha_{13}) M^2_{1,0,0} (\alpha_{23}) \nonumber \\
  & & - \, M_{0,0,0} (\alpha_{23}) 
  M_{1,0,0} (\alpha_{13}) M_{1,0,0} (\alpha_{23}) \bigg] \, ,
  \label{F123_2} \\
  F^{(3)}_{(1,2,3), \, 3} (\alpha_{13},\alpha_{23}) & = & -\, \frac{1}{12} \, r(\alpha_{13}) 
  r^2(\alpha_{23}) \bigg[ - \frac12 \, M_{2,0,0} (\alpha_{13}) M_{0,0,0}^2 (\alpha_{23}) 
  \nonumber \\
  & & + \, \frac12 \Big( M_{2,0,0} (\alpha_{23})
  - 8 M_{0,2,0} (\alpha_{23}) \Big) M_{0,0,0} (\alpha_{13}) M_{0,0,0} (\alpha_{23}) 
   \label{F123_3}\\
  & & - \, M_{0,0,0} (\alpha_{13}) M_{1,0,0}^2 (\alpha_{23}) + M_{0,0,0} (\alpha_{23}) 
  M_{1,0,0} (\alpha_{13}) M_{1,0,0} (\alpha_{23}) \bigg] \, , \nonumber \\
  F_{(1,2,3), \, 4}^{(3)} (\alpha_{13},\alpha_{23}) & = & \frac{4}{3} \, r (\alpha_{13}) 
  r^2 (\alpha_{23}) \bigg[  M_{0,1,1} (\alpha_{23}) M_{1,0,0} (\alpha_{13}) \\
  & & + \frac18 \, \Big(  
  M_{1,0,0}^2 (\alpha_{23}) - M_{0,0,0} (\alpha_{23}) M_{2,0,0} (\alpha_{23})  
  - \frac{1}{12} \, M_{0,0,0}^4 (\alpha_{23}) \nonumber \\
  & & + \, 2 \, M_{0,0,0} (\alpha_{23}) M_{0,2,0} (\alpha_{23}) \Big) 
  M_{0,0,0} (\alpha_{13}) 
  \bigg] \, \label{F123_4}. \nonumber 
\eeqa
\end{subequations}
As for the (2,2,2) web, and the MGEWs discussed in the previous section, these results are 
fully consistent with our expectations: the kinematic functions entering the anomalous 
dimension take the form of a sum of products of polylogarithmic functions of individual 
cusp angles, consistently with the factorization conjecture of Ref.~\cite{Gardi:2013saa},
and these functions all belong to the basis of \eqn{eq:Mbasis}. The way this is realised in the 
case of the (1,2,3) web is non-trivial: this web includes two cusp angles and these are 
entangled in individual diagrams due to three Heaviside functions. We find again that 
polylogarithms appear in the kernel of individual diagrams and in the unsubtracted web, 
but not in the subtracted web kernel. Furthermore, the only Heaviside function surviving 
in the subtracted web kernel ${\cal G}_{(1,2,3),4}^{(3)}$ in \eqn{calG123} relates two of 
the angular integrations associated with the same kinematic variable $\alpha_{23}$, and
therefore is consistent with the factorization property. 

\begin{figure}[htb]
\begin{center}
\scalebox{0.5}{\includegraphics{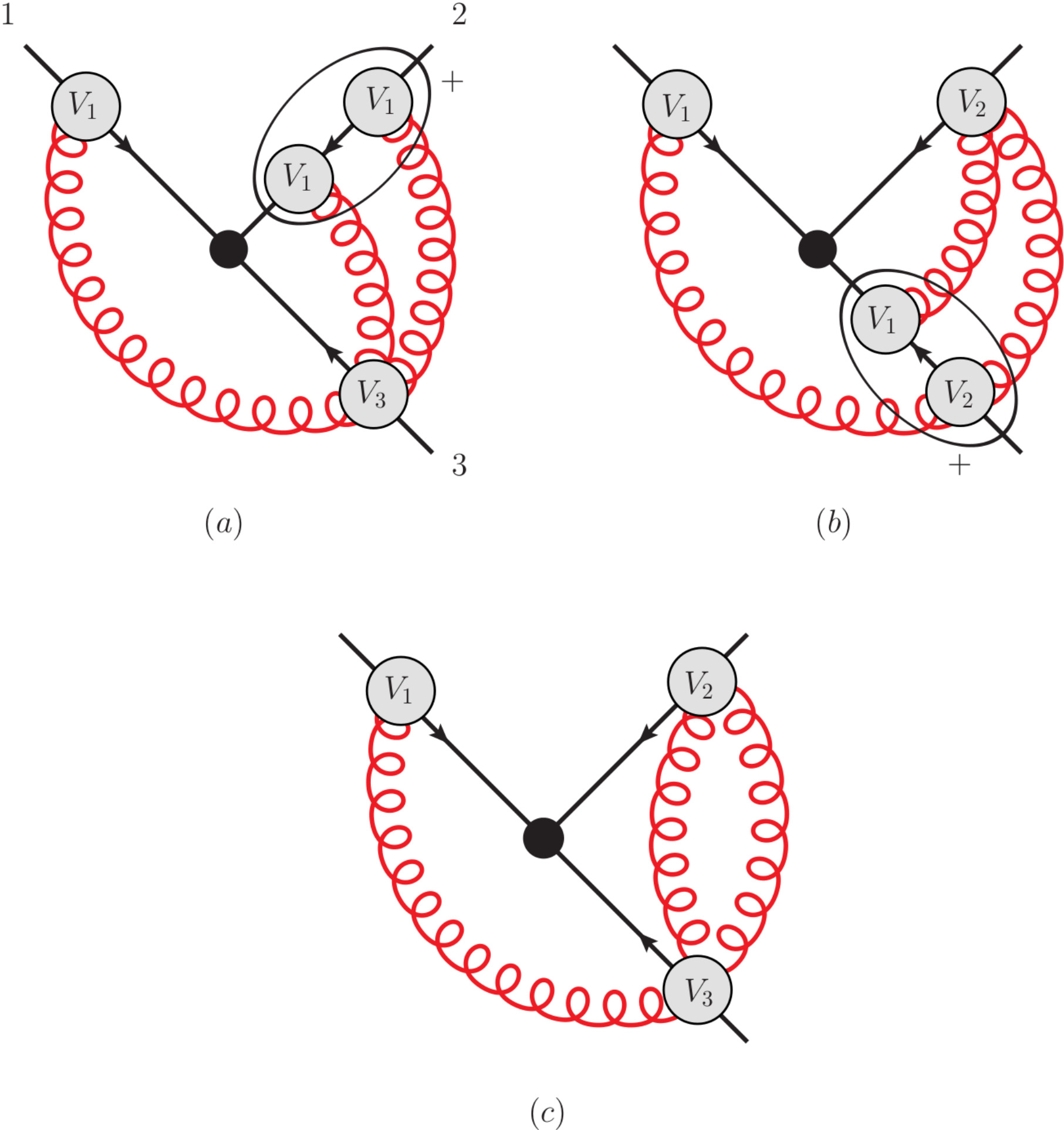}}
\caption{The three components of the (1,2,3) web using the effective vertex formalism. 
The components described by the two upper diagrams can be obtained via collinear 
reduction of: (a) the (1,1,3,1) web; (b) the (1,2,2,1) web. Diagram (c) shows the
connected colour factor that features one vertex on each line and cannot be determined 
by collinear reduction.}
\label{fig:123eff}
\end{center}
\end{figure}

We conclude by discussing the constraints provided by collinear reduction. As for the 
(2,2,2) web discussed above, one may obtain certain components of the (1,2,3) web 
from collinear reductions of webs connecting four Wilson lines. Two of the three 
components of \eqn{W123res_sub} can be obtained this way: the component 
associated with the colour factor $c_2^{(3)}$ corresponds, in the effective vertex 
description, to diagram (a) in Fig.~\ref{fig:123eff}, and can be obtained by collinear 
reduction from the (1,1,3,1) web, while the component associated with colour factor 
$c_3^{(3)}$ corresponds to diagram (b) in Fig.~\ref{fig:123eff}, and can be obtained 
by collinear reduction from the (1,2,2,1) web. The component of $c_4^{(3)}$ corresponds, 
in turn, to diagram (c); since the latter has a single effective vertex on each line, it cannot 
be obtained through collinear reduction. 

Let us now examine the two components that can be deduced from four-line webs. 
In the case of diagram (a) in Fig.~\ref{fig:123eff}, one may first permute lines 3 and 4 
in the result of the (1,1,1,3) web, so that the line carring the $V_3$ vertex will be 
line $3$, matching our conventions for the (1,2,3) web. Following this permutation, 
we apply the collinear reduction by identifying line $4$ with line $2$. To match diagram 
(a) in Fig.~\ref{fig:123eff} we must also include a symmetry factor\footnote{For a 
detailed discussion of the Feynman rules in the vertex effective theory, see 
Ref.~\cite{Gardi:2013ita}. The specific example of the (1,2,3) web was also 
analysed there, see eqs. (63) and (64).} of $1/2$, associated with the exchange 
of the two gluons, both propagating between the $V_3$ vertex on line 3 and the 
$V_1$ vertices on line $2$: this symmetry factor is absent in the original (1,1,3,1) 
web, where the two gluons attach to different lines. We thus get
\beqa
 \label{1113gen_to_123}
  \overline{w}^{(3)}_{(1,1,3,1)} \left( \alpha_{13}, \alpha_{23}, \alpha_{34} \right)
  \begin{array}{c} 
  \\ \longrightarrow \\ _{4||2}
  \end{array}
  & & - \frac12 \times \frac16 \, \, T_1^a \, \frac12 \, \big\{ T_2^b, T_2^d \big\} \, 
  T_3^c \left( \frac{1}{4 \pi} \right)^3 r(\alpha_{13}) \, r^2(\alpha_{23}) \nonumber \\
  & & \hspace{-4cm} \times \, \Big[ f_{ace} f^e_{\phantom{e}bd} \, G_{(1,1,1,3)} 
  \left( \alpha_{13}, \alpha_{23}, \alpha_{23} \right) + f_{ade} f^e_{\phantom{e}bc} \,
  G_{(1,1,1,3)} \left( \alpha_{23}, \alpha_{13}, \alpha_{23} \right) \Big]  
  \\
  & & \hspace*{-120pt} = \, - \frac16 \, T_1^a \, \frac12 \, \big\{ T_2^b,T_2^d \big\} \, 
  T_3^c \left( \frac{1}{4 \pi} \right)^3 r(\alpha_{13}) \, r^2(\alpha_{23}) \, 
  f_{ade} f^e_{\phantom{e}bc} \,
  G_{(1,1,1,3)}  \left( \alpha_{23}, \alpha_{13}, \alpha_{23} \right) \, ,
  \nonumber
\eeqa
where in the last line we kept only the second term, noting that the first vanishes owing 
to the contraction of the colour tensors. It is straightforward to check, using \eqn{1113}, 
that \eqn{1113gen_to_123} reproduces the $c_2^{(3)}$ component of the (1,2,3) web 
in \eqn{W123res_sub}, with the kinematic function given by \eqn{F123_2}.

Finally, consider diagram (b) in Fig.~\ref{fig:123eff}.  This diagram can be obtained 
from the collinear reduction of the (1,2,2,1) web of \eqn{1221gen} by identifying lines 
$3$ and $1$, and then renaming $4$ as $1$, to match our conventions for the (1,2,3) 
web. One finds
\beqa
\label{1221gen_to_123}
  \overline{w}^{(3)}_{(1,2,2,1)} \left( \alpha_{12}, \alpha_{23}, \alpha_{34} \right) 
  \begin{array}{c}
  \\ \longrightarrow \\  _{1||3}
  \end{array}
  & & - \frac16  \, f_{abe} f^e_{\phantom{e}cd} \, T_1^d \,T_2^b  \, 
  \frac12 \, \big\{T_3^a, T_3^c \big\}  
  \left( \frac{1}{4 \pi} \right)^3  r^2(\alpha_{23}) \, r(\alpha_{13})
  \nonumber \\
  && \hspace{2cm} \times \,G_{(1,2,2,1)} \left( \alpha_{23}, \alpha_{23}, \alpha_{13} 
  \right) \, .
\eeqa
One may verify, using \eqn{1221}, that this equation exactly reproduces \eqn{W123res_sub}, 
with the kinematic function given by \eqn{F123_3}. 

In this section, we have calculated the three-loop, three-line MGEWs that are needed 
for the three-loop soft anomalous dimension. In the remainder of the paper, we examine 
how the methods developed here, and in Ref.~\cite{Gardi:2013saa}, can be applied 
beyond the three-loop order. We begin by studying a particular four-loop example in 
the following section.


\section{A four-loop, five-line web}
\label{Fourloop}

The method developed in Ref.~\cite{Gardi:2013saa} and in Sections~\ref{Webs}-\ref{Basis} 
above allows us to compute high-order webs in the MGEW class with relatively little effort. 
It is then worthwhile to look for interesting examples beyond three loops: this will provide 
non-trivial checks of our conjectures on the analytic structure of subtracted webs, and on 
the relevant basis of functions.

\begin{figure}[htb]
\begin{center}
\scalebox{0.8}{\includegraphics{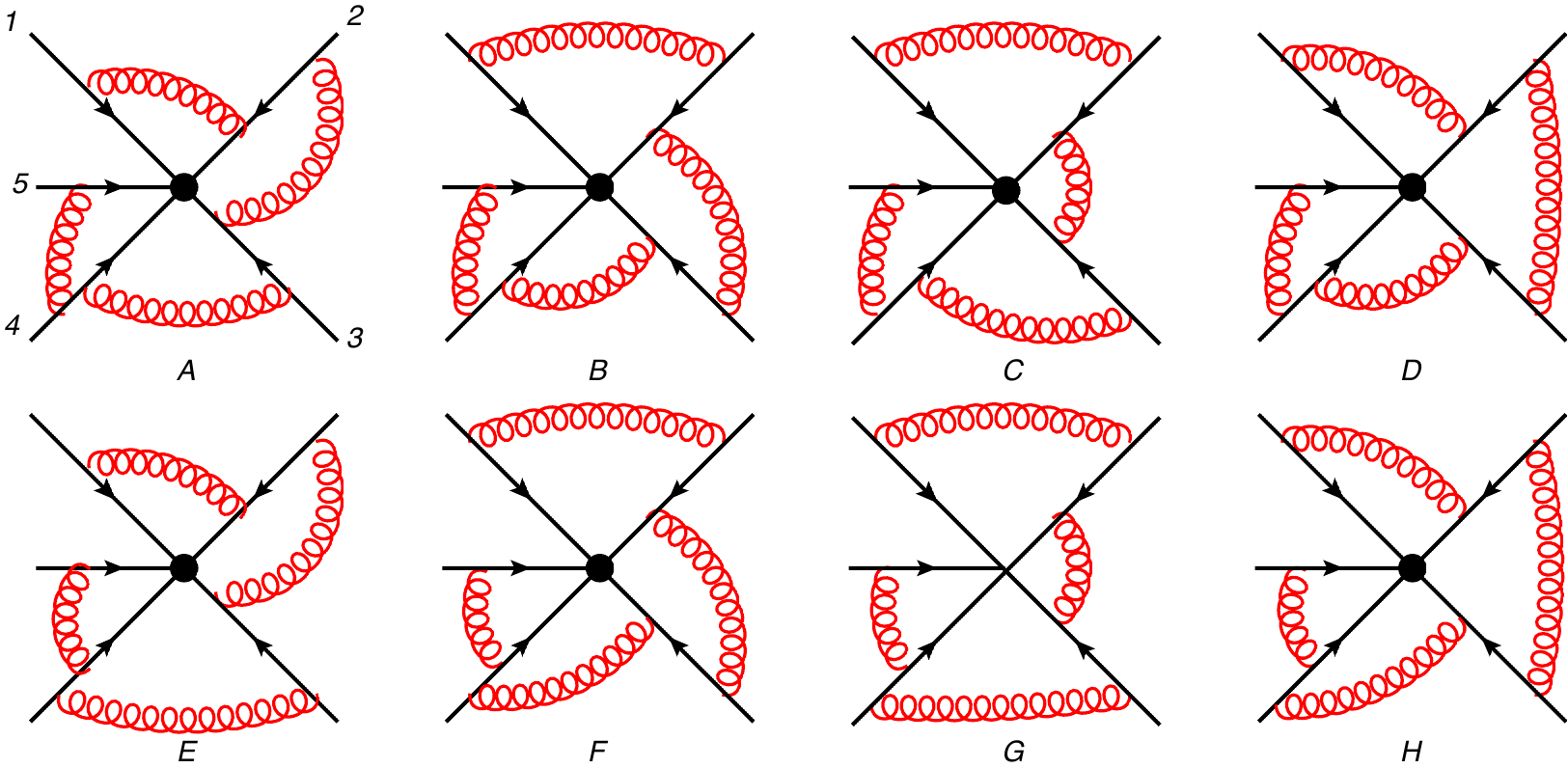}}
\caption{The (1,2,2,2,1) web connecting five Wilson lines at four loops.}
\label{fig:12221}
\end{center}
\end{figure}

In this Section, we present for the first time a complete calculation of a fully 
subtracted four-loop web. As our example, we choose to consider the (1,2,2,2,1)
web, connecting five Wilson lines at four loops, and consisting of the diagrams 
depicted in Fig.~\ref{fig:12221}. The (1,2,2,2,1) web is particularly interesting because 
of its simple colour structure, and because, spanning five legs, it will allow one to 
determine certain components of other webs at the same order, spanning a smaller 
number of Wilson lines but having more than one effective vertex on at least one line.
Furthermore, the (1,2,2,2,1) web is the third member of the infinite series of MGEWs
(1,2,2,$\cdots$,2,1), connecting $n + 1$ lines at $n$ loops. All the webs in this 
class have a single colour structure, and the general solution of the corresponding 
web mixing matrices for any $n$ have been obtained using combinatorial methods
in~\cite{Dukes:2013wa,Dukes:2013gea}, while the kinematic functions have been 
determined in~\cite{Gardi:2013saa} for the cases $n = 2, 3$. One may hope that a 
completely explicit answer for the first three elements of this collection could provide 
some insights for an all-order answer.

The pattern of subtractions at the four-loop level is particularly intricate, as can be
seen from \eqn{Gamres}. For example, in the specific case of the web (1,2,2,2,1), 
we need to consider the commutators of the webs (1,2,2,1), (1,2,1) and (1,1) connecting 
the five lines. In this section we organize and discuss the result, while further details 
are given in Appendix~\ref{Calc12221}.

The first step in the construction of the subtracted web is the determination of the
colour structure. In the case of the (1,2,2,2,1) web, depicted in Fig.~\ref{fig:12221}, 
there is only one colour structure, which can be written in terms of ordinary colour 
generators as 
\beq
  c_1^{(5)} \, = \, 
  T^{a}_1 \, \big[ T^a_2, T^b_2 \big] \, \big[T^b_3, T^c_3 \big] \big[ T^c_4, T^d_4 \big] 
  T^d_5 \, .
\label{colour12221}
\eeq
The corresponding combination of integrals can be constructed from the appropriate 
web mixing matrix, which is known~\cite{Dukes:2013gea}. Expressing the result in terms 
of the kinematic Feynman integrals of the individual diagrams in Fig.~\ref{fig:12221}, one finds
\beqa
\label{kin:12221}
  {\cal F}^{(4)}_{(1,2,2,2,1), \, 1} \left( \alpha_{ij}, \epsilon \right) & = & 
  \frac{1}{24} \, \left[6 \left( {\cal F}^{(4)} (F) - {\cal F}^{(4)} (A) \right) + 
  2 \left({\cal F}^{(4)} (C) + {\cal F}^{(4)} (D) + {\cal F}^{(4)} (E) \right) 
  \right. \nonumber \\  & & \left. 
  - 2 \left( {\cal F}^{(4)} (B) + {\cal F}^{(4)} (G) + {\cal F}^{(4)} (H) \right) \right] \, . 
\eeqa
It is convenient to work at the level of the integrand of the diagrams, by taking 
directly the combination in \eqn{kin:12221} of the functions $\phi_X^{(4)}$ given 
in Appendix~\ref{Calc12221}. The unsubtracted web is lengthy, and, much like 
the (1,2,2,1) web of Ref.~\cite{Gardi:2013saa}, contains polylogarithms, so 
upon integration it does not yield a factorised function of the cusp angles, but 
rather a lengthy sum of Goncharov polylogarithms involving different kinematic 
variables.
 
According to the factorisation conjecture, we expect that the commutators of the 
webs at lower orders will cancel all the correlations between different cusp angles, 
as well as all polylogarithmic functions in the kernel. We find that indeed the integrand 
of the subtracted web becomes much simpler, and the integrated result is factorised 
as expected. The subtracted web kernel, in terms of the functions defined in 
\eqn{logdef}, is given by
\beqa
\label{Gbar12221}
  {\cal G}^{(4)}_{(1,2,2,2,1), \, 1} \big( x_i, q (x_i, \alpha_i) \big) & = & - \, \frac{1}{144} 
  \bigg\{ L_{12}^3 - 3 L_{23}^3 + 3 L_{34}^3 - L_{45}^3 \nonumber \\ 
  && \hspace{-3cm} + \, \,
  3 L_{12}^2 \bigg[ L_{23} + L_{34} - 3 L_{45} \bigg] - 
  3 L_{45}^2 \bigg[ L_{23} + L_{34} - 3 L_{12} \bigg] \nonumber \\ 
  && \hspace{-3cm} + \,
  3 L_{23}^2 \bigg[ L_{12} - 3 L_{34} + 5 L_{45} \bigg] - 
  3 L_{34}^2 \bigg[ L_{45} - 3 L_{23} + 5 L_{12} \bigg]  \\  
  && \hspace{-3cm} + \, 
  6 \bigg[ L_{12} L_{23} L_{34} - 3 L_{12} L_{23} L_{45} + 3 L_{12} L_{34} L_{45} - 
  L_{23} L_{34} L_{45} \bigg] \nonumber \\ 
  && \hspace{-3cm} + \,  
  24 \bigg[ R_2^2 \, \bigg( L_{12} + L_{23} + L_{34} - 3 L_{45} \bigg) -
  R_3^2 \, \bigg( L_{23} + L_{34} + L_{45} - 3 L_{12} \bigg) \bigg] \bigg\} 
  \nonumber \, .
\eeqa
By looking at the subtracted web kernel in \eqn{Gbar12221}, we immediately note 
that the result is already expressed in terms of the functions of the basis in \eqn{eq:Mbasis}. 
More precisely, the functions $M_{3,0,0} (\alpha_{ij})$ and $M_{1,2,0} (\alpha_{ij})$ at 
weight four are the only new functions appearing at this order. Upon performing
the $x_k$ integrals, and in the notations of \eqn{subtracted_web_mge_kin}, we find 
that the contribution of this web to the anomalous dimension is given by
\beqa
  &&  F^{(4)}_{(1,2,2,2,1), 1} \left( \alpha_{ij} \right)  =  - \, \frac{1}{144} \, 
  r (\alpha_{12}) r (\alpha_{23}) r (\alpha_{34}) r (\alpha_{45}) \times \\
  && \times
  \bigg\{ \bigg[ 6 \bigg( M_{1,0,0}(\alpha_{12}) M_{1,0,0}(\alpha_{23}) M_{1,0,0}(\alpha_{34}) 
  M_{0,0,0} (\alpha_{45}) \nonumber \\
  && \hspace{.5cm} - \, \, 3 M_{1,0,0}(\alpha_{12}) M_{1,0,0}(\alpha_{23}) M_{1,0,0}(\alpha_{45}) 
  M_{0,0,0}(\alpha_{34}) \bigg) \nonumber \\
  && \hspace{.5cm} + \, \, \bigg(M_{3,0,0} (\alpha_{12}) M_{0,0,0} (\alpha_{45}) - 
  9 M_{2,0,0} (\alpha_{12}) M_{1,0,0} (\alpha_{45}) \bigg) 
  M_{0,0,0} (\alpha_{23}) M_{0,0,0}(\alpha_{34}) \nonumber \\
  && \hspace{.5cm} - \, \, 3 \bigg(M_{3,0,0}(\alpha_{23})M_{0,0,0}(\alpha_{34}) + 
  3 M_{2,0,0}(\alpha_{23})M_{1,0,0} (\alpha_{34})\bigg)
  M_{0,0,0}(\alpha_{12})M_{0,0,0}(\alpha_{45}) \nonumber \\
  && \hspace{.5cm} + \, \, 3 M_{2,0,0} (\alpha_{12})M_{0,0,0}(\alpha_{45})
  \bigg(M_{1,0,0}(\alpha_{23})M_{0,0,0}  
  (\alpha_{34})+M_{1,0,0}(\alpha_{34})M_{0,0,0}(\alpha_{23})\bigg)\nonumber \\
  && \hspace{.5cm} + \, \, 3 M_{2,0,0}(\alpha_{23})M_{0,0,0}(\alpha_{34})
  \bigg(M_{1,0,0}(\alpha_{12})
  M_{0,0,0} (\alpha_{45})+5M_{1,0,0}(\alpha_{45})M_{0,0,0}(\alpha_{12})\bigg) \nonumber \\
  && \hspace{.5cm} + \, \, 24  M_{0,2,0}(\alpha_{23}) \bigg(M_{1,0,0}(\alpha_{12}) 
  M_{0,0,0}(\alpha_{34}) M_{0,0,0} (\alpha_{45})
\nonumber \\
  && \hspace{.5cm}
+ M_{1,0,0}(\alpha_{34}) M_{0,0,0}(\alpha_{12}) 
  M_{0,0,0}(\alpha_{45})  - \, \, 3 M_{1,0,0}(\alpha_{45}) M_{0,0,0}(\alpha_{12}) 
  M_{0,0,0}(\alpha_{34})\bigg)+ 
\nonumber \\
  && \hspace{.5cm}
  24 M_{1,2,0}(\alpha_{23}) M_{0,0,0}(\alpha_{12}) M_{0,0,0}(\alpha_{34}) 
  M_{0,0,0}(\alpha_{45})\bigg] \nonumber \\
  && \hspace{.5cm} - \, \, \bigg[(\alpha_{12} \leftrightarrow \alpha_{45}),
  (\alpha_{23} \leftrightarrow \alpha_{34}) \bigg] \bigg\} \, . \nonumber 
\eeqa
As expected, we find a factorized function of uniform transcendental weight ${\rm w} = 7$, 
expressed as a sum of products of our basis functions, each one depending on a single 
cusp angle, and satisfying the symbol conjecture. Through various collinear limits, this 
will also yield information on other four-loop webs, involving less than five Wilson lines, 
but with more than one effective vertex on a given line.


\section{The Escher Staircase and (2, 2, \ldots, 2) webs}
\label{Escher}

In this section we would like to display the flexibility and the reach
of the formalism that we have developed by computing a certain class
of diagrams contributing to a specific series of MGEW's to all orders
in perturbation theory. The results of the calculation are not of immediate 
physical relevance, since we will not be computing complete webs, much 
less subtracted webs, to all orders. Nevertheless, the calculation of these 
particular diagrams will allow us to prove a general statement about this 
series of webs. Moreover, the feasibility of this calculation suggests that
all-order calculations of MGEWs are possible. In addition the simplicity 
of the result, which by itself is properly factorized into functions belonging 
to our basis, provides further evidence for our conjectures.

\begin{figure}[htb]
\begin{center}
\scalebox{0.8}{\includegraphics{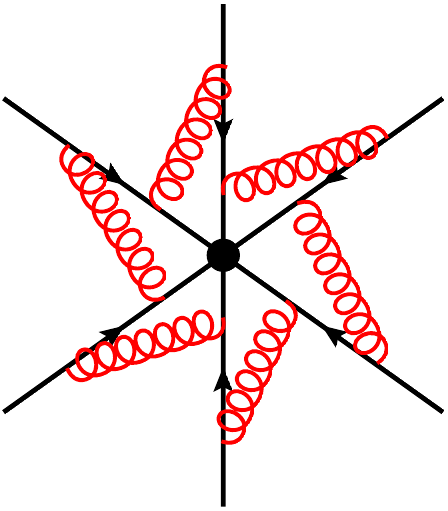}}
\caption{Example of an Escher staircase with six external legs. There are two 
staircases at arbitrary order, related by reflection.}
\label{EscherFig}
\end{center}
\end{figure}

Following Ref.~\cite{Gardi:2010rn}, we dub the class of diagrams we
will compute `Escher staircases', for reasons that should be apparent
from their graphical structure. An example with $n = 6$ is displayed
in Fig.~\ref{EscherFig}. These diagrams are the most symmetric members
of the $n$-loop $(2,2,\ldots,2)$ webs: each such web contains $2^n$ diagrams, 
two of which are of the form we are studying, differing by clockwise or 
counterclockwise orientations. The staircase diagrams are particularly 
interesting, not only because of their graphical symmetry, but also 
because they do not have subdivergences, so they do not require 
commutator counterterms. As a consequence, they should satisfy 
the alphabet and factorization conjectures by themselves, and
indeed we will find that they do.
Interestingly, we also find that a non-trivial cancellation takes place when the 
kinematic factors of staircase diagrams are combined to build up contributions 
to the $n$-loop soft anomalous dimension. Indeed, as verified in \sect{222} for 
the (2,2,2) web, the two staircases are the only diagrams to contribute to the colour
structure composed of only $V_2$ effective vertices, denoted by $c_4^{(3)}$ in 
\sect{222}. Their contributions to that colour structure however cancel exactly, as 
noted in \eqn{gamints222c}.  The (2,2,2) staircases of course do contribute to the 
other three colour structures, $c_i^{(3)}$, with $i = 1,2,3$. One should keep in
mind, however, that the kinematic functions of those colour structures can be obtained 
also from collinear reduction of the $(1,2,2,1)$ web. A similar story is played out for 
the $(2,2,\ldots,2)$ web at arbitrary order, and we provide a general argument for 
this in what follows.

Turning now to the evaluation of the Escher staircase at $n$ loops, we
begin by noticing that the diagram kernel in \eqn{phi_D} can be
explicitly written down, for the staircase, as 
\beqa 
\label{phiS}
  \phi_S^{(n)} \left( x_i; \, \epsilon \right) & = & \int_0^1 \prod_{k = 1}^{n - 1} d
  y_k \, \left(1 - y_k \right)^{- 1 + 2 \epsilon} \, y_k ^{- 1 + 2 k
  \epsilon} \, \nonumber \\ && \times \, \prod_{i = 1}^n \, \theta
  \left[ \left(1 - x_i \right) \left( 1 - y_{i - 1} \right) \prod_{j = i}^{n - 1} y_j - x_{i + 1} 
  \left(1 - y_i \right) \prod_{j = i + 1}^{n - 1} y_j \right] \, ,
\eeqa
where the $i$-th Heaviside function guarantees that the absorption point of gluon 
$i - 1$ is further away (along Wilson line $i$) than the emission point of gluon $i$.
To properly define the product of $\theta$ functions, we set $1 - y_0 = 1 - y_n = 1$
when they occur, as well as $\prod_{j = n + 1}^{n - 1} y_j = \prod_{j = 1}^{n - 1} y_j$ 
in the last factor, and $\prod_{j = n}^{n - 1} y_j = 1$ in the penultimate factor; similarly, 
we set $x_{n + 1} = x_1$ when it occurs. To illustrate the notation, we write explicitly 
the products of $\theta$ functions for small $n$, as
\beqa
\label{thetas}
  n = 2 \quad & \rightarrow& \quad \theta \left[ (1 - x_1) \, y_1 - x_2 \, (1 - y_1) \right]
  \theta \left[ (1 - x_2) (1 - y_1) - x_1 \, y_1 \right] \, , \nonumber \\
  n = 3 \quad & \rightarrow& \quad \theta \left[ (1 - x_1) \, y_1 \, y_2 - 
  x_2 \, (1 - y_1) \, y_2 \right] \, \theta \left[ (1 - x_2) (1 - y_1) \, y_2 - 
  x_3 \, (1 - y_2) \right] \nonumber \\
  && \quad \theta \left[ (1 - x_3) (1 - y_2) - x_1 \, y_1 \, y_2 \right] 
  \, , \\
  n = 4 \quad & \rightarrow& \quad \theta \left[ (1 - x_1) \, y_1 \, y_2 \, y_3 - 
  x_2 \, (1 - y_1) \, y_2 \, y_3 \right] \nonumber \\
  && \quad \theta \left[ (1 - x_2) (1 - y_1) \, y_2 \, y_3 - 
  x_3 \, (1 - y_2) \, y_3 \right] \, \theta \left[ (1 - x_3) (1 - y_2) \, y_3 - x_4 \, (1 -  y_3) 
  \right] \nonumber \\
  && \quad \theta \left[ (1 - x_4) (1 - y_3) - x_1 \, y_1 \, y_2 \, y_3 \right] \, .
  \nonumber 
\eeqa
Using the fact that $0 < x_i < 1$, and $0 < y_i < 1$, it is easy to realize that 
the $\theta$ functions are more naturally expressed by changing variables 
to $\xi_i \equiv y_i/(1 - y_i)$, so that $0 < \xi_i < \infty$. In this way one gets, 
for small $n$, factors of
\beqa
\label{althetas}
  n = 2 \quad & \rightarrow& \quad  
  \theta \left[ \frac{x_2}{1 - x_1} < \xi_1 < \frac{1 - x_2}{x_1} \right] \, 
  \theta \left[ \frac{(1 - x_1)(1 - x_2)}{x_1 x_2} > 1 \right] \, , \nonumber \\
  n = 3 \quad & \rightarrow& \quad
  \theta \left[ \frac{x_3}{1 - x_2} < \xi_2 < \frac{1 - x_3}{x_1 \xi_1} \right] \, 
  \theta \left[ \frac{x_2}{1 - x_1} < \xi_1 < \frac{(1 - x_2)(1 - x_3)}{x_1 x_3} \right] 
  \nonumber \\
  && \quad \theta \left[ \frac{(1 - x_1)(1 - x_2)(1 - x_3)}{x_1 x_2 x_3} 
  > 1\right] \, , 
   \\
  n = 4 \quad & \rightarrow& \quad 
  \theta \left[ \frac{x_4}{1 - x_3} < \xi_3 < \frac{(1 - x_4)(1 + \xi_1)}{x_1 \xi_1 \xi_2} \right] \, 
  \theta \left[ \frac{x_3}{1 - x_2} < \xi_2 < \frac{(1 - x_3)(1 - x_4)}{x_1 x_4 \xi_1} \right] 
  \nonumber \\
  && \quad \theta \left[ \frac{x_2}{1 - x_1} < \xi_1 
  < \frac{(1 - x_2)(1 - x_3)(1 - x_4)}{x_1 x_3 x_4} \right] \nonumber \\
  && \quad \theta \left[ \frac{(1 - x_1)(1 - x_2)(1 - x_3)(1 - 
  x_4)}{x_1 x_2 x_3 x_4} > 1 \right] \, , \nonumber 
\eeqa
where we used the notation $\theta(a < x < b)$ to denote the product $\theta(b - x) 
\, \theta(x - a)$. Furthermore, the last $\theta$ function for each $n$ must be present 
since the variables $x_i$ must also later be integrated in the interval $0 < x_i < 1$; 
its meaning is clear: it distinguishes the `clockwise' staircase diagram from the 
`counterclockwise' one, which carries the complementary $\theta$ function. 
We define therefore
\beq
  \theta_+ (n) \, \equiv \, \theta \Big( S_n (x_i) - 1 \Big) \, 
\label{theta+}
\eeq
where
\beq
  S_n (x_i) \, \equiv \, \prod_{i = 1}^n \frac{1 - x_i}{x_i} \, .
\label{Sn}
\eeq
The $n$-loop `clockwise' staircase diagram carries a factor of $\theta_+ (n)$, while
its `counterclockwise' image carries a factor of $\theta_- (n) = 1 - \theta_+(n)$.
Upon further inspection of \eqn{althetas} one sees also that the $\xi_i$ integrals 
are all bounded from above and from below: since poles in $\epsilon$ could only 
arise from the limits $\xi_i \to \{0, \infty\}$, this implies, as expected, that the staircase 
graph has only a single overall ultraviolet divergence, given by $\Gamma(2 n \epsilon)$ in 
\eqn{gendiag3}, and no subdivergences. It is not difficult to work out the generalization
to all $n$ of the constraints in \eqn{althetas}: one can then write \eqn{phiS} as
\beqa
 \label{phiS2} 
  \phi_S^{(n)} \left( x_i; \, \epsilon \right) & = &  \int_0^\infty \prod_{k = 1}^{n - 1} 
  d \xi_k \left(1 + \xi_k \right)^{- 2 (k + 1) \epsilon} \, 
  \xi_k ^{- 1 + 2 k \epsilon} \, \, \Theta_S \big[ \left\{ x_k, \xi_k \right\} \big] 
  \\ & = & \theta_+ (n) 
  \int_{A_1}^{B_1} \frac{d \xi_1}{\xi_1} \, \int_{A_2}^{B_2 \left(\xi_1 \right)} 
  \frac{d \xi_2}{\xi_2} \, \, \ldots \,  
  \int_{A_{n - 1}}^{B_{n - 1} \left(\xi_1, \xi_2, \ldots, \xi_{n - 2} \right)} 
  \frac{d \xi_{n - 1}}{\xi_{n - 1}} \, + \, {\cal O} \left( \epsilon \right)  \, . \nonumber 
\eeqa 
Importantly, the resulting integral is explicitly of a `$d \log$' form, in agreement with the 
considerations of Refs.~\cite{Henn:2013wfa,Gardi:2013saa}. The remaining difficulty is 
that (also as expected) the integrals are nested and not completely factorized. One can 
however work out explicitly the limits of integration, generalizing \eqn{althetas}. One finds
\beq
  A_k \, = \, \frac{x_{k + 1}}{1 - x_k} \, , \qquad B_k \left( \xi_1, \ldots \xi_{k - 1} \right) 
  \, = \, \frac{\prod_{j = k + 1}^n (1 - x_j)}{\prod_{j = k + 2}^{n+1} x_j} \,
  \frac{\prod_{j = 1}^{k - 2} (1 + \xi_j)}{\prod_{j = 1}^{k - 1} \xi_j} \, ,
\label{limint}
\eeq
where we define $x_{n + 1} = x_1$, and all products running over empty sets of 
integers are set equal to one. The final integration is now trivial, and one finds the
remarkably simple result
\beq
  \phi_S^{(n)} \left( x_i; \, 0 \right) \, = \, \frac{1}{(n - 1)!} \, \, \theta_+ (n)  \, 
  \bigg( \log \Big[ S_n (x_i) \Big] \bigg)^{n - 1} \, .
\label{finstair}
\eeq
The result is Bose symmetric and completely factorized, and, when
integrated to give the kinematic function ${\cal F}^{(n)} (S)$ of \eqn{gendiag3},
manifestly expressible in terms of our basis functions.

Note now that by sending $x_i \leftrightarrow 1 - x_i$ in \eqn{finstair}, the result 
has the same form, but with a factor $(-1)^{n - 1}$ from the power of the logarithm, 
and $\theta_+$ replaced by $\theta_-$. Denoting the clockwise and anticlockwise 
staircase diagrams by $S_+$ and $S_-$ respectively, one thus finds
\beq
  {\cal F}(S_+) \, = \, (-1)^{n-1} {\cal F}(S_-) \, .
\label{FS}
\eeq
This leads to the cancellation observed at three loops ($n=3$) in \sect{222}: the two 
staircase diagrams have identical kinematic factors, which however combine to form 
the coefficient of the colour structure $c_3^{(4)}$ with the same weight and opposite 
signs. Since the other diagrams in the (2,2,2) web do not contribute to $c_3^{(4)}$, 
the corresponding coefficient vanishes. We now argue that a similar argument applies 
in an arbitrary $(2,2,\ldots,2)$ web.

Let us define a colour basis for $n$-loop webs connecting $n$ lines, generalising 
the 3-loop basis of \eqn{colfacs3}. The requirement in constructing this basis is that 
it should allow one to express all the colour components of the (2,2,$\ldots$,2) web. 
The dimension of this basis is\footnote{This is the rank of the  mixing matrix of the 
(2,2,$\ldots$,2) web, as proved in Theorem 8.2 of Ref.~\cite{Dukes:2013gea}.} $n+1$,
and its elements have a transparent interpretation (see Fig.~\ref{fig:222tot} for the 
three-loop case) upon considering the (2,2,$\ldots$,2) web in the effective vertex 
formalism of Ref.~\cite{Gardi:2013ita}: $c_i^{(n)}$ ($1\leq i\leq n$) is the colour 
factor one obtains upon having two $V_1$ vertices on line $i$, corresponding to 
an anticommutator of colour generators, and a $V_2$ vertex, corresponding to 
a commutator, on all the other lines, yielding
\beq
  c_i^{(n)} \, = \, \left[ T_1^{a_1},T_1^{a_2}\right] \left[T_2^{a_2},T_2^{a_3}\right]
  \ldots \left\{T_i^{a_i},T_i^{a_{i+1}}\right\} \ldots \left[T_{n-1}^{a_{n-1}},T_{n-1}^{a_{n}}\right]  
  \left[T_n^{a_{n}},T_n^{a_1}\right] \, .
\label{cindef}
\eeq
Finally, the $n+1$-st basis element is the fully antisymmetric colour factor corresponding 
to a $V_2$ vertex on each of the lines, 
\beq
  c_{n+1}^{(n)} \, = \, \left[T_1^{a_1},T_1^{a_2}\right] \left[T_2^{a_2},T_2^{a_3}\right]
  \ldots \left[T_{n-1}^{a_{n-1}},T_{n-1}^{a_{n}}\right] \left[T_n^{a_{n}},T_n^{a_1}\right] \, .
\label{cin+1def}
\eeq
It is this latter component of the $(2,2,\ldots,2)$ web that will receive a contribution 
from the kinematic integrals of the staircase diagrams only, which will ultimately turn
out to vanish. First, one notes that the fact that staircase diagrams are irreducible 
means that the web mixing matrix has the generic form
\begin{displaymath}
  \left(\begin{array}{ccc} 1& 0& \ldots\\ 0& 1&\ldots\\ 0 & 0 & 
\ldots \\ \vdots & \vdots & \ldots\\ 0 & 0 &\ldots \end{array}\right),
\end{displaymath}
where the first two columns arise from the staircase diagrams. This form follows 
from the replica trick analysis of Ref.~\cite{Gardi:2010rn}, which dictates that the 
exponentiated colour factor of diagram $D$ receives no contributions from diagrams 
$D'$ which are more reducible (less irreducible) than $D$. From the above form, 
it then follows that any such mixing matrix has right-eigenvectors\footnote{Right 
eigenvectors of the mixing matrix correspond to $Y_{D,j}^{-1}$ in \eqn{W_diag}, 
dictating the linear combination of integrals associated with a given connected 
colour factor $c_j$.} of the form
\beq
  \left(\begin{array}{cccccc} a & b & 0 & 0 & \ldots & 0 \end{array} \right) \, ,
\label{eigenvec}
\eeq
for arbitrary $a$ and $b$. Two special cases are $a = \pm b = 1$, corresponding 
to the symmetric and antisymmetric combinations of $S_+$ and $S_-$. These 
combinations are special in that they have definite parity under a ``flipping 
transformation'' that interchanges all pairs of gluons on all lines simultaneously. 
Such a transformation exchanges $S_{+}\rightarrow S_{-}$, and so the symmetric 
(anti-symmetric) combination has flipping parity $+$ ($-$) respectively. The contribution
from the entire web must be invariant under this transformation, as it maps the total 
web to itself. Given that the basis of colour factors in \Eqns{cindef}{cin+1def} is 
linearly independent, each separate colour factor multiplied by the corresponding 
kinematic function must also be invariant under the flipping transformation. The 
basis we have chosen is particularly convenient in this regard, as each colour factor 
has a definite flipping parity: $(-1)^{n - 1}$ for $c_i^{(n)}$ ($1 \leq i \leq n$) and
$(-1)^n$ for the fully antisymmetric colour factor $c_{n + 1}^{(n)}$. One then finds 
that the contribution to the latter colour factor contains the combination
\begin{displaymath}
  c_{n + 1}^{(n)} \left[{\cal F}(S_+) - {\cal F}(S_-) + \ldots \right] \, ,
\end{displaymath}
for odd $n$, and
\begin{displaymath}
  c_{n + 1}^{(n)} \left[{\cal F}(S_+) + {\cal F}(S_-) + \ldots \right] \, ,
\end{displaymath}
for even $n$, where the ellipsis in each case denotes possible contributions from 
non-staircase diagrams. In fact, such contributions are not present, which can be 
seen as follows. At the $n$-loop order, there are $n+1$ colour factors, $n$ of which 
have parity $(-1)^{n - 1}$, and only one of which has parity $(-1)^n$. It follows that of 
the $(n + 1)$ right eigenvectors with eigenvalue~1, $n$ must correspond to kinematic 
functions with flipping parity $(-1)^{n - 1}$, and only one to parity $(-1)^n$. Since we 
know that the eigenvector of \eqn{eigenvec} with $a = (-1)^n b$ has parity $(-1)^n$, 
this must be the only possibility, and there can be no other contributions. Having 
established \eqn{FS} above, this completes the proof that the contribution to the fully 
antisymmetric colour factor from the $(2,2,\ldots,2)$ web vanishes.

Returning to consider the (2,2,$\ldots$ 2) $n$-th order web as a whole, we now see 
that all of its non-vanishing components belong to the colour structures in  \eqn{cindef}, 
where one of the lines features an anticommutator of two $V_1$ emission vertices. 
This means that it may be determined from the corresponding $n$-loop $n + 1$ line 
(1,2,2,$\ldots$ 2,1) web through collinear reduction, just as the (2,2,2) web was 
obtained from the (1,2,2,1) web in \eqn{1221gen}. Specifically, at four-loops we 
essentially have the result for the (2,2,2,2) web based on the calculation of the 
(1,2,2,2,1) web in \sect{Fourloop}. 

In this section, we have shown that is possible to calculate a particular type of web 
diagram to all orders in perturbation theory. As seen from the explicit calculations 
of other webs, the computations for diagrams with sudivergences are considerably 
more intricate, and further technical developments will be needed before a complete 
calculation of an all-order class of subtracted webs can be completed. The present 
example however testifies to the underlying simplicity of the structure of MGEWs, 
and suggests that the problem of computing this class of webs might at some point 
be completely solved.

\section{Conclusion}
\label{Conclu}

In this paper, we have extended the programme of Refs.~\cite{Gardi:2010rn,Gardi:2011yz,
Gardi:2013ita,Gardi:2013saa}, which established a diagrammatic approach for studying 
infrared singularities in QCD scattering amplitudes (see also~\cite{Mitov:2010rp,
Vladimirov:2014wga}). We have done this by computing the ultraviolet singularities of 
products of semi-infinite Wilson lines in terms of webs. In the multi-line case, webs are 
sets of diagrams, each closed under the group of gluon permutations on the Wilson lines, 
whose contribution to the \emph{exponent} of the Wilson line correlator consists of
a sum of terms, each involving a fully connected colour factor multiplying a linear 
combination of Feynman integrals of diagrams belonging to the set. As explained in detail 
in Refs.~\cite{Gardi:2011yz,Gardi:2013ita,Gardi:2013saa}, and reviewed here, each web 
contains ultraviolet subdivergences in general, which must be removed by appropriate 
counterterms involving commutators of subdiagrams. This makes the case of multi-leg 
scattering qualitatively different to the familiar case of a Wilson loop; we call the combination 
of an unrenormalised web and its counterterms  a \emph{subtracted web}.

Specifically, we have focused on the contribution to the soft anomalous dimension from 
diagrams consisting of multiple gluon exchanges between the Wilson lines. These Multi-Gluon 
Exchange Webs (MGEWs) are the simplest diagrams at any given order; however, they
also contain the highest number of ultraviolet subdivergences. Thus, the web language, 
coupled with a suitable infrared regulator for calculating individual diagrams, is extremely 
useful in order to cleanly isolate their contribution to the soft anomalous dimension.

MGEWs connecting four Wilson lines at three-loop order were already considered 
in Ref.~\cite{Gardi:2013saa}, which also analysed the analytic structure of MGEWs in 
general. It was conjectured that: (i) the contributions of MGEWs to the soft anomalous
dimension take the form of sums of products of polylogarithmic functions, each involving 
a single cusp angle; (ii) the symbol alphabet of these functions consists of $\{\alpha_{ij},
\alpha/(1-\alpha_{ij}^2)\}$, where $\alpha_{ij}$ is defined in \eqn{alphadef} as the 
exponent of the cusp angle formed between lines $i$ and $j$. 

In this paper we provided significant additional evidence supporting these conjectures. 
Moreover, we took a step forward in understanding MGEWs by constructing a basis of 
functions, motivated by the alphabet conjecture as well as specific calculations, that 
we conjecture can be used to express MGEWs connecting any number of Wilson lines 
at arbitrary order in perturbation theory. The basis $M_{k,l,n} (\alpha)$, is defined in 
\eqn{eq:Mbasis} as a single parameter integral over the gluon emission angle. The 
integrand consists of a product of three types of logarithms (depending on the gluon 
emission angle and the corresponding cusp angle) raised to non-negative integer 
powers $k, l $ and $n$, respectively. The basis functions are consistent with the 
alphabet conjecture, and they are conveniently characterised by their symbols; these 
are listed in Table~\ref{tab:basis} up to weight 5. The functions may also be explicitly 
evaluated in terms of Harmonic polylogarithms with indices $0$ and~$1$: all independent 
functions up to weight 5 are listed in Appendix~\ref{appBasis}. The three logarithms 
appearing in the integrands of $M_{k,l,n} (\alpha)$ have been identified in two-loop 
calculations of the (2,2) and (1,2,1) webs, where functions up to weight 3 appear, 
that is $M_{k,l,n} (\alpha)$ with $k + n + l\leq 2$. The basis passed all tests at three 
loops, where basis elements up weight $5$ appear, corresponding to $k + n + l \leq 4$. 
These tests include the four-line webs of Ref.~\cite{Gardi:2013saa}, namely the (1,2,2,1) 
and (1,1,1,3) webs, the three-line webs of \sect{Threeloop} above, namely the (2,2,2) 
and (1,2,3) webs, as well as the two-line (3,3) web of \sect{ToCusp}. Further tests 
are provided by the five-line four-loop (1,2,2,2,1) web computed in \sect{Fourloop}, 
and by the all order analysis of the Escher Staircase diagrams and the (2,2,$\ldots$,2) 
webs in \sect{Escher}. 
  
Another important result of the present paper is the discovery of the collinear reduction 
relations between webs connecting different numbers of Wilson lines. These relations, 
which we briefly summarise below, are discussed in more detail in \sect{Webs}, and 
then illustrated in several examples throughout the paper. The idea is formulated using 
the effective vertex language of Ref.~\cite{Gardi:2013ita}, which provides a convenient 
colour basis for webs. In this language, the emission of $K$ gluons from a Wilson line, 
associated with a tree-graph colour factor (a fully nested commutator) is described by 
an effective vertex $V_K$. Considering a given web, different components may be 
described as connected graphs made out of such vertices, as shown for example in
Figs.~\ref{fig:33_effective_vertices}, \ref{fig:222tot} and \ref{fig:123eff}. These graphs 
may in general feature one or more effective vertices on a given line. However, when 
multiple vertices appear on a line, they are not ordered: the integrals range over the 
entire Wilson line independently of each other, and in the colour factor one takes the 
symmetric combination of all possible orderings. As a consequence, the calculation of 
such a graph maps directly into the calculation of a graph where the Wilson line that 
features several vertices is replaced by a set of collinear lines, each featuring 
only one of these vertices. The upshot is that starting with a web that features a single 
effective vertex on each line, one may deduce various components of webs connecting 
fewer Wilson lines, but featuring more than one vertex on some of the lines. Collinear 
reduction provides stringent checks of the two- and three-line webs computed in 
this paper: it allowed us to determine one of the two components of the (3,3) web 
in \eqn{double_coll_redu_of_1221gen}, and the entire (2,2,2) web in \eqn{1221gen_into_222}, 
from the (1,2,2,1) web, as well as two of the three components of the (1,2,3) web in
\Eqns{1221gen_to_123}{1113gen_to_123}, using, respectively, the (1,2,2,1) and 
(1,1,3,1) webs. We note that the only colour component that cannot be deduced by 
collinear reduction for the (1,2,3) web is the one corresponding to the fully antisymmetric 
colour factor $c_4^{(3)}$ in \eqn{colfacs3}. Given that the same diagrams enter both 
the components that can be deduced by collinear reduction and those that cannot, 
webs connecting less than the maximal number of lines at a given order are strongly 
constrained, providing us with high confidence in the results presented in this paper. 

As an additional check on the basis of functions we propose, as well as to illustrate the 
power and general applicability of the web language, we have also calculated the 
(1,2,2,2,1) four-loop web, and the Escher Staircase diagrams to arbitrary order in 
perturbation theory. The latter enter the (2,2,\ldots,2) web, and are especially simple 
because they do not contain ultraviolet subdivergences. Furthermore, we were able 
to show that the component corresponding to the fully antisymmetric colour factor of 
the (2,2,\ldots,2) web vanishes. This was proven using the fact that this colour structure 
is associated exclusively with the Escher Staircase diagrams, and these two diagrams 
are related by a parity transformation as in \eqn{FS}. The conclusion is rather striking: 
the entire contribution of the (2,2,\ldots,2) web family to the exponent can be deduced 
from the corresponding (1,2,2,\ldots,2,1) webs through collinear reduction. Specifically, 
at four loops, the result for the (2,2,2,2) web can be directly read off the results of 
\sect{Fourloop} for the (1,2,2,2,1)  web.

The analysis and explicit calculations performed in this paper promote our 
understanding of an entire class of contributions to the renormalization of Wilson 
line correlators. The progress achieved in understanding the analytic properties of 
the result~\cite{Gardi:2013saa}, and the specific class of functions by which they 
may be expressed, is a step towards translating the entire calculation of an 
arbitrary MGEW into a combinatorial problem: given the factorization conjecture 
and the basis of functions, along with the expected transcendental weight, one 
may write down an ansatz for the answer where only rational numerical coefficients 
need to be fixed. This, along with the progress already made on webs~\cite{Gardi:2010rn,
Gardi:2011yz,Gardi:2013ita} and their combinatorics~\cite{Gardi:2011wa,Dukes:2013wa,
Dukes:2013gea} may facilitate all-order calculations in the future.

The results of this paper also constitute a further step forward in assembling all 
necessary ingredients for the soft anomalous dimension of massive partons at 
three-loop order. In order to complete this programme, one needs to include 
MGEWs in which gluons can be emitted and absorbed from the same Wilson line. 
Furthermore, one needs to include diagrams containing a single three-gluon vertex 
off the Wilson lines, and those fully connected graphs which contain two three-gluon 
vertices or a single four-gluon vertex. Work in these directions is ongoing~\cite{Gardi:2014kpa}.


\section*{Acknowlegments}
EG and CDW are supported by the Science and Technologies Facilities Council 
(STFC). MH is supported by the University of Edinburgh.  CDW, LM and GF thank 
the Higgs Centre for Theoretical Physics for warm hospitality, and MH thanks the
University of Torino and INFN, Torino for support and hospitality. Work supported 
by MIUR (Italy), under contract 2010YJ2NYW$\_$006 and by the University of 
Torino and the Compagnia di San Paolo under contract ORTO11TPXK. The authors 
would like to thank Jenni Smillie for help and collaboration on related topics, and
Claude Duhr and \O{}yvind Almelid for making available their
codes for the treatment of polylogarithmic functions, and for useful discussions.


\appendix


\section{Basis Functions}
\label{appBasis}

In this appendix we list the explicit expressions for the basis functions defined
in \eqn{eq:Mbasis}, in terms of polylogarithms and harmonic polylogarithms, up 
to transcendental weight five. Harmonic polylogarithms are defined as in 
\cite{Remiddi:1999ew}.

\begin{itemize}

\item Weight one.
    \begin{align}
          \begin{split}
          M_{0,0,0}(\alpha) \, = \, 2 \log(\alpha) \, . 
          \end{split} 
    \end{align}

\item Weight two.
    \begin{align}
          \begin{split}
          M_{1,0,0}(\alpha) \, = \, 2 \, \text{Li}_2 (\alpha^2) + 4 \log(\alpha)
          \log \left(1 - \alpha^2 \right) - 2 \log^2 (\alpha) - 2 \, \zeta(2) .
          \end{split}
    \end{align}

\item Weight three.
    \begin{align}
          \begin{split}
          M_{0,0,2}(\alpha) \, = \, \frac{8}{3} \log^3(\alpha) \, ,
          \end{split} 
    \end{align}
    \begin{align}
          \begin{split}
          M_{0,1,1}(\alpha) \, = \, 2 \, \text{Li}_3 (\alpha^2) - 2 \log(\alpha) 
          \bigg[ \text{Li}_2 (\alpha^2) + \frac{\log^2 (\alpha)}{3}  + \zeta(2) \bigg]
          - 2 \, \zeta(3) \, ,  
          \end{split} 
    \end{align}
    \begin{align}
          \begin{split}
          M_{0,2,0}(\alpha) \, = \, \frac{2}{3} \log^3 (\alpha) + 4 \, \zeta(2) \, 
          \log(\alpha) \, ,
          \end{split}
    \end{align}
    \begin{align}
          \begin{split}
          M_{2,0,0}(\alpha) & = - \, 4 \, \bigg[ \text{Li}_3 (\alpha^2) + 2 \text{Li}_3 
          \left(1 - \alpha^2 \right) \bigg] - 8 \log \left(1 - \alpha^2 \right) \log^2(\alpha) \\
          & + \, \frac{8}{3} \, \log^3(\alpha) + 8 \, \zeta(2) \, \log(\alpha) 
          + 4 \, \zeta(3) \, .
          \end{split}
    \end{align}

\item Weight four.
    \begin{align}
	  \begin{split}
	  M_{3,0,0} (\alpha) & = \, 12 \, \bigg[ \text{Li}_4 (\alpha^2) - 4 \, \text{Li}_4 
	  \left(1 - \alpha^2 \right) \bigg] - 24 \, S_{2,2} (\alpha^2) \\ &
	  - \, 24 \, \log \left(1 - \alpha^2 \right) \, \text{Li}_3 (\alpha^2) 
	  - 24 \, \log^2 \left(1 - \alpha^2 \right) \, \log^2 (\alpha) \\ & 
	  + \, 16 \, \log \left(1 - \alpha^2 \right) \, \log^3(\alpha) - \, 4 \, \log^4 (\alpha) \\ &
	  - \, 24 \, \zeta(2) \, \log(\alpha) \, 
	  \log \left[ \frac{\alpha}{\left(1 - \alpha^2 \right)^2} \right] \\ & 
	  + \, 24 \, \zeta(3) \, \log \left[ \alpha \left(1 - \alpha^2 \right) \right] - 
	  \, 6 \, \zeta(4) \, ,
	  \end{split} 
    \end{align}
    \begin{align}
	  \begin{split} \hspace{-3mm}
	  M_{1,2,0} (\alpha) & = \, 4 \, \text{Li}_4 (\alpha^2) - 4 \, \log (\alpha) \, 
	  \text{Li}_3 (\alpha^2) + 2 \, \log^2(\alpha) \, \text{Li}_2 (\alpha^2) \\ &
	  + \, \frac{4}{3} \, \log^3 (\alpha) \, \log \left(1 - \alpha^2 \right) 
	  - \, \frac{2}{3} \, \log^4(\alpha) \\ & 
	  + \, \zeta(2) \, \Big[ 8 \, \log(\alpha) \log \left(1 - \alpha^2 \right) + 
	  4 \, \text{Li}_2 (\alpha^2) - 6 \, \log^2 (\alpha) \Big] \\ &
	  + \, 4 \, \zeta(3) \, \log(\alpha) - 14 \, \zeta(4) \, ,
	  \end{split} 
    \end{align}
    \begin{align}
	  \begin{split}
	  M_{1,0,2} (\alpha) & = \, 4 \, \text{Li}_4 (\alpha^2) - 8 \, \log(\alpha) \, 
	  \text{Li}_3 (\alpha^2) + 8 \, \log^2 (\alpha) \, \text{Li}_2 (\alpha^2) \\ &
	  + \frac{16}{3} \, \log^3(\alpha) \, \log \left(1 - \alpha^2 \right)
	  - \, \frac{4}{3} \, \log^4 (\alpha) - 4 \, \zeta(4) \, , 
	  \end{split}
    \end{align}
    \begin{align}
          \begin{split} \hspace{-3mm}
          M_{1,1,1} (\alpha) & = - \, 4 \, \text{Li}_4 (\alpha^2) + 4 \, S_{2,2} (\alpha^2) + 
          2 \, \log \left[ \alpha \left(1 - \alpha^2 \right)^2 \right] \, \text{Li}_3 (\alpha^2) \\ & 
          + \,4 \, \log (\alpha) \, \text{Li}_3 \left(1 - \alpha^2 \right) \\ &
          - \, \frac{4}{3} \, \log (\alpha)^2 \, \log \left(1 - \alpha^2 \right) \bigg[
          \log (\alpha) - 3 \, \log \left(1 - \alpha^2 \right) \bigg] \\ & 
          - \, 8 \, \zeta(2) \, \log(\alpha)\log(1-\alpha^2) 
          - 2 \, \zeta(3) \, \log \left[ \alpha \left(1 - \alpha^2 \right)^2 \right] + 3 \, \zeta(4) \, . 
          \end{split}
    \end{align}

\item Weight five
    \begin{align}
          \begin{split}
          M_{0,0,4} (\alpha) \, = \, \frac{32}{5} \, \log^5(\alpha) \, ,
          \end{split}
    \end{align}
    \begin{align}
	  \begin{split}
          M_{4,0,0} (\alpha) & = \, 48 \, H_{0,0,0,1,0} (\alpha^2) 
          + 96 \, H_{1,0,0,1,0} (\alpha^2) + 96 \, H_{0,1,0,1,0} (\alpha^2) \\ &
          + \, 48 \, H_{0,0,1,0,0} (\alpha^2) + 96 \, H_{0,0,1,1,0} (\alpha^2) 
          + 192 \, H_{1,1,0,1,0} (\alpha^2) \\ & 
          + \, 96 \, H_{1,0,1,0,0} (\alpha^2) + 192 \, H_{1,0,1,1,0} (\alpha^2)
          + 48 \, H_{0,1,0,0,0} (\alpha^2) \\ & 
          + \, 96 \, H_{0,1,1,0,0} (\alpha^2) + 192 \, H_{0,1,1,1,0} (\alpha^2) 
          + 48 \, H_{1,0,0,0,0} (\alpha^2) \\ &
          + \, 96 \, H_{1,1,0,0,0} (\alpha^2) + 192 \, H_{1,1,1,0,0} (\alpha^2)
          + 384\, H_{1,1,1,1,0} (\alpha^2) \\ & 
          + \, \frac{32}{5} \, \log^5 (\alpha) 
          + 64 \, \zeta(2) \, \log^3 \left( \frac{\alpha}{1 - \alpha^2} \right) \\ &
          - 96 \, \zeta(3) \bigg[\zeta(2) + \log^2 \left( \frac{\alpha}{1 - \alpha^2} \right) \bigg] \\ &
          + \, 432 \, \zeta(4) \, \log \frac{\alpha}{1 - \alpha^2}  - 144 \, \zeta(5) \, ,
          \end{split}
    \end{align}
    \begin{align}
	  \begin{split}
          M_{0,4,0} (\alpha) & =  \, \frac{2}{5} \, \log^5 (\alpha) + 8 \, \zeta(2) \, \log^3(\alpha) 
          + 84 \, \zeta(4) \, \log(\alpha) \, , 
          \end{split}
    \end{align}
    \begin{align}
	  \begin{split}
	  M_{0,1,3} (\alpha) & = \, - \, 6 \, H_{0,0,0,1,0} (\alpha^2) 
	  - 6 \, H_{0,0,1,0,0} (\alpha^2) - 6 \, H_{0,1,0,0,0} (\alpha^2) \\ &
	  - \, \frac{12}{5} \, \log^5 (\alpha) - 8 \, \zeta(2) \, \log^3(\alpha) 
	  - 12 \, \zeta(3) \, \log^2(\alpha) \\ &
	  - \, 12 \, \zeta(4) \, \log(\alpha) - 12 \, \zeta(5) \, ,
          \end{split}
    \end{align}
    \begin{align}
	  \begin{split}
          M_{0,3,1} (\alpha) & = \, - \frac{3}{2} \, \bigg[ H_{0,0,0,1,0} (\alpha^2) + 
          H_{0,1,0,0,0} (\alpha^2) \bigg]  - \frac{2}{5} \, \log^5 (\alpha) \\ &
          + \, 6 \, \zeta(2) \, \Big[ - \log^3(\alpha) - 2 \, \log(\alpha) \, \text{Li}_2 (\alpha^2)
          + 2 \, \text{Li}_3 (\alpha^2) \Big] \\ &
          - \, 6 \, \zeta(3) \, \log^2(\alpha) 
          - 42 \, \zeta(4) \, \log(\alpha) - 12 \, \zeta(2) \, \zeta(3) - 12 \, \zeta(5) \, , 
	  \end{split}
    \end{align}
    \begin{align}
          \begin{split}
	  M_{2,2,0} (\alpha) & = \, 4 \, H_{0,0,0,1,0} (\alpha^2) + 4 \, H_{1,0,0,1,0} (\alpha^2)
	  + 4 \, H_{0,0,1,1,0} (\alpha^2) \\ & 
	  + \, 4 \, H_{1,0,0,0,0} (\alpha^2)
	  + 4 \, H_{1,1,0,0,0} (\alpha^2) + 2 \, H_{0,1,0,0,0} (\alpha^2) \\ &
	  + \, 2 \, H_{0,0,1,0,0} (\alpha^2) + \, \frac{16}{15} \, \log^5(\alpha) \\ &
	  + \, 8 \, \zeta(2) \, \bigg[ \log^2 (\alpha) \bigg( 2 \, \log(\alpha) 
	  - 3 \, \log \left(1 - \alpha^2 \right) \bigg) \\ &
	  \hspace{18mm} - \, \log(\alpha) \, \text{Li}_2 (\alpha^2) 
	  - 2 \, \text{Li}_3 \left(1 - \alpha^2 \right) \bigg] \\ & + 
	  \, 4 \, \zeta(3) \, \bigg[ 2 \, \text{Li}_2 (\alpha^2) + \log(\alpha) \, \bigg(4 \, 
	  \log \left(1 - \alpha^2 \right) - 3 \, \log(\alpha) \bigg) \bigg] \\ &
	  + \, 2 \, \zeta(4) \, \bigg[ 67 \, \log(\alpha) - 8 \, \log \left(1 - \alpha^2 \right) 
	  \bigg] -16 \, \zeta(5) \, ,
          \end{split}
    \end{align}
    \begin{align}
          \begin{split}
          M_{2,0,2} (\alpha) & = \, 8 \, H_{0,1,0,0,0} (\alpha^2) 
          + 8 \, H_{1,0,0,0,0} (\alpha^2) + 16 \, H_{1,1,0,0,0} (\alpha^2) \\ &
          + \, \frac{16}{15} \, \log^5 (\alpha)
          + 16 \, \zeta(4) \, \log \left( \frac{\alpha}{1 - \alpha^2} \right) \\ &
          + \, 16 \, \zeta(2) \zeta(3) - 8 \, \zeta(5) \, ,
          \end{split}
    \end{align}
    \begin{align}
          \begin{split}
          M_{2,1,1} (\alpha) & = \, - \, 4 \, H_{0,0,0,1,0} (\alpha^2) 
          - 8 \, H_{1,0,0,1,0} (\alpha^2) - 8 \, H_{0,1,0,1,0} (\alpha^2) \\ &
          - 2 \, H_{0,0,1,0,0} (\alpha^2) - 8 \, H_{0,0,1,1,0} (\alpha^2) 
          - 8 \, H_{1,1,0,1,0} (\alpha^2) \\ & 
          - 4 \, H_{1,0,1,0,0} (\alpha^2) - 8 \, H_{1,0,1,1,0} (\alpha^2)
          - 4 \, H_{0,1,0,0,0} (\alpha^2) \\ & 
          - 4 \, H_{0,1,1,0,0} (\alpha^2) - 8 \, H_{0,1,1,1,0} (\alpha^2)
          - 4 \, H_{1,1,0,0,0}(\alpha^2) + \frac{4}{15} \, \log^5(\alpha) \\ &
          - 8 \, \zeta(2) \bigg[ \log(\alpha) \bigg( 2 \, \text{Li}_2 (\alpha^2)
          + \, \log^2 \left(1 - \alpha^2 \right) \bigg) - \text{Li}_3 (\alpha^2) \bigg] \\ &
          - 4 \, \zeta(3) \bigg[ 3 \, \log^2(\alpha) + 2 \, \log^2 \left(1 - \alpha^2 \right) 
          - 2 \, \log(\alpha) \, \log \left(1 - \alpha^2 \right) \bigg] \\ &
          - 6 \, \zeta(4) \, \Big[ 5 \, \log(\alpha) - 2 \, \log \left(1 - \alpha^2 \right) \Big] 
          - 8 \, \zeta(2) \, \zeta(3) - 20 \, \zeta(5) \, ,  
          \end{split}
    \end{align}
    \begin{align}
          \begin{split} \hspace{-5mm}
          M_{0,2,2} (\alpha) & = \, 2 \, H_{0,0,0,1,0} (\alpha^2) 
          + 4 \, H_{0,1,0,1,0} (\alpha^2) + 2 \, H_{0,1,0,0,0} (\alpha^2) \\ &
          + \frac{16}{15} \, \log^5(\alpha) + 8 \, \zeta(2) \, \bigg[ \log^3 (\alpha)
          + \log(\alpha) \, \text{Li}_2 (\alpha^2) - \text{Li}_3 (\alpha^2) \bigg] \\ &  
          + 8 \, \zeta(3) \bigg[ \log^2 (\alpha) + \text{Li}_2 (\alpha^2) \bigg] 
          + 22 \, \zeta(4) \, \log(\alpha) + 8 \, \zeta(2) \, \zeta(3) + 8 \, \zeta(5) \, .
\end{split} 
\end{align}
\end{itemize}


\section{Calculation of the (2,2,2) web}
\label{App222}

In this appendix, we provide more details regarding the calculation of the (2,2,2) web,
the results of which are presented in \sect{222}. 


\subsection{Unsubtracted web}
\label{Unsubtra222}

The general method for calculating a given web diagram has been presented 
in \sect{MGEWs}. Here we present the kernels, defined in \eqn{gendiag3} and 
\eqn{phi_D}, for each diagram appearing in the (2,2,2) web. To simplify our notations 
slightly, we relabel the variables used in the \sect{222} as $\{x_1, x_2, x_3\} \to \{x,y,z\}$. 
Applying the relevant definitions to the kinematic factor of diagrams $A$ and $B$ in 
Fig.~\ref{fig:222} one finds,
\beq
  \phi^{(3)}_A(x,y,z;\epsilon) \, = \, \phi^{(3)}_B(x,y,z;\epsilon) \, = \, \frac{1}{4}
  \log^2 \left[ \frac{x}{1 - x} \, \frac{y}{1 - y} \, \frac{z}{1 - z} \right] + \mathcal{O}(\epsilon) \, ,
\label{FA222b}
\eeq
where we used \Eqns{prefac}{propafu}. Note that it is necessary to use the symmetry of the 
function to remove the factor,
\begin{align}
  \theta_+(n = 3) \, = \, \theta \bigg( \frac{1 - x}{x} \frac{1 - y}{y} \frac{1 - z}{z} > 1 \bigg) \, .
\end{align}
The other diagrams in Fig.~\ref{fig:222} have subdivergences, therefore the kernels have 
to be computed including higher orders in their $\epsilon$ expansion. For example, from 
diagram $C$ one finds
\beqa
\label{FC222c}
  \phi^{(3)}_C(x,y,z;\epsilon) & = & \frac{1}{8 \epsilon^2}
  \bigg\{ 1 + 2 \, \epsilon \log \left[ \frac{(1 - y)^2 (1 - z)}{z^2 x} \right] + \epsilon^2
  \left[ 24 \, \dilog{1 - y}{z} \right. \nonumber \\
  & &\left. + \, 2 \, \log^2 \left( \frac{(1 - y)^2 (1 - z)}{z^2 x} \right) - 2 \log^2 
  \left(\frac{(1 - x)(1 - y)(1 - z)}{xyz} \right) \right] \\ & & +~ \mathcal{O}(\epsilon^3) \bigg\}
  \, . \nonumber 
\eeqa
The remaining web kernel contributions can then be found by permuting the variables $x$, 
$y$ and $z$ in the integrand, with the results
\beqa
\label{FD-H}
  \phi^{(3)}_D(x,y,z;\epsilon) & = & \phi^{(3)}_C (z, x, y; \epsilon) \, , \nonumber \\
  \phi^{(3)}_E(x,y,z;\epsilon) & = & \phi^{(3)}_C (x, z, y; \epsilon) \, , \nonumber \\
  \phi^{(3)}_F(x,y,z;\epsilon) & = & \phi^{(3)}_C (y, x, z; \epsilon) \, ,  \\
  \phi^{(3)}_G(x,y,z;\epsilon) & = & \phi^{(3)}_C (z, y, x; \epsilon) \, , \nonumber \\
  \phi^{(3)}_H(x,y,z;\epsilon) & = & \phi^{(3)}_C (y, z, x; \epsilon) \, . \nonumber
\eeqa
From these functions one can compute the unsubtracted web, 
\begin{align}
  \begin{split}
  W_{(2,2,2)} \, = \, \sum_{j = 1}^4 c_j^{(3)} \mathcal{F}_{(2,2,2); \, j}^{(3)} \, ,
  \end{split}
\end{align}
using \eqn{webcolco} and taking the specific coefficients $Y_{X,j}^{-1}$ given in 
Table \ref{tab:222}. One finds
\begin{align}
  \begin{split}
  \mathcal{F}_{(2,2,2),j}^{(3)} & = \kappa^3 \, \Gamma(6 \epsilon) \int_0^1 
  dx \, dy \, dz \, p_\epsilon(x,\alpha_{12}) p_\epsilon(y,\alpha_{23}) 
  p_\epsilon(z,\alpha_{13}) \sum_{X \in (2,2,2)} Y_{X,j}^{-1}~ \phi_X^{(3)} (x,y,z; \epsilon) \\
  & = \kappa^3 \, \Gamma(6 \epsilon) \int_0^1 dx \, dy \, dz \, p_\epsilon(x,\alpha_{12}) 
  p_\epsilon(y,\alpha_{23}) p_\epsilon(z,\alpha_{13})~ \phi_{(2,2,2),j}^{(3)}(x,y,z; \epsilon) \, .
  \end{split}
\label{eq:app222unsub}
\end{align}
 

\subsection{Subtracted web}
\label{Subtra222}

As explained in \sect{Webs}, the anomalous dimension is obtained from subtracted 
webs. Thus the single pole of \eqn{eq:app222unsub} must be combined with suitable 
subtraction terms, consisting of commutators of lower-order webs, as prescribed 
\eqn{websub}. Recall that simplifications occur~\cite{Gardi:2013saa} when this subtraction 
is performed under the integral over variables $(x,y,z)$ corresponding to the gluon emission 
angles. In particular, the symbol alphabet and factorization constraints which allow us to 
use the basis of functions introduced in section \ref{Basis} only apply to the subtracted web.
Consider then the integrand of \eqn{websub},
\begin{align}
\label{sub_web_222}
 \begin{split}
  \! \overline{w}_{(2,2,2),i}^{(3,-1)} \! = \bigg(\frac{1}{4\pi}\bigg)^3 \! 
  \int_0^1 \!dx dy dz \, p_0(x,\alpha_{12}) p_0(y,\alpha_{23}) p_0(z,\alpha_{13}) 
  \bigg( \mathcal{G}_{0,i} + \sum_k \Delta\mathcal{G}_{k,i} \bigg) \, ,
 \end{split}
\end{align}
where we defined $\mathcal{G}_{0,i}$ as the contribution proportional to $c_i^{(3)}$ 
from the unsubtracted web, obtained by expanding the integrand of \eqn{eq:app222unsub}; 
$\Delta\mathcal{G}_{k,i}$, in turn, are the contributions from the commutators of lower order 
webs to this colour factor, where $k$ runs over the relevant commutator terms in \eqn{websub}  
(note that $\Delta\mathcal{G}_{k,i}$ include the numerical factors appearing there). We recall 
that each of these contributions to $\mathcal{G}_{(2,2,2),i}$ depends on $x, y$ and $z$ as 
well as on $q(x,\alpha_{12}), q(y,\alpha_{23})$ and $q(z,\alpha_{13})$. Using the lower-order 
web results collected in Ref.~\cite{Gardi:2013saa}, the contributions to the integrand of 
$\overline{w}_{(2,2,2)}^{(3)}$ multiplying the colour factor $c_1^{(3)}$ are
\beqa
 \label{comms1}
  \Delta\mathcal{G}_{\left[ w^{(1,0)}, w^{(2,-1)} \right],1} \, & = & \,
   \, \frac{1}{4} \, \bigg[ \log \big[ q(x, \alpha_{12}) \big] \, 
  \log \left( \frac{y}{z} \right) \nonumber \\ 
  & & \hspace{1.2cm} - \, \log \big[ q(z,\alpha_{13}) \big] 
  \log \left( \frac{x}{y} \right) \bigg] \, , \nonumber \\
  \Delta\mathcal{G}_{\left[ w^{(2,0)}, w^{(1,-1)} \right],1} \, & = & \,
   -\frac12 \bigg[ \dilog{y}{z} - \dilog{z}{y} 
  \nonumber \\ & & + \, \dilog{y}{x} - \dilog{x}{y} 
  \\
  & & + \, \frac{1}{2} \log \left( \frac{y}{z} \right)
  \left( \log \big[ q(y,\alpha_{23}) \big] + \log \big[ q(z,\alpha_{13})] \right)
  \nonumber \\
  & & - \, \frac{1}{2} \log \left( \frac{x}{y} \right)
  \left( \log \big[ q(x,\alpha_{12}) \big] + \log \big[ q(y,\alpha_{23}) \big]
  \right) \bigg] \, , \nonumber \\
  \Delta\mathcal{G}_{\big[w^{(1,0)}, [ w^{(1,-1)}, w^{(1,0)}] \big],1} \,
  & = & \,
 - \frac{1}{12} \, \Big[ \log \big[ q(x,\alpha_{12}) \big] 
  \log \big[ q(y,\alpha_{23}) \big] \nonumber \\
  & & \hspace{-25mm}+\log \big[ q(z,\alpha_{13}) \big] 
  \log \big[ q(y,\alpha_{23}) \big] - \, 2 \log \big[ q(z,\alpha_{13}) \big] 
  \log \big[ q(x,\alpha_{12}) \big] \Big] \, , \nonumber \\
  \Delta\mathcal{G}_{\big[ w^{(1,-1)}, [ w^{(1,1)}, w^{(1,-1)}] \big],1} \, 
  & = & \, 
  -\frac{1}{24} \, \Big[ \log^2 \big[ q(z,\alpha_{13}) \big] - 2 
  \log^2 \big[ q(y,\alpha_{23}) \big] \nonumber \\
  & & \hspace{9mm} + \, \log^2 \big[ q(x,\alpha_{12}) \big] \Big] \, . \nonumber
\eeqa
The contributions to the coefficient of the colour factor $c^{(3)}_2$ are
\beqa
\label{comms2}
  \Delta\mathcal{G}_{\left[ w^{(1,0)}, w^{(2,-1)} \right],2} \, & = & \,
   \, \frac{1}{4} \, \bigg[ - \log \big[ q(x, \alpha_{12}) \big] \, 
  \log \left( \frac{y}{z} \right)\,+\,
  \, \log \big[ q(y,\alpha_{23}) \big] 
  \log \left( \frac{z}{x} \right) \bigg] \, , \nonumber \\
  \Delta\mathcal{G}_{\left[ w^{(2,0)}, w^{(1,-1)} \right],2} \, & = & \,
  -\frac12 \bigg[ \dilog{z}{x} - \dilog{x}{z} 
  + \, \dilog{z}{y} - \dilog{y}{z} \nonumber \\
  & & + \, \frac{1}{2} \log \left( \frac{z}{x} \right)
  \left( \log \big[ q(z,\alpha_{13}) \big] + \log \big[ q(x,\alpha_{12})] \right)
  \nonumber \\  & & 
  - \, \frac{1}{2} \log \left( \frac{y}{z} \right)
  \left( \log \big[ q(y,\alpha_{23}) \big] + \log \big[ q(z,\alpha_{13}) \big]
  \right) \bigg] \, , \nonumber \\
  \Delta\mathcal{G}_{\big[w^{(1,0)}, [ w^{(1,-1)}, w^{(1,0)}] \big],2} \,
  & = & \,
 - \frac{1}{12} \, \Big[ \log \big[ q(x,\alpha_{12}) \big] 
  \log \big[ q(z,\alpha_{13}) \big]   \\
  & & \hspace{-25mm}+\log \big[ q(z,\alpha_{13}) \big] 
  \log \big[ q(y,\alpha_{23}) \big] - \, 2 \log \big[ q(y,\alpha_{23}) \big] 
  \log \big[ q(x,\alpha_{12}) \big] \Big] \, , \nonumber \\
  \Delta\mathcal{G}_{\big[ w^{(1,-1)}, [ w^{(1,1)}, w^{(1,-1)}] \big],2} \,
  & = & \, 
 - \frac{1}{24} \, \Big[ \log^2 \big[ q(y,\alpha_{23}) \big] - 2 
  \log^2 \big[ q(z,\alpha_{13}) \big] \nonumber \\
  & & \hspace{9mm} + \, \log^2 \big[ q(x,\alpha_{12}) \big] \Big] \, . \nonumber
\eeqa
Finally, the contributions to the coefficient of the colour factor $c_3^{(3)}$ are 
found to be
\beqa
\label{comms3}
  \Delta\mathcal{G}_{\left[ w^{(1,0)}, w^{(2,-1)} \right],3} \, & = & \,
   \, \frac{1}{4} \, \bigg[ - \log \big[ q(z, \alpha_{13}) \big] \, 
  \log \left( \frac{x}{y} \right) \nonumber \\ 
  & & \hspace{1.2cm} + \, \log \big[ q(y,\alpha_{23}) \big] 
  \log \left( \frac{z}{x} \right) \bigg] \, , \nonumber \\
  \Delta\mathcal{G}_{\left[ w^{(2,0)}, w^{(1,-1)} \right],3} \, & = & \,
  -\frac12 \bigg[ \dilog{z}{x} - \dilog{x}{z} 
  \nonumber \\ & & + \, \dilog{y}{x} - \dilog{x}{y} 
   \\
  & & + \, \frac{1}{2} \log \left( \frac{z}{x} \right)
  \left( \log \big[ q(z,\alpha_{13}) \big] + \log \big[ q(x,\alpha_{12})] \right)
  \nonumber \\
  & & - \, \frac{1}{2} \log \left( \frac{x}{y} \right)
  \left( \log \big[ q(x,\alpha_{12}) \big] + \log \big[ q(y,\alpha_{23}) \big]
  \right) \bigg] \, , \nonumber \\
  \Delta\mathcal{G}_{\big[w^{(1,0)}, [ w^{(1,-1)}, w^{(1,0)}] \big],3} \,
  & = & \,
  \, \frac{1}{12} \, \Big[ \log \big[ q(x,\alpha_{12}) \big] 
  \log \big[ q(y,\alpha_{23}) \big] \nonumber \\
  & & \hspace{-25mm}+\log \big[ q(z,\alpha_{13}) \big] 
  \log \big[ q(x,\alpha_{12}) \big] - \, 2 \log \big[ q(y,\alpha_{23}) \big] 
  \log \big[ q(z,\alpha_{13}) \big] \Big] \, , \nonumber \\
  \Delta\mathcal{G}_{\big[ w^{(1,-1)}, [ w^{(1,1)}, w^{(1,-1)}] \big],3} \,
  & = & \, 
   \, \frac{1}{24} \, \Big[ \log^2 \big[ q(z,\alpha_{13}) \big] - 2 
  \log^2 \big[ q(x,\alpha_{12}) \big] \nonumber \\
  & & \hspace{9mm} + \, \log^2 \big[ q(y,\alpha_{23}) \big] \Big] \, . \nonumber
\eeqa
There are no commutator counterterms contributing to the fully antisymmetric 
colour factor $c_4^{(3)}$, as there are no colour factors of lower order webs that 
commute to produce the desired structure. This is consistent with the fact that the 
kinematic function associated with $c^{(3)}_4$ involves only staircase diagrams, which 
are irreducible, and as such do not contain subdivergences.

Combining all terms, using appropriate dilogarithm identities, and the symmetry of 
$p_0(x, \alpha)$ under $x \leftrightarrow 1 - x$, and finally using the definition given 
in \eqn{eq:Mbasis}, one finds the results presented in \Eqns{gamints222}{gamints222c}. 
As a check of these results, one may verify that the ${\cal O} (\epsilon^{-2})$ pole of 
the (2,2,2,) web vanishes, according to the web consistency relation discussed
in Ref.~\cite{Gardi:2011yz},
\beq
  \left. w^{(3,-2)} \right|_{\beta_0 = 0} \, = \, - \, \frac{1}{6}
  \left[w^{(1,-1)},w^{(2,-1)} \right] \, .
\label{eps2}
\eeq
As a further check, the sum of all web diagrams must give a product of one-loop 
graphs, in accordance with \eqn{Abelian_sum}. We have explicitly confirmed that 
both of these criteria are satisfied.


\section{Calculation of the (1,2,3) web}
\label{App123}

The calculation of the (1,2,3) web, shown in Fig.~\ref{fig:123}, proceeds similarly to 
the (2,2,2) case considered in the previous Appendix. 


\subsection{Unsubtracted web}
\label{Unsubtra123}

Using the method of \sect{MGEWs}, the result for the contribution to the web kernel
$\phi_{(1,2,3)}^{(3)}$ from diagram $A$ is
\beq
\phi_A^{(3)}  =  \frac{1}{4} \log^2 \left[\frac{y (1 - x)}{x(1 - y)} \right] 
\label{FA123c}
\eeq
Carrying out a similar exercise for the other diagrams of the (1,2,3) web gives the 
results
\beqa
  \phi_B^{(3)} & = & \frac{1}{8 \epsilon^2} \, \theta(x - y) \, 
  \left( \frac{1 - x}{z} \right)^{2 \epsilon} \bigg[ \left( \frac{1 - y}{1 - x} \right)^{4 \epsilon}
  - \left( \frac{y}{x} \right)^{4 \epsilon} 
  \nonumber \\  & & \hspace{2cm} 
  + \, 24 \, \epsilon^2 \,\left( \dilog{x}{y} - \, 
  \dilog{y}{x} \right)
  +{\cal O}(\epsilon^3) \bigg] \, ,
  \nonumber \\
  \phi_C^{(3)} & = & \frac{1}{2\epsilon} \, \theta(x - y) \left[ \frac{\Gamma(2 \epsilon) 
  \Gamma(4 \epsilon)}{\Gamma(6 \epsilon)} - \frac{1}{2 \epsilon}
  \left( \frac{1 - y}{z} \right)^{2 \epsilon} \right] 
  \left[ \left( \frac{1 - y}{1 - x} \right)^{2 \epsilon} - \left( \frac{y}{x} \right)^{2 \epsilon}
  \right] \, , \nonumber \\
  \phi_D^{(3)} & = & \frac{1}{8 \epsilon^2} \left( \frac{1 - y}{z} \right)^{2 \epsilon} 
  \left[ \theta(y - x) \left( \frac{x}{y} \right)^{4 \epsilon} + \theta(x - y)
  \left( \frac{1 - x}{1 - y} \right)^{4 \epsilon} \right. \nonumber \\ & & 
  \left. + 24 \, \epsilon^2 \, \theta(y - x) \, \dilog{x}{y}
  + \, 24 \, \epsilon^2 \, \theta(x - y) \, \dilog{1 - x}{1 - y}+{\cal O}(\epsilon^3) \right] \, , \nonumber \\
  \phi_E^{(3)} & = & \frac{1}{8 \epsilon^2} \left\{ \theta(x - y) \left[ 2 \left( 
  \frac{1 - x}{z} \right)^{2 \epsilon} \left( \frac{1 - y}{1 - x} \right)^{- 2 \epsilon}
  - \left( \frac{1 - y}{z} \right)^{2 \epsilon} \left( \frac{1 - y}{1 - x} \right)^{- 4 \epsilon}
  \right] \right. \nonumber \\  & & \hspace{8mm} 
  \left. + \, \theta(y - x) \left[ 2 \left( \frac{1 - x}{z} \right)^{2 \epsilon}
  \left( \frac{y}{x} \right)^{- 2 \epsilon} - \left( \frac{1 - y}{z} \right)^{2 \epsilon}
  \left( \frac{y}{x} \right)^{ - 4 \epsilon} \right] \right. \nonumber \\ & & \left. \hspace{8mm}
  + \, 24 \, \epsilon^2 \, \dilog{1 - x}{z} +{\cal O}(\epsilon^3)\right\} \, ,
  \label{diags123}  \\
  \phi_F^{(3)} & = & \frac{1}{2 \epsilon}
  \Bigg\{ \theta(x - y) \left( \frac{1 - y}{1 - x} \right)^{- 2 \epsilon}
  \left[ \frac{\Gamma(4 \epsilon) \Gamma(2 \epsilon)}{\Gamma(6 \epsilon)}
  \left( 1 + 8 \epsilon^2 \dilog{1 - x}{1 - y} \right) \right. \nonumber \\
  & & \hspace{6mm} \left. - \frac{1}{2 \epsilon} \left( \frac{1 - x}{z}
  \right)^{2 \epsilon} \left( 1 + 12 \epsilon^2 \dilog{1 - x}{1 - y} \right) \right]
  \nonumber \\
  & & \hspace{6mm} + \, \theta(y - x) \left( \frac{y}{x} \right)^{- 2 \epsilon}
  \left[\frac{\Gamma(4 \epsilon) \Gamma(2 \epsilon)}{\Gamma(6 \epsilon)}
  \left( 1+ 8 \epsilon^2 \dilog{x}{y} \right) 
  - \frac{1}{2 \epsilon} \left( \frac{1 - x}{z} \right)^{2 \epsilon} \right. \nonumber \\
  & & \hspace{6mm} \left. \times \left( 1 + 12 \epsilon^2 
  \dilog{x}{y} \right) \right]  -  6 \epsilon\left( \frac{1 - x}{z}
  \right)^{2 \epsilon} \dilog{1 - x}{z}  +{\cal O}(\epsilon^2)\Bigg\} \, . \nonumber 
\eeqa
These expression must of course be expanded in powers of $\epsilon$, up to ${\cal O}
\left( \epsilon^0 \right)$. Notice that there is a difference with respect to the (2,2,2) 
web, in that Heaviside functions survive in the integrand. This can ultimately be traced 
to the fact that the (1,2,3) web contains a crossed-gluon pair spanning a single cusp 
angle, and a Heaviside function is needed in order to implement the crossing condition.
From \eqn{diags123}, the unsubtracted web, 
\begin{align}
  \begin{split}
  W_{(1,2,3)} \, = \, \sum_{j = 1}^4 c_j^{(3)} \mathcal{F}_{(1,2,3),j}^{(3)} \, ,
  \end{split}
\end{align}
is obtained through \eqn{webcolco} by taking the specific coefficients $Y_{X,j}^{-1}$ 
given in table \ref{tab:123}, as
\begin{align}
\label{F123_i_}
  \begin{split}
  \! \mathcal{F}_{(1,2,3),j}^{(3)} & = \kappa^3 \, \Gamma(6 \epsilon) \int_0^1 dx \, dy \, dz \,
  p_{\epsilon}(x,\alpha_{23}) p_{\epsilon}(y,\alpha_{23}) p_{\epsilon}(z,\alpha_{13}) \!\!
  \sum_{X \in (1,2,3)} Y_{X,j}^{-1} \, \phi_X^{(3)}(x,y,z;\epsilon) \\
  & = \kappa^3 \, \Gamma(6 \epsilon) \int_0^1 dx \, dy \, dz \, p_{\epsilon}(x,\alpha_{23}) 
  p_{\epsilon}(y,\alpha_{23}) p_{\epsilon}(z,\alpha_{13}) \, 
  \phi_{(1,2,3),j}^{(3)}(x,y,z;\epsilon) \, ,
  \end{split}
\end{align}
where we recall that the contribution to the $i=1$ colour component vanishes. In the following we compote the (1,2,3) subtracted web for $i=2,\,3$ and $4$.


\subsection{Subtracted web}
\label{Subtra123}

As for the (2,2,2) web, it is useful to collect results for the subtraction terms separated 
according to the colour factor they contribute to. We once again split the subtracted 
web kernel into contributions from the unsubtracted web, $\mathcal{G}_{0,i}$ originating 
in the expansion of the integrand in \eqn{F123_i_}, and the commutators of lower order 
webs, $\Delta\mathcal{G}_{j,i}$, so that the coefficient of the colour factors $c_i^{(3)}$
can be written as
\begin{align}
  \begin{split}
  \overline{w}_{(1,2,3),i}^{(3,-1)} & = \bigg(\frac{1}{4\pi}\bigg)^3 \! \int_0^1 \! 
  dx \, dy \,dz \, p_0(x,\alpha_{23}) p_0(y,\alpha_{23}) p_0(z,\alpha_{13}) 
  \bigg( \mathcal{G}_{0,i} + \sum_k \Delta \mathcal{G}_{k,i} \bigg) \, .
  \end{split}
\end{align}
We recall that each of these contributions to  $\mathcal{G}_{(1,2,3),i}$ depends on $x,y$ and 
$z$ as well as on $q(x,\alpha_{23}), q(y,\alpha_{23})$ and $q(z,\alpha_{13})$. We will see that 
after cancellations only logarithmic dependence on these arguments will survive.
The results of the commutators with the colour factors $c_i^{(3)}$ with $i=2$ and $3$ are
\beqa
\label{comms1123}
  \Delta\mathcal{G}_{\left[ w^{(1,0)}, w^{(2,-1)} \right],i} 
  \, & = & \, 
   \, \frac{1}{4} \, \log \left( \frac{x}{z} \right) 
  \log \big[ q(y,\alpha_{23}) \big] \, , \nonumber \\
  \Delta\mathcal{G}_{\left[ w^{(2,0)}, w^{(1,-1)} \right],i} 
  \, & = & \, 
  \, -\frac12 \left[ \dilog{x}{z} - \dilog{z}{x} \right. \nonumber \\
  & & \hspace{6mm} \left. + \, \frac{1}{2} \log \left( \frac{x}{z} \right)
  \Big( \log \big[ q(x,\alpha_{23}) \big] + \log \big[ q(z,\alpha_{13}) 
  \big] \Big) \right] \, , \nonumber \\
  \Delta\mathcal{G}_{\big[ w^{(1,0)}, [ w^{(1,-1)}, w^{(1,0)}] \big],i}
  \, & = & \, 
  -\frac{1}{12} \, \log \big[ q(x,\alpha_{23}) \big] 
  \Big( \log \big[ q(y,\alpha_{23}) \big] - \log \big[ q(z,\alpha_{13}) \big] \Big) 
  \, , \nonumber \\
  \Delta\mathcal{G}_{\big[ w^{(1,-1)}, [ w^{(1,1)}, w^{(1,-1)}] \big],i} 
  \, & = & \,
 - \frac{1}{24} \, \Big( \log^2 \big[ q(z,\alpha_{13}) \big]
  - \log^2 \big[ q(x,\alpha_{23}) \big] \Big) \, . 
\eeqa
Finally, there are contributions to the fully antisymmetric colour factor $c_4^{(3)}$, given by
\beqa
\label{comms3123}
  \Delta\mathcal{G}_{\left[ w^{(1,0)}, w^{(2,-1)} \right],4}
  \, & = & \,  \,  \, \theta(x - y) \,
  \log \left( \frac{y}{x} \right) \log \big[ q(z,\alpha_{13}) \big] 
  \, ,  \\
  \Delta\mathcal{G}_{\left[ w^{(2,0)}, w^{(1,-1)} \right],4}
  \, & = & \,  \, 
  - \theta(x - y) \bigg[ 4 \, \dilog{y}{x} + \log^2 \left( \frac{y}{x} \right)
  + 2 \zeta(2) \nonumber \\
  & & \hspace{6mm} - \, \log \left( \frac{x}{y} \right) 
  \Big( \log \big[ q(x,\alpha_{23}) \big] + \log \big[ q(y,\alpha_{23}) \big] \Big)
  \bigg] \, . \nonumber
\eeqa
This colour factor had no commutator contributions in the (2,2,2) case, as discussed
above. In this case they can be generated due to the fact that one of the lower order 
webs contains a crossed gluon pair. Combining, as prescribed by \eqn{websub},
the non-subtracted web and the commutator counterterms, one finds again that the 
integrated results can be expressed using the basis functions of \eqn{eq:Mbasis}, 
and they are given in \eqn{results123b}. Finally, as for the (2,2,2) web, we have 
verified the cancellation of the double pole, and also the Abelian sum rule given
in \eqn{Abelian_sum}.


\section{Calculation of the (1,2,2,2,1) web}
\label{Calc12221}


\subsection{Unsubtracted web}
\label{Unsubtra12221}

In this appendix we write the integrand $\phi^{(4)}_X$ for each diagram $X$ of 
the (1,2,2,2,1) unsubtracted web. We begin by noting that the diagrams of 
Fig.~\ref{fig:12221} are pairwise related by a symmetry under permutations 
mapping a clockwise orientation into an anticlockwise one, thus swapping lines 
$1 \leftrightarrow 5$ along with $2 \leftrightarrow 4$. Because of this symmetry 
argument, we need to report only four out of the eight diagram kernels, 
while the remaining ones can be obtained through
\beq
  \phi_X^{(4)}(x_1,x_2,x_3,x_4) \, = \, \phi_{X^{\prime}}^{(4)}(1 - x_4,1 - x_3,1 - x_2,1 - x_1) \, . 
\label{symm12221}
\eeq
where $X$ and $X^{\prime}$ are any two diagrams related to each other by the 
symmetry. In order to express the results in compact form, it is useful to define the 
function
\beq
  \mathcal{L}_3 (a,  b, -x) \, = \, a \, b \, \Big( \log(x) \, \text{Li}_2 (- x) - 
  \text{Li}_3 (- x) \Big) \, + \, b^2 \, S_{1,2} (- x) \, ,
\label{L3def}
\eeq
which arises in the expansion of integrals of the form
\beqa
  I (m, n; a, \epsilon) & \equiv & \int_0^a d \zeta \, \zeta^{- 1 +  m \epsilon} \,
  (1 + \zeta)^{- n \epsilon} \nonumber  \\ 
  & = & \frac{a^{m \epsilon}}{m \epsilon} + n \epsilon \, \text{Li}_2 (- a) + 
  \epsilon^2 \mathcal{L}_3 (m,  n, - a) + {\cal O} \left( \epsilon^3 \right) \, ,
\eeqa
as well as the integral
\beqa
  \mathcal{I}_{3} (a,b) & = & \int_0^a \, \frac{d \zeta}{\zeta} \,
  \bigg[ \text{Li}_2 \big(- b (1 + \zeta) \big) - \text{Li}_2 \left(- b \right) \bigg] 
  \, = \, \int_0^b \frac{d \zeta}{\zeta} \, \, \text{Li}_2 \left(- b \, \frac{\zeta}{1 + \zeta} 
  \right) \nonumber \\
  & = & \log(1+ b) \, \text{Li}_2 (- a) - G \left(0, - 1, - \frac{1 + b}{b}; a \right) \, ,
\eeqa
where $G(a_1, \dots, a_n; z)$ is the generalised polylogarithm defined by the 
iterated integral \cite{Duhr:2012fh}
\beq
  G(a_1, \dots, a_n; z) \, = \, \int_0^z \frac{dt}{t - a_1} G(a_2, \dots, a_n; t) \, ,
\label{GHPL}
\eeq
with $G(z) = 1$ and $a_i$, $z \in \mathbb{C}$. Using these definitions, the results 
for the first four diagrams in Fig.~\ref{fig:12221} are given by
\beqa
\label{12221a}
  \phi^{(4)}_A (x_i; \epsilon) & = & \frac{1}{48 \epsilon^3} 
  \left( \frac{x_4}{1 - x_3} \right)^{6 \epsilon} 
  \left( \frac{x_3}{1 - x_2} \right)^{4 \epsilon}
  \left( \frac{x_2}{1 - x_1} \right)^{2 \epsilon} \nonumber \\
  && + \, \frac{1}{\epsilon} 
  \left( \frac{x_3}{1 - x_2} \right)^{4 \epsilon}
  \left( \frac{x_2}{1 - x_1} \right)^{2 \epsilon}
  \text{Li}_2 \left( - \frac{x_4}{1 - x_3} \right)   \\
  & & + \, 4 \, \mathcal{I}_3 \left( \frac{x_3}{1 - x_2}, \frac{x_4}{1 - x_3} \right) + 
  \frac{1}{8} \, \mathcal{L}_3 \left( 6, 8, - \frac{x_4}{1 - x_3} \right) \, , \nonumber
\eeqa
\beqa
\label{12221b} 
  \phi^{(4)}_B (x_i; \epsilon) & = & - \frac{1}{48 \epsilon^3}
  \left( \frac{1 - x_3}{x_4} \right)^{2 \epsilon}
  \left( \frac{1 - x_2}{x_3} \right)^{4 \epsilon}
  \left( \frac{1 - x_1}{x_2} \right)^{6 \epsilon} + 
  \frac{1}{12 \epsilon^3}
  \left( \frac{1 - x_2}{x_3} \right)^{2 \epsilon}
  \left( \frac{1 - x_1}{x_2} \right)^{4 \epsilon} \nonumber \\
  & & + \, \frac{2}{\epsilon}
  \left( \frac{1 - x_2}{x_3} \right)^{2 \epsilon}
  \text{Li}_2 \left( - \frac{1 - x_1}{x_2} \right) - 
  \frac{\zeta(2)}{\epsilon} \left( \frac{1 - x_2}{x_3} \right)^{2 \epsilon}
  \left( \frac{1 - x_1}{x_2} \right)^{4 \epsilon} \\
  & & -  \, \frac{1}{\epsilon}
  \left( \frac{1 - x_3}{x_4} \right)^{2 \epsilon}
  \left( \frac{1 - x_2}{x_3} \right)^{4 \epsilon}\text{Li}_2\left(-\frac{1-x_1}{x_2}\right)
  + \frac{1}{3} \mathcal{L}_3 \left(4, 6, - \frac{1 - x_1}{x_2} \right) \nonumber \\ && - 
  \frac{1}{8} \mathcal{L}_3 \left(6, 8, - \frac{1 - x_1}{x_2} \right)+8\zeta(3) \, , \nonumber 
\eeqa
\beqa
  \phi^{(4)}_D (x_i; \epsilon) & = & \frac{1}{6 \epsilon^3}
  \left( \frac{1 - x_2}{x_3} \right)^{2 \epsilon}
  \left( \frac{1 - x_1}{x_2} \right)^{- 2 \epsilon} - 
  \frac{1}{16 \epsilon^3}
  \left( \frac{1 - x_3}{x_4} \right)^{2 \epsilon} 
  \left( \frac{1 - x_2}{x_3} \right)^{4 \epsilon} 
  \left( \frac{1 - x_1}{x_2} \right)^{- 2 \epsilon} \nonumber \\ 
  & & \hspace{-1cm} + \, \frac{2}{\epsilon} 
  \left( \frac{1 - x_1}{x_2} \right)^{- 2 \epsilon} 
  \text{Li}_2 \left( - \frac{1 - x_2}{x_3} \right) + 
  \frac{2}{\epsilon} 
  \left( \frac{1 - x_2}{x_3} \right)^{2 \epsilon} 
  \text{Li}_2 \left( - \frac{x_2}{1 - x_1} \right) \nonumber \\
  & & \hspace{-1cm} - \, \frac{2}{\epsilon} 
  \left( \frac{1 - x_3}{x_4} \right)^{2 \epsilon} 
  \left( \frac{1 - x_1}{x_2} \right)^{ - 2 \epsilon} 
  \text{Li}_2 \left( - \frac{1 - x_2}{x_3} \right) -
  \frac{2 \zeta(2)}{\epsilon} 
  \left( \frac{1 - x_2}{x_3} \right)^{2 \epsilon} 
  \left( \frac{1 - x_1}{x_2} \right)^{- 2 \epsilon} \nonumber \\
  & & \hspace{-1cm} - \, \frac{1}{\epsilon}
  \left( \frac{1 - x_3}{x_4} \right)^{2 \epsilon} 
  \left( \frac{1 - x_2}{x_3} \right)^{4 \epsilon} 
  \text{Li}_2 \left( - \frac{x_2}{1 - x_1} \right) \nonumber \\
  & & \hspace{-1cm} + \, \frac{1}{3} \, \mathcal{L}_3 \left(2, 6, - \frac{x_2}{1 - x_1} \right)
  + \frac{1}{3} \, \mathcal{L}_3 \left(2, 6, - \frac{1 - x_2}{x_3} \right) - 
  \frac{1}{4} \mathcal{L}_3 \left( 4, 8, \frac{1 - x_2}{x_3} \right) \nonumber \\
  & & \hspace{-1cm} -  \, \frac{1}{8} \, \mathcal{L}_3 \left( 2, 8, \frac{x_2}{1 - x_1} \right)
  - 4 \, \mathcal{I}_3 \left( \frac{1 - x_3}{x_4}, \frac{1 - x_2}{x_3} \right) + 16 \zeta(3 ) \, ,
\eeqa
\beqa
  \phi^{(4)}_E (x_i; \epsilon) & = & \frac{1}{16 \epsilon ^3}
  \left( \frac{x_2}{1 - x_1} \right)^{2 \epsilon } 
  \left( \frac{x_3}{1 - x_2} \right)^{4 \epsilon } 
  \left( \frac{1 - x_3}{x_4} \right)^{2 \epsilon } \nonumber \\ 
  & & + \,  \frac{2}{\epsilon }
  \left( \frac{x_2}{1 - x_1} \right)^{2 \epsilon } 
  \left( \frac{1 - x_3}{x_4} \right)^{2 \epsilon } 
  \text{Li}_2 \left( - \frac{x_3}{1 - x_2} \right) \nonumber \\ 
  & & + \, \frac{1}{\epsilon }
  \left( \frac{x_2}{1 - x_1} \right)^{2 \epsilon } 
  \left( \frac{x_3}{1 - x_2} \right)^{4 \epsilon } 
  \text{Li}_2 \left( - \frac{1 - x_3}{x_4} \right) \nonumber \\
  & & - \, 4 \, \text{Li}_{1, 2} \left( - \frac{1 - x_3}{x_4} \right) - 
  4 \log \left( \frac{x_3}{1 - x_2} \right) \text{Li}_2 
  \left( - \frac{1 - x_3}{x_4} \right) \nonumber \\
& & + \, \frac{1}{8} \mathcal{L}_3 \left(2, 8, - \frac{1 - x_3}{x_4} \right)
  + \frac{1}{4} \mathcal{L}_3 \left(4, 8, - \frac{x_3}{1 - x_2} \right) \\
  & & + \, 4 \, \mathcal{I}_3 \left( \frac{x_2}{1 - x_1}, \frac{x_3}{1 - x_2} \right)
  - \, 4 \, \mathcal{I}_3 \left( \frac{1 - x_3}{x_4}, \frac{1 - x_2}{x_3} \right) \nonumber \, ,
\eeqa 
where, as above, these expressions must be expanded to order ${\cal O} \left(
\epsilon^0 \right)$. The remaining four diagrams can be obtained using symmetry, as
\beqa
  \phi^{(4)}_C (x_1, x_2, x_3, x_4; \epsilon) & = & 
  \phi^{(4)}_B (1 - x_4, 1 - x_3, 1 - x_2, 1 - x_1; \epsilon) \, , \nonumber \\
  \phi^{(4)}_F (x_1, x_2, x_3, x_4; \epsilon) & = & 
  \phi^{(4)}_A (1 - x_4, 1 - x_3, 1 - x_2, 1 - x_1; \epsilon) \, , \\ 
  \phi^{(4)}_G (x_1, x_2, x_3, x_4; \epsilon) & = & 
  \phi^{(4)}_D (1 - x_4, 1 - x_3, 1 - x_2, 1 - x_1; \epsilon) \, , \nonumber \\ 
  \phi^{(4)}_H (x_1, x_2, x_3, x_4; \epsilon) & = & 
  \phi^{(4)}_E (1 - x_4, 1 - x_3, 1 - x_2, 1 - x_1; \epsilon) \, .  \nonumber
\eeqa


\subsection{Subtracted web}
\label{Subtra12221}

The subtracted (1,2,2,2,1) web involves a sum of commutators of lower-order
webs, comprising subdiagrams of the original unsubtracted web, and given in
\eqn{Gamres}. The relevant webs are the (1,1) one-loop web, which is needed
to ${\cal O} (\epsilon^2)$ and can be taken from Ref.~\cite{Gardi:2013saa}, the 
(1,2,1) two-loop web, which is needed to ${\cal O} (\epsilon)$, and the (1,2,2,1)
three-loop web, which is needed to ${\cal O} (\epsilon^0)$. The commutators of
these webs have precisely the same colour structure $c^{(4)}_1$ as the 
non-subtracted (1,2,2,2,1) three-loop web. In order to complete the calculation, 
we list here the kernels for the (1,2,2,1) and the (1,2,1) webs, in a form which
is appropriate to be expanded to the relevant order. 

The non-subtracted (1,2,2,1) web~\cite{Gardi:2013saa} is given by the combination
\beq
  \phi_{(1,2,2,1)}^{(3)} \, = \, \frac{1}{6} \bigg( \phi_A^{(3)} - 2 \phi_B^{(3)}
  - 2 \phi_C^{(3)} + \phi_D^{(3)} \bigg) \, ,
\label{comb1221}
\eeq
where the diagrams are labelled as in Fig.~\ref{fig:1221}. The expansions of the 
kinematic integrands $\phi_X^{(3)}$ up to ${\cal O} (\epsilon)$ can be obtained from
\beqa
  \phi_A^{(3)} (x_i; \epsilon) & = &  B (2 \epsilon, 4 \epsilon) \, 
  \Bigg[ B (2 \epsilon, 2 \epsilon) - \frac{1}{2 \epsilon}
  \left( \frac{1 - x_2}{x_3} \right)^{2 \epsilon} \nonumber \\
  & & \hspace{1cm} - \, 4 \, \epsilon \, \text{Li}_2 \left( - \frac{1 - x_2}{x_3} \right)
  - \epsilon^2 \mathcal{L}_3 \left(2, 4, - \frac{1 - x_2}{x_3} \right) \Bigg] \nonumber \\
  & &  \hspace{1cm} \, - \, \frac{1}{2 \epsilon}
  \left( \frac{x_1}{x_2} \right)^{2 \epsilon}
  \Bigg[ B(2 \epsilon, 4 \epsilon) - \frac{1}{2 \epsilon} 
  \left( \frac{1 - x_2}{x_3} \right)^{2 \epsilon} \nonumber \\
  & &  \hspace{1cm} - \, 6 \, \epsilon \, \text{Li}_2 \left( - \frac{1 - x_2}{x_3} \right)
  - \epsilon^2 \mathcal{L}_3 \left(2, 6, - \frac{1 - x_2}{x_3} \right) \Bigg] \nonumber \\
  & &  \hspace{1cm} - \, 6 \, \epsilon \, 
  \mathcal{I}_3 \left( \frac{x_2}{x_1}, \frac{x_3}{1 - x_2} \right) \, ,
\eeqa
\beqa
  \phi_B^{(3)} (x_i; \epsilon) & = & \frac{1}{2 \epsilon} 
  \left( \frac{x_2}{x_1} \right)^{2 \epsilon}
  \Bigg[ B(2 \epsilon, 4 \epsilon) - \frac{1}{2 \epsilon}
  \left( \frac{1 - x_2}{x_3} \right)^{2 \epsilon} \nonumber \\
  & & \hspace{1cm} - \, 6 \, \epsilon \, \text{Li}_2 \left( - \frac{1 - x_2}{x_3} \right) - 
  \epsilon^2 \mathcal{L}_3 \left( - \frac{1 - x_2}{x_3} \right) \Bigg] \nonumber \\
  & & \hspace{1cm} + \, 6 \, \epsilon \, 
  \mathcal{I}_3 \left( \frac{x_2}{x_1}, \frac{x_3}{1 - x_2} \right) \, ,
\eeqa
\beqa
  \phi_D^{(3)} (x_i; \epsilon) & = & \frac{1}{2 \epsilon}
  \left( \frac{x_2}{x_1} \right)^{2 \epsilon} \Bigg[ \frac{1}{2 \epsilon} 
  \left( \frac{1 - x_2}{x_3} \right)^{2 \epsilon} + 6 \epsilon \,
  \text{Li}_2 \left( - \frac{1 - x_2}{x_3} \right)
  + \epsilon^2 \mathcal{L}_3 \left(2, 6, - \frac{1 - x_2}{x_3} \right) \Bigg] \nonumber \\
  & & \hspace{-15mm} + \, 3 \, \text{Li}_2 \left(- \frac{x_2}{x_1} \right) - 6 \epsilon \,
  \Bigg[ \mathcal{I}_3 \left(\frac{x_2}{x_1},\frac{x_3}{1 - x_2}\right) + 
  \text{Li}_{1,2} \left( - \frac{x_2}{x_1} \right) \Bigg] + 
  \frac{\epsilon}{2} \, \mathcal{L}_3 \left(2, 6, - \frac{x_2}{x_1} \right) .
\eeqa 
Again, we observe that diagram B and diagram C are related by the exchange of the 
gluon labels. We obtain diagram C by means of the relation
\beq
  \phi_C^{(3)} (x_1, x_2, x_3; \epsilon) \, = \, \phi_B^{(3)} (x_3, 1 - x_2, x_1; \epsilon) \, . 
\label{symBC}
\eeq
We finally need the integrand of the non-subtracted two-loop (1,2,1) web. It is given by
\beqa
  \phi^{(2)}_{(1,2,1)} (x_1, x_2) & = & \frac{1}{4 \epsilon} \bigg[ 
  \left( \frac{x_1}{x_2} \right)^{2 \epsilon} - \left( \frac{x_2}{x_1} \right)^{2 \epsilon} \bigg]
  + 2 \, \epsilon \, \bigg[ \text{Li}_2 \left( - \frac{x_1}{x_2} \right) - 
  \text{Li}_2 \left( - \frac{x_2}{x_1} \right) \bigg] \nonumber \\
  & & \hspace{-4mm} + \, \, \frac{\epsilon^2}{2} \, \bigg[ \mathcal{L}_3 
  \left(2, 4, - \frac{x_1}{x_2} \right) - \mathcal{L}_3 \left(2, 4, - \frac{x_2}{x_1} 
  \right) \bigg] \, . 
\eeqa
Note that, as in all other cases we examined, both the unsubtracted webs and the 
lower-order webs entering \eqn{Gamres} are not factorised integrals, because the 
functions $\mathcal{I}_3$, $\mathcal{L}_3$ and the other polylogarithms entering 
the web depend on ratios of different integration variables $x_i$. All such functions,
however, cancel in the sum in \eqn{Gamres}, and the resulting expression for the 
integrand is factorised, as reported in the text, \eqn{Gbar12221}.


\bibliographystyle{JHEP}
\bibliography{refs.bib}

\providecommand{\href}[2]{#2}\begingroup\raggedright\begin{thebibliography}{10}

\bibitem{Catani:1998bh}
S.~Catani, {\it {The Singular behavior of QCD amplitudes at two loop order}},
  {\em Phys.Lett.} {\bf B427} (1998) 161--171,
  [\href{http://xxx.lanl.gov/abs/hep-ph/9802439}{{\tt hep-ph/9802439}}].

\bibitem{Sterman:2002qn}
G.~F. Sterman and M.~E. Tejeda-Yeomans, {\it {Multi-loop amplitudes and
  resummation}},  {\em Phys. Lett.} {\bf B552} (2003) 48--56,
  [\href{http://xxx.lanl.gov/abs/hep-ph/0210130}{{\tt hep-ph/0210130}}].

\bibitem{Aybat:2006wq}
S.~M. Aybat, L.~J. Dixon, and G.~F. Sterman, {\it {The two-loop anomalous
  dimension matrix for soft gluon exchange}},  {\em Phys. Rev. Lett.} {\bf 97}
  (2006) 072001, [\href{http://xxx.lanl.gov/abs/hep-ph/0606254}{{\tt
  hep-ph/0606254}}].

\bibitem{Aybat:2006mz}
S.~M. Aybat, L.~J. Dixon, and G.~F. Sterman, {\it {The two-loop soft anomalous
  dimension matrix and resummation at next-to-next-to leading pole}},  {\em
  Phys. Rev.} {\bf D74} (2006) 074004,
  [\href{http://xxx.lanl.gov/abs/hep-ph/0607309}{{\tt hep-ph/0607309}}].

\bibitem{Becher:2009cu}
T.~Becher and M.~Neubert, {\it {Infrared singularities of scattering amplitudes
  in perturbative QCD}},  {\em Phys. Rev. Lett.} {\bf 102} (2009) 162001,
  [\href{http://xxx.lanl.gov/abs/0901.0722}{{\tt 0901.0722}}].

\bibitem{Becher:2009qa}
T.~Becher and M.~Neubert, {\it {On the Structure of Infrared Singularities of
  Gauge-Theory Amplitudes}},  {\em JHEP} {\bf 06} (2009) 081,
  [\href{http://xxx.lanl.gov/abs/0903.1126}{{\tt 0903.1126}}].

\bibitem{Gardi:2009qi}
E.~Gardi and L.~Magnea, {\it {Factorization constraints for soft anomalous
  dimensions in QCD scattering amplitudes}},  {\em JHEP} {\bf 0903} (2009) 079,
  [\href{http://xxx.lanl.gov/abs/0901.1091}{{\tt 0901.1091}}].

\bibitem{Yennie:1961ad}
D.~R. Yennie, S.~C. Frautschi, and H.~Suura, {\it {The infrared divergence
  phenomena and high-energy processes}},  {\em Ann. Phys.} {\bf 13} (1961)
  379--452.

\bibitem{Sterman:1981jc}
G.~F. Sterman, {\it Infrared divergences in perturbative {QCD}. (talk)},  {\em
  AIP Conf. Proc.} 22--40.

\bibitem{Gatheral:1983cz}
J.~G.~M. Gatheral, {\it {Exponentiation of eikonal cross-sections in nonabelian
  gauge theories}},  {\em Phys. Lett.} {\bf B133} (1983) 90.

\bibitem{Frenkel:1984pz}
J.~Frenkel and J.~C. Taylor, {\it {Nonabelian eikonal exponentiation}},  {\em
  Nucl. Phys.} {\bf B246} (1984) 231.

\bibitem{Magnea:1990zb}
L.~Magnea and G.~F. Sterman, {\it {Analytic continuation of the Sudakov
  form-factor in QCD}},  {\em Phys. Rev.} {\bf D42} (1990) 4222--4227.

\bibitem{Magnea:2000ss}
L.~Magnea, {\it {Analytic resummation for the quark form-factor in QCD}},  {\em
  Nucl.Phys.} {\bf B593} (2001) 269--288,
  [\href{http://xxx.lanl.gov/abs/hep-ph/0006255}{{\tt hep-ph/0006255}}].

\bibitem{Gardi:2010rn}
E.~Gardi, E.~Laenen, G.~Stavenga, and C.~D. White, {\it {Webs in multiparton
  scattering using the replica trick}},  {\em JHEP} {\bf 1011} (2010) 155,
  [\href{http://xxx.lanl.gov/abs/1008.0098}{{\tt 1008.0098}}].

\bibitem{Mitov:2010rp}
A.~Mitov, G.~Sterman, and I.~Sung, {\it {Diagrammatic Exponentiation for
  Products of Wilson Lines}},  {\em Phys. Rev.} {\bf D82} (2010) 096010,
  [\href{http://xxx.lanl.gov/abs/1008.0099}{{\tt 1008.0099}}].

\bibitem{Gardi:2011wa}
E.~Gardi and C.~D. White, {\it {General properties of multiparton webs: Proofs
  from combinatorics}},  {\em JHEP} {\bf 1103} (2011) 079,
  [\href{http://xxx.lanl.gov/abs/1102.0756}{{\tt 1102.0756}}].

\bibitem{Gardi:2011yz}
E.~Gardi, J.~M. Smillie, and C.~D. White, {\it {On the renormalization of
  multiparton webs}},  {\em JHEP} {\bf 1109} (2011) 114,
  [\href{http://xxx.lanl.gov/abs/1108.1357}{{\tt 1108.1357}}].

\bibitem{Dukes:2013wa}
M.~Dukes, E.~Gardi, E.~Steingrimsson, and C.~D. White, {\it {Web worlds,
  web-colouring matrices, and web-mixing matrices}},  {\em J. Comb. Theory Ser.
  A} {\bf 120} (2013) 1012--1037,
  [\href{http://xxx.lanl.gov/abs/1301.6576}{{\tt 1301.6576}}].

\bibitem{Dukes:2013gea}
M.~Dukes, E.~Gardi, H.~McAslan, D.~J. Scott, and C.~D. White, {\it {Webs and
  Posets}},  \href{http://xxx.lanl.gov/abs/1310.3127}{{\tt 1310.3127}}.

\bibitem{Gardi:2013ita}
E.~Gardi, J.~M. Smillie, and C.~D. White, {\it {The Non-Abelian Exponentiation
  theorem for multiple Wilson lines}},
  \href{http://xxx.lanl.gov/abs/1304.7040}{{\tt 1304.7040}}.

\bibitem{Polyakov:1980ca}
A.~M. Polyakov, {\it {Gauge Fields as Rings of Glue}},  {\em Nucl. Phys.} {\bf
  B164} (1980) 171--188.

\bibitem{Arefeva:1980zd}
I.~Y. Arefeva, {\it {Quantum contour field equations}},  {\em Phys. Lett.} {\bf
  B93} (1980) 347--353.

\bibitem{Dotsenko:1979wb}
V.~S. Dotsenko and S.~N. Vergeles, {\it {Renormalizability of Phase Factors in
  the Nonabelian Gauge Theory}},  {\em Nucl. Phys.} {\bf B169} (1980) 527.

\bibitem{Brandt:1981kf}
R.~A. Brandt, F.~Neri, and M.-a. Sato, {\it {Renormalization of Loop Functions
  for All Loops}},  {\em Phys. Rev.} {\bf D24} (1981) 879.

\bibitem{Korchemsky:1985xj}
G.~P. Korchemsky and A.~V. Radyushkin, {\it Loop space formalism and
  renormalization group for the infrared asymptotics of {QCD}},  {\em Phys.
  Lett.} {\bf B171} (1986) 459--467.

\bibitem{Korchemsky:1985xu}
G.~Korchemsky and A.~Radyushkin, {\it Infrared asymptotics of perturbative
  {QCD}: {R}enormalization properties of the wilson loops in higher orders of
  perturbation theory},  {\em Sov. J. Nucl. Phys.} {\bf 44} (1986) 877.

\bibitem{Korchemsky:1987wg}
G.~Korchemsky and A.~Radyushkin, {\it {Renormalization of the Wilson Loops
  Beyond the Leading Order}},  {\em Nucl. Phys.} {\bf B283} (1987) 342--364.

\bibitem{Naculich:2009cv}
S.~G. Naculich and H.~J. Schnitzer, {\it {IR divergences and Regge limits of
  subleading-color contributions to the four-gluon amplitude in N=4 SYM
  Theory}},  {\em JHEP} {\bf 0910} (2009) 048,
  [\href{http://xxx.lanl.gov/abs/0907.1895}{{\tt 0907.1895}}].

\bibitem{Naculich:2011pd}
S.~G. Naculich, H.~Nastase, and H.~J. Schnitzer, {\it {Applications of
  Subleading Color Amplitudes in N=4 SYM Theory}},  {\em Adv. High Energy
  Phys.} {\bf 2011} (2011) 190587,
  [\href{http://xxx.lanl.gov/abs/1105.3718}{{\tt 1105.3718}}].

\bibitem{Naculich:2013xa}
S.~G. Naculich, H.~Nastase, and H.~J. Schnitzer, {\it {All-loop
  infrared-divergent behavior of most-subleading-color gauge-theory
  amplitudes}},  {\em JHEP} {\bf 1304} (2013) 114,
  [\href{http://xxx.lanl.gov/abs/1301.2234}{{\tt 1301.2234}}].

\bibitem{Alday:2007hr}
L.~F. Alday and J.~M. Maldacena, {\it {Gluon scattering amplitudes at strong
  coupling}},  {\em JHEP} {\bf 0706} (2007) 064,
  [\href{http://xxx.lanl.gov/abs/0705.0303}{{\tt 0705.0303}}].

\bibitem{Basso:2007wd}
B.~Basso, G.~P. Korchemsky, and J.~Kotanski, {\it {Cusp anomalous dimension in
  maximally supersymmetric Yang- Mills theory at strong coupling}},  {\em Phys.
  Rev. Lett.} {\bf 100} (2008) 091601,
  [\href{http://xxx.lanl.gov/abs/0708.3933}{{\tt 0708.3933}}].

\bibitem{Correa:2012nk}
D.~Correa, J.~Henn, J.~Maldacena, and A.~Sever, {\it {The cusp anomalous
  dimension at three loops and beyond}},  {\em JHEP} {\bf 1205} (2012) 098,
  [\href{http://xxx.lanl.gov/abs/1203.1019}{{\tt 1203.1019}}].

\bibitem{Henn:2012qz}
J.~M. Henn and T.~Huber, {\it {Systematics of the cusp anomalous dimension}},
  {\em JHEP} {\bf 1211} (2012) 058,
  [\href{http://xxx.lanl.gov/abs/1207.2161}{{\tt 1207.2161}}].

\bibitem{Correa:2012at}
D.~Correa, J.~Henn, J.~Maldacena, and A.~Sever, {\it {An exact formula for the
  radiation of a moving quark in N=4 super Yang Mills}},  {\em JHEP} {\bf 1206}
  (2012) 048, [\href{http://xxx.lanl.gov/abs/1202.4455}{{\tt 1202.4455}}].

\bibitem{Henn:2013wfa}
J.~M. Henn and T.~Huber, {\it {The four-loop cusp anomalous dimension in N=4
  super Yang-Mills and analytic integration techniques for Wilson line
  integrals}},  \href{http://xxx.lanl.gov/abs/1304.6418}{{\tt 1304.6418}}.

\bibitem{Erdogan:2011yc}
O.~Erdogan and G.~Sterman, {\it {Gauge Theory Webs and Surfaces}},
  \href{http://xxx.lanl.gov/abs/1112.4564}{{\tt 1112.4564}}.

\bibitem{Cherednikov:2012yd}
I.~Cherednikov, T.~Mertens, and F.~Van~der Veken, {\it {Evolution of cusped
  light-like Wilson loops and geometry of the loop space}},  {\em Phys. Rev.}
  {\bf D86} (2012) 085035, [\href{http://xxx.lanl.gov/abs/1208.1631}{{\tt
  1208.1631}}].

\bibitem{Cherednikov:2012qq}
I.~Cherednikov, T.~Mertens, and F.~Van~der Veken, {\it {Cusped light-like
  Wilson loops in gauge theories}},  {\em Phys. Part. Nucl.} {\bf 44} (2013)
  250--259, [\href{http://xxx.lanl.gov/abs/1210.1767}{{\tt 1210.1767}}].

\bibitem{Korchemsky:1992xv}
G.~P. Korchemsky and G.~Marchesini, {\it {Structure function for large x and
  renormalization of Wilson loop}},  {\em Nucl. Phys.} {\bf B406} (1993)
  225--258, [\href{http://xxx.lanl.gov/abs/hep-ph/9210281}{{\tt
  hep-ph/9210281}}].

\bibitem{Korchemsky:1993uz}
G.~P. Korchemsky and G.~Marchesini, {\it {Resummation of large infrared
  corrections using Wilson loops}},  {\em Phys. Lett.} {\bf B313} (1993)
  433--440.

\bibitem{Korchemskaya:1994qp}
I.~Korchemskaya and G.~Korchemsky, {\it {High-energy scattering in QCD and
  cross singularities of Wilson loops}},  {\em Nucl. Phys.} {\bf B437} (1995)
  127--162, [\href{http://xxx.lanl.gov/abs/hep-ph/9409446}{{\tt
  hep-ph/9409446}}].

\bibitem{Korchemskaya:1992je}
I.~Korchemskaya and G.~Korchemsky, {\it {On lightlike Wilson loops}},  {\em
  Phys. Lett.} {\bf B287} (1992) 169--175.

\bibitem{Akhoury:2011kq}
R.~Akhoury, R.~Saotome, and G.~Sterman, {\it {Collinear and Soft Divergences in
  Perturbative Quantum Gravity}},  {\em Phys. Rev.} {\bf D84} (2011) 104040,
  [\href{http://xxx.lanl.gov/abs/1109.0270}{{\tt 1109.0270}}].

\bibitem{Naculich:2011ry}
S.~G. Naculich and H.~J. Schnitzer, {\it {Eikonal methods applied to
  gravitational scattering amplitudes}},  {\em JHEP} {\bf 1105} (2011) 087,
  [\href{http://xxx.lanl.gov/abs/1101.1524}{{\tt 1101.1524}}].

\bibitem{Miller:2012an}
D.~Miller and C.~White, {\it {The Gravitational cusp anomalous dimension from
  AdS space}},  {\em Phys. Rev.} {\bf D85} (2012) 104034,
  [\href{http://xxx.lanl.gov/abs/1201.2358}{{\tt 1201.2358}}].

\bibitem{White:2011yy}
C.~D. White, {\it {Factorization Properties of Soft Graviton Amplitudes}},
  {\em JHEP} {\bf 1105} (2011) 060,
  [\href{http://xxx.lanl.gov/abs/1103.2981}{{\tt 1103.2981}}].

\bibitem{Melville:2013qca}
S.~Melville, S.~Naculich, H.~Schnitzer, and C.~White, {\it {Wilson line
  approach to gravity in the high energy limit}},
  \href{http://xxx.lanl.gov/abs/1306.6019}{{\tt 1306.6019}}.

\bibitem{Kidonakis:2009ev}
N.~Kidonakis, {\it {Two-loop soft anomalous dimensions and NNLL resummation for
  heavy quark production}},  {\em Phys. Rev. Lett.} {\bf 102} (2009) 232003,
  [\href{http://xxx.lanl.gov/abs/0903.2561}{{\tt 0903.2561}}].

\bibitem{Grozin:2014axa}
A.~Grozin, J.~M. Henn, G.~P. Korchemsky, and P.~Marquard, {\it {The $n_{f}$
  terms of the QCD cusp anomalous dimension}},
  \href{http://xxx.lanl.gov/abs/1406.7828}{{\tt 1406.7828}}.

\bibitem{Ferroglia:2009ep}
A.~Ferroglia, M.~Neubert, B.~D. Pecjak, and L.~L. Yang, {\it {Two-loop
  divergences of scattering amplitudes with massive partons}},  {\em Phys. Rev.
  Lett.} {\bf 103} (2009) 201601,
  [\href{http://xxx.lanl.gov/abs/0907.4791}{{\tt 0907.4791}}].

\bibitem{Ferroglia:2009ii}
A.~Ferroglia, M.~Neubert, B.~D. Pecjak, and L.~L. Yang, {\it {Two-loop
  divergences of massive scattering amplitudes in non-abelian gauge theories}},
   {\em JHEP} {\bf 11} (2009) 062,
  [\href{http://xxx.lanl.gov/abs/0908.3676}{{\tt 0908.3676}}].

\bibitem{Mitov:2010xw}
A.~Mitov, G.~F. Sterman, and I.~Sung, {\it {Computation of the Soft Anomalous
  Dimension Matrix in Coordinate Space}},  {\em Phys. Rev.} {\bf D82} (2010)
  034020, [\href{http://xxx.lanl.gov/abs/1005.4646}{{\tt 1005.4646}}].

\bibitem{Chien:2011wz}
Y.-T. Chien, M.~D. Schwartz, D.~Simmons-Duffin, and I.~W. Stewart, {\it {Jet
  Physics from Static Charges in AdS}},  {\em Phys. Rev.} {\bf D85} (2012)
  045010, [\href{http://xxx.lanl.gov/abs/1109.6010}{{\tt 1109.6010}}].

\bibitem{Mitov:2009sv}
A.~Mitov, G.~Sterman, and I.~Sung, {\it {The Massive Soft Anomalous Dimension
  Matrix at Two Loops}},  {\em Phys. Rev.} {\bf D79} (2009) 094015,
  [\href{http://xxx.lanl.gov/abs/0903.3241}{{\tt 0903.3241}}].

\bibitem{Becher:2009kw}
T.~Becher and M.~Neubert, {\it {Infrared singularities of QCD amplitudes with
  massive partons}},  {\em Phys. Rev.} {\bf D79} (2009) 125004,
  [\href{http://xxx.lanl.gov/abs/0904.1021}{{\tt 0904.1021}}].

\bibitem{Beneke:2009rj}
M.~Beneke, P.~Falgari, and C.~Schwinn, {\it {Soft radiation in heavy-particle
  pair production: all- order colour structure and two-loop anomalous
  dimension}},  {\em Nucl. Phys.} {\bf B828} (2010) 69--101,
  [\href{http://xxx.lanl.gov/abs/0907.1443}{{\tt 0907.1443}}].

\bibitem{Czakon:2009zw}
M.~Czakon, A.~Mitov, and G.~F. Sterman, {\it {Threshold Resummation for
  Top-Pair Hadroproduction to Next-to-Next-to-Leading Log}},  {\em Phys. Rev.}
  {\bf D80} (2009) 074017, [\href{http://xxx.lanl.gov/abs/0907.1790}{{\tt
  0907.1790}}].

\bibitem{Chiu:2009mg}
J.-y. Chiu, A.~Fuhrer, R.~Kelley, and A.~V. Manohar, {\it {Factorization
  Structure of Gauge Theory Amplitudes and Application to Hard Scattering
  Processes at the LHC}},  {\em Phys. Rev.} {\bf D80} (2009) 094013,
  [\href{http://xxx.lanl.gov/abs/0909.0012}{{\tt 0909.0012}}].

\bibitem{Ferroglia:2010mi}
A.~Ferroglia, M.~Neubert, B.~D. Pecjak, and L.~L. Yang, {\it {Infrared
  Singularities and Soft Gluon Resummation with Massive Partons}},  {\em Nucl.
  Phys. Proc. Suppl.} {\bf 205-206} (2010) 98--103,
  [\href{http://xxx.lanl.gov/abs/1006.4680}{{\tt 1006.4680}}].

\bibitem{Gardi:2013saa}
E.~Gardi, {\it {From Webs to Polylogarithms}},  {\em JHEP} {\bf 1404} (2014)
  044, [\href{http://xxx.lanl.gov/abs/1310.5268}{{\tt 1310.5268}}].

\bibitem{Dixon:2008gr}
L.~J. Dixon, L.~Magnea, and G.~Sterman, {\it {Universal structure of subleading
  infrared poles in gauge theory amplitudes}},  {\em JHEP} {\bf 08} (2008) 022,
  [\href{http://xxx.lanl.gov/abs/0805.3515}{{\tt 0805.3515}}].

\bibitem{Dixon:2009gx}
L.~J. Dixon, {\it {Matter Dependence of the Three-Loop Soft Anomalous Dimension
  Matrix}},  {\em Phys. Rev.} {\bf D79} (2009) 091501,
  [\href{http://xxx.lanl.gov/abs/0901.3414}{{\tt 0901.3414}}].

\bibitem{Dixon:2009ur}
L.~J. Dixon, E.~Gardi, and L.~Magnea, {\it {On soft singularities at three
  loops and beyond}},  {\em JHEP} {\bf 02} (2010) 081,
  [\href{http://xxx.lanl.gov/abs/0910.3653}{{\tt 0910.3653}}].

\bibitem{Gardi:2009zv}
E.~Gardi and L.~Magnea, {\it {Infrared singularities in QCD amplitudes}},  {\em
  Nuovo Cim.} {\bf 032C} (2009) 137--157,
  [\href{http://xxx.lanl.gov/abs/0908.3273}{{\tt 0908.3273}}].

\bibitem{Gehrmann:2010ue}
T.~Gehrmann, E.~Glover, T.~Huber, N.~Ikizlerli, and C.~Studerus, {\it
  {Calculation of the quark and gluon form factors to three loops in QCD}},
  {\em JHEP} {\bf 1006} (2010) 094,
  [\href{http://xxx.lanl.gov/abs/1004.3653}{{\tt 1004.3653}}].

\bibitem{Bret:2011xm}
V.~Del~Duca, C.~Duhr, E.~Gardi, L.~Magnea, and C.~D. White, {\it {An infrared
  approach to Reggeization}},  {\em Phys.Rev.} {\bf D85} (2012) 071104,
  [\href{http://xxx.lanl.gov/abs/1108.5947}{{\tt 1108.5947}}].

\bibitem{DelDuca:2011ae}
V.~Del~Duca, C.~Duhr, E.~Gardi, L.~Magnea, and C.~D. White, {\it {The Infrared
  structure of gauge theory amplitudes in the high-energy limit}},  {\em JHEP}
  {\bf 1112} (2011) 021, [\href{http://xxx.lanl.gov/abs/1109.3581}{{\tt
  1109.3581}}].

\bibitem{Ahrens:2012qz}
V.~Ahrens, M.~Neubert, and L.~Vernazza, {\it {Structure of Infrared
  Singularities of Gauge-Theory Amplitudes at Three and Four Loops}},  {\em
  JHEP} {\bf 1209} (2012) 138, [\href{http://xxx.lanl.gov/abs/1208.4847}{{\tt
  1208.4847}}].

\bibitem{Caron-Huot:2013fea}
S.~Caron-Huot, {\it {When does the gluon reggeize?}},
  \href{http://xxx.lanl.gov/abs/1309.6521}{{\tt 1309.6521}}.

\bibitem{DelDuca:2013ara}
V.~Del~Duca, G.~Falcioni, L.~Magnea, and L.~Vernazza, {\it {High-energy QCD
  amplitudes at two loops and beyond}},  {\em Phys.Lett.} {\bf B732} (2014)
  233--240, [\href{http://xxx.lanl.gov/abs/1311.0304}{{\tt 1311.0304}}].

\bibitem{Laenen:2008gt}
E.~Laenen, G.~Stavenga, and C.~D. White, {\it {Path integral approach to
  eikonal and next-to-eikonal exponentiation}},  {\em JHEP} {\bf 03} (2009)
  054, [\href{http://xxx.lanl.gov/abs/0811.2067}{{\tt 0811.2067}}].

\bibitem{Vladimirov:2014wga}
A.~Vladimirov, {\it {Generating function for web diagrams}},
  \href{http://xxx.lanl.gov/abs/1406.6253}{{\tt 1406.6253}}.

\bibitem{Gardi:2014kpa}
E.~Gardi, {\it {Infrared singularities in multi-leg scattering amplitudes}},
  \href{http://xxx.lanl.gov/abs/1407.5164}{{\tt 1407.5164}}.

\bibitem{Goncharov.A.B.:2009tja}
A.~Goncharov, {\it {A simple construction of Grassmannian polylogarithms}},
  \href{http://xxx.lanl.gov/abs/0908.2238}{{\tt 0908.2238}}.

\bibitem{Goncharov:2010jf}
A.~B. Goncharov, M.~Spradlin, C.~Vergu, and A.~Volovich, {\it {Classical
  Polylogarithms for Amplitudes and Wilson Loops}},  {\em Phys.Rev.Lett.} {\bf
  105} (2010) 151605, [\href{http://xxx.lanl.gov/abs/1006.5703}{{\tt
  1006.5703}}].

\bibitem{Duhr:2011zq}
C.~Duhr, H.~Gangl, and J.~R. Rhodes, {\it {From polygons and symbols to
  polylogarithmic functions}},  {\em JHEP} {\bf 1210} (2012) 075,
  [\href{http://xxx.lanl.gov/abs/1110.0458}{{\tt 1110.0458}}].

\bibitem{Duhr:2012fh}
C.~Duhr, {\it {Hopf algebras, coproducts and symbols: an application to Higgs
  boson amplitudes}},  {\em JHEP} {\bf 1208} (2012) 043,
  [\href{http://xxx.lanl.gov/abs/1203.0454}{{\tt 1203.0454}}].

\bibitem{Remiddi:1999ew}
E.~Remiddi and J.~Vermaseren, {\it {Harmonic polylogarithms}},  {\em
  Int.J.Mod.Phys.} {\bf A15} (2000) 725--754,
  [\href{http://xxx.lanl.gov/abs/hep-ph/9905237}{{\tt hep-ph/9905237}}].

\bibitem{Bassetto:1984ik}
A.~Bassetto, M.~Ciafaloni, and G.~Marchesini, {\it {Jet Structure and Infrared
  Sensitive Quantities in Perturbative QCD}},  {\em Phys. Rept.} {\bf 100}
  (1983) 201--272.

\bibitem{Catani:1996vz}
S.~Catani and M.~H. Seymour, {\it {A general algorithm for calculating jet
  cross sections in NLO QCD}},  {\em Nucl. Phys.} {\bf B485} (1997) 291--419,
  [\href{http://xxx.lanl.gov/abs/hep-ph/9605323}{{\tt hep-ph/9605323}}].

\end{thebibliography}\endgroup
\end{document}